\documentclass[fleqn,usenatbib]{mnras}

\usepackage{newtxtext,newtxmath}

\usepackage[T1]{fontenc}

\DeclareRobustCommand{\VAN}[3]{#2}
\let\VANthebibliography\thebibliography
\def\thebibliography{\DeclareRobustCommand{\VAN}[3]{##3}\VANthebibliography}

\usepackage{graphicx}	
\usepackage{amsmath}	
\usepackage{float}
\usepackage{graphicx}
\usepackage{epsfig}
\usepackage{lineno}
\usepackage{url}
\usepackage{cleveref}
\usepackage[all]{hypcap}
\usepackage[frozencache,cachedir=_minted-exoplasim_mnras_arxiv]{minted}

\newcommand*\mean[1]{\overline{#1}}

\title[ExoPlaSim]{ExoPlaSim: Extending the Planet Simulator for Exoplanets}

\author[Paradise et al.]{
Adiv Paradise,$^{1}$\thanks{E-mail: paradise@astro.utoronto.ca}
Evelyn Macdonald,$^{2}$
Kristen Menou,$^{1,2,3}$
Christopher Lee,$^{2}$
and Bo Lin Fan$^{1}$
\\
$^{1}$David A. Dunlap Department of Astronomy \& Astrophysics, University of Toronto, Toronto, ON M5S 3H4, Canada\\
$^{2}$Department of Physics, University of Toronto, Toronto, ON M5S 1A7, Canada\\
$^{3}$Physics \& Astrophysics Group, Department of Physical and Environmental Sciences, University of Toronto, Scarborough, ON M1C 1A4, Canada
}

\date{Accepted XXX. Received YYY; in original form ZZZ}

\pubyear{2021}

\begin{document}
\label{firstpage}
\pagerange{\pageref{firstpage}--\pageref{lastpage}}
\maketitle

\begin{abstract}
The discovery of a large number of terrestrial exoplanets in the habitable zones of their stars, many of which are qualitatively different from Earth, has led to a growing need for fast and flexible 3D climate models, which could model such planets and explore multiple possible climate states and surface conditions. We respond to that need by creating ExoPlaSim, a modified version of the Planet Simulator (PlaSim) that is designed to be applicable to synchronously rotating terrestrial planets, planets orbiting stars with non-solar spectra, and planets with non-Earth-like surface pressures. In this paper we describe our modifications, present validation tests of ExoPlaSim's performance against other GCMs, and demonstrate its utility by performing two simple experiments involving hundreds of models. We find that ExoPlaSim agrees qualitatively with more-sophisticated GCMs such as ExoCAM, LMDG, and ROCKE-3D, falling within the ensemble distribution on multiple measures. The model is fast enough that it enables large parameter surveys with hundreds to thousands of models, potentially enabling the efficient use of a 3D climate model in retrievals of future exoplanet observations. We describe our efforts to make ExoPlaSim accessible to non-modellers, including observers, non-computational theorists, students, and educators through a new Python API and streamlined installation through \texttt{pip}, along with \href{https://exoplasim.readthedocs.io/en/latest/}{online documentation}.
\end{abstract}




\section{Introduction}\label{pysec:intro}

A significant challenge facing the exoplanet climate modeling community is that many general circulation models (GCMs) were built for Earth-like climates \citep[e.g.][]{Anderson2004,Collins2004,Fraedrich2005,Neal2010,Wordsworth2010,Mayne2014,Way2017}, but many terrestrial exoplanets in the habitable zones of their stars differ qualitatively from Earth. Observations from space and from the ground have revealed that terrestrial habitable zone planets have a range of radii, masses, compositions, and orbital properties, and occupy a wide range of stellar environments \citep[e.g.][]{Anglada-Escude2016,Cloutier2017,Gillon2017,Luger2017,Kopparapu2018,Benneke2019,Gilbert2020}. Similarly, theorists have begun exploring the ways in which the details of the planet's surface and atmosphere can qualitatively affect the climate \citep[e.g.][]{wk97,Segura2010,Abe2011,Pierrehumbert2011,Arney2016,Menou2013,Yang2013,Yang2014,Linsenmeier2015,Cullum2016,Forgan2016,Owen2016,Turbet2018,Cowan2018,Ramirez2018,Koll2019}. Terrestrial planets appear to be common \citep{Burke2015}, so we are likely to encounter much of this diversity in the observed population. However, while new techniques and new instruments are poised to begin probing the atmospheres and climates of terrestrial habitable zone planets \citep[e.g.][]{Gaidos2004,Segura2005,Kaltenegger2009,Esteves2014,Lovis2017,Snellen2017,Kempton2017,Morley2017,Rauscher2017,OMalley2019}, GCM studies of the diverse range of planets to be characterized are hampered by the computational expense and time-consuming nature of many GCMs \citep[e.g.][]{Adcroft2004,Collins2004,Wordsworth2010,Wordsworth2010a,Forget2013,Way2017}, resulting in studies that are limited in the breadth and number of models they can include \citep[e.g.][]{Fujii2017,Turbet2018,Yang2019b,Yan2020}. In addition, even the most sophisticated GCMs still show substantial disagreement on certain aspects of the climate, such as clouds \citep{Medeiros2008,Yang2014,Eyring2016,Yang2019,Fauchez2020}. This results in a sizable gap between the community's modeling capability and the need for theoretical models to underpin interpretation of observations \citep{Ramirez2018}.

Intermediate-complexity GCMs, which model the climate in 3D complete with coupled land, ocean, and ice models, but simplify certain parameterizations and processes such as radiative transfer or surface hydrology, can help to bridge this gap by producing 3D models of the climate at lower computational cost \citep{Jacob1997,Poulsen2001,Fraedrich2005,Paradise2017,Vallis2018}. In this paper, we modify one such model, PlaSim, for use with a broader range of exoplanets, including synchronously rotating planets around M dwarfs. PlaSim is an intermediate-complexity Earth GCM that includes a slab ocean, sea ice, land surface hydrology, snow, and a coupled atmosphere that includes moist processes and a simple three-band radiation scheme \citep{Fraedrich2005}. PlaSim has been used in the past to study modern Earth climate \citep{Dekker2010,Garreaud2010,Haberkorn2012,Nowajewski2018}, paleoclimate and snowball climate dynamics \citep{Lucarini2010,Boschi2013,Linsenmeier2015,Paradise2017,Paradise2019}, and synchronously rotating and slow-rotating planets \citep{Checlair2017,Abbot2018,Checlair2019b,Paradise2021}. PlaSim is a fast GCM, able to model a year of climate in under a minute of wall-time \citep{Paradise2017}. 

PlaSim uses a spectral core, and is typically used at T21 (32 latitudes and 64 longitudes) or T42 (64 latitudes and 128 longitudes) resolution, with 5 or 10 vertical layers. The top of the atmosphere can be optionally damped with a sponge layer and/or Rayleigh drag (the latter is the default). Vertical mixing is accomplished via vertical diffusion (representing unresolved turbulence), Kuo-type deep convection \citep{Kuo1965,Kuo1974}, shallow convection following \citet{Tiedke1983}, and a dry convective adjustment. The radiation scheme uses a simple two-band shortwave parameterization with gray water and ozone absorption in the red and blue bands respectively, gray cloud scattering, and a single-band longwave radiation scheme with gray absorption from water, CO$_2$, and clouds. 

This paper presents modifications to PlaSim, called ExoPlaSim, designed to accommodate diverse rotation states, surface pressures, and stellar types. We validate our model against several other GCMs through two different sets of benchmarks \citep{Yang2019,Fauchez2020,THAI2021a,THAI2021b,THAI2021c}, and demonstrate the model's utility and speed by confirming an existing result \citep{Kopparapu2013,Yang2014,Kopparapu2016} and testing a prediction from \citet{Paradise2021}. The main components of this paper are structured as follows:
\begin{enumerate}
\item A description of the modifications to PlaSim necessary to create ExoPlaSim (\autoref{pysec:methods}).
\item ExoPlaSim's computational performance (\autoref{pysec:speed}), and comparison to other GCMs (\autoref{pysec:comparison}).
\item A demonstration of the model's potential (\autoref{pysec:demonstration}) by replicating the finding that the habitable zone moves to lower fluxes around cool stars \citep{Kopparapu2013,Yang2014,Kopparapu2016}, and confirming that Rayleigh scattering is less important around cool stars \citep{Paradise2021}.
\item Description of known issues and caveats (\autoref{pysec:problems}) and future features (\autoref{pysec:plans}).
\item Description of ExoPlaSim's new Python API (\autoref{pysec:Python}).
\end{enumerate}
In Appendix \ref{pysec:appendix}, we provide more detail on how appropriate physics filters were chosen for \autoref{pysec:comparison}.

\section{Modifications to PlaSim}\label{pysec:methods}

\subsection{Synchronous and Slow Rotation}

The study of habitable exoplanet climates faces a challenging mismatch between theoretical capabilities and observational data. Most climate models have origins in studies of Earth's climate \citep[e.g.][]{Anderson2004,Collins2004,Fraedrich2005,Vallis2018}, and so are best-suited for studying the climates of exoplanets that most resemble Earth. The Sun, however, is in the galactic minority---most stars are much lower-mass \citep{Bochanski2010}. Most planetary systems that may host potentially-habitable terrestrial planets therefore orbit at much smaller separations and orbital periods than Earth does, and are thus likely to be synchronously rotating, with one side always facing the star \citep{Peale1977,Kasting1993,Dobrovolskis2009,Edson2011,Shields2016b}. The bias towards such planets is even more pronounced in the sample of observed potentially-habitable planets, as the long periods of true Earth analogs make them significantly harder to detect and characterize with our current technology\citep{Nottale2004,Jones2004,Gill2020}. The task of modeling the climates of the exoplanets being discovered therefore requires that models developed for Earth-like climates be used to model planets whose climate dynamics may differ qualitatively from Earth's \citep{HaqqMisra2016,Koll2016,Ramirez2018,Hammond2021}. One of our main goals therefore, in extending PlaSim to model a broad range of habitable exoplanets, is to be able to model synchronously rotating planets around low-mass stars. 

Implementing synchronous rotation in PlaSim is fairly straightforward; simply fixing the longitude of the Sun in the sky produces a synchronous instellation pattern. It is tempting to produce synchronous rotation by simply setting the length of the sidereal day equal to the length of the sidereal year; this would for example permit slightly eccentric synchronously rotating orbits. However, properly-defined, the solar day (which can be derived from the sidereal day and orbital period) is undefined/infinite for synchronously rotating planets \citep{Abbot2018}, and PlaSim uses the solar day as part of several timekeeping functions, requiring workarounds when the sidereal day matches the orbital period. These workarounds interfere with the correct computation of the Sun's position in the sky in the case of 1:1 rotation. We therefore simply fix the Sun in place, with a longitude that can be specified by the user at runtime. 

More broadly, PlaSim's definition and use of the solar day poses a challenge for modeling planets other than Earth, as PlaSim also uses the solar day for unit conversions related to the strength of numerical hyperdiffusion, various calendar functions, and as a shortcut to set the sidereal day and angular frequency ($\omega$). The latter shortcut is appropriate when the length of sidereal year is much longer than the sidereal day, but can produce significant inaccuracies for longer rotation periods. For example, if the rotation speed (\texttt{rotspd}) were set 10 times lower, this would produce a solar day ten times longer (which would be incorrect if the sidereal year were 10 days). This would produce a sidereal day only 9.12 times longer, and an angular frequency similarly 9.12 times faster, rather than 10 times faster. Simply changing \texttt{rotspd} via the input namelists will thus produce incorrect results for slow rotators. We have therefore reduced the use of the solar day in PlaSim's code, replacing it with a standard 24-hour conversion factor (86400 seconds) for unit conversions, and favoring the sidereal day for dynamical contexts.

Additionally, PlaSim's dynamical core uses nondimensional copies of the primary variables in grid-space. The process of reducing the physical variables into their nondimensional versions depends on the rotation rate:
\begin{align}
u' &= \frac{u\cos\phi}{r_p\omega} \\
v' &= \frac{v\cos\phi}{r_p\omega} \\
T' &= \frac{R(T-T_0)}{(r_p\omega)^2} \\
q' &= q\frac{p}{p_s}
\end{align}
Reducing the angular frequency, $\omega$, will therefore increase the magnitude of the nondimensional variables (with the exception of humidity, $q$). In the case of temperature, this effect is very large for slow rotators ($\geq$90 days). This would ordinarily not be a problem for the dynamical core, but PlaSim has stability criteria that check for nondimensional variables whose values exceed allowed limits, as a check to ensure the model remains physical. In the original version of PlaSim, these limits are hard-coded. We therefore replace the temperature limits with computed nondimensional counterparts of 0 K and 1000 K---if physical temperatures go outside these bounds, something is very wrong, and the model should abort. We note that neither extreme is within the range we expect PlaSim to model accurately; this criterion is used solely to abort obviously unphysical models, and users should still be mindful of the limitations of using an Earth-based climate model in radiative regimes far from Earth-like conditions.

\subsection{Vertical Discretization}\label{pysec:vertical}

PlaSim's default vertical discretization is mostly-linear in pressure, with a slight nonlinearity near the surface to provide near-surface resolution. PlaSim uses a sigma vertical coordinate, where $\sigma_n=p_n/p_s$, such that $p_n$ is the layer pressure and $p_s$ is the surface pressure. The interface at the bottom of each layer, $\sigma_{h,n}$, is defined such that
\begin{equation}
\sigma_{h,n} = 0.75\left(\frac{n}{N}\right)+1.75\left(\frac{n}{N}\right)^3-1.5\left(\frac{n}{N}\right)^4
\end{equation}
This discretization performs well for an Earth-like 1 bar atmosphere. However, if the surface pressure increases, so too does the pressure of every other layer in the model. This is not a problem for capturing lower-atmosphere dynamics, but for some radiative processes, optical depth from the top of the model is more important, and thus these processes do not strongly depend on surface pressure. For surface pressures significantly higher than Earth's, such as 10 bars, this can mean the top layer of the model spans an entire bar, and thus is unable to resolve these radiative processes. 

\begin{figure}
\begin{center}
\includegraphics[width=3in]{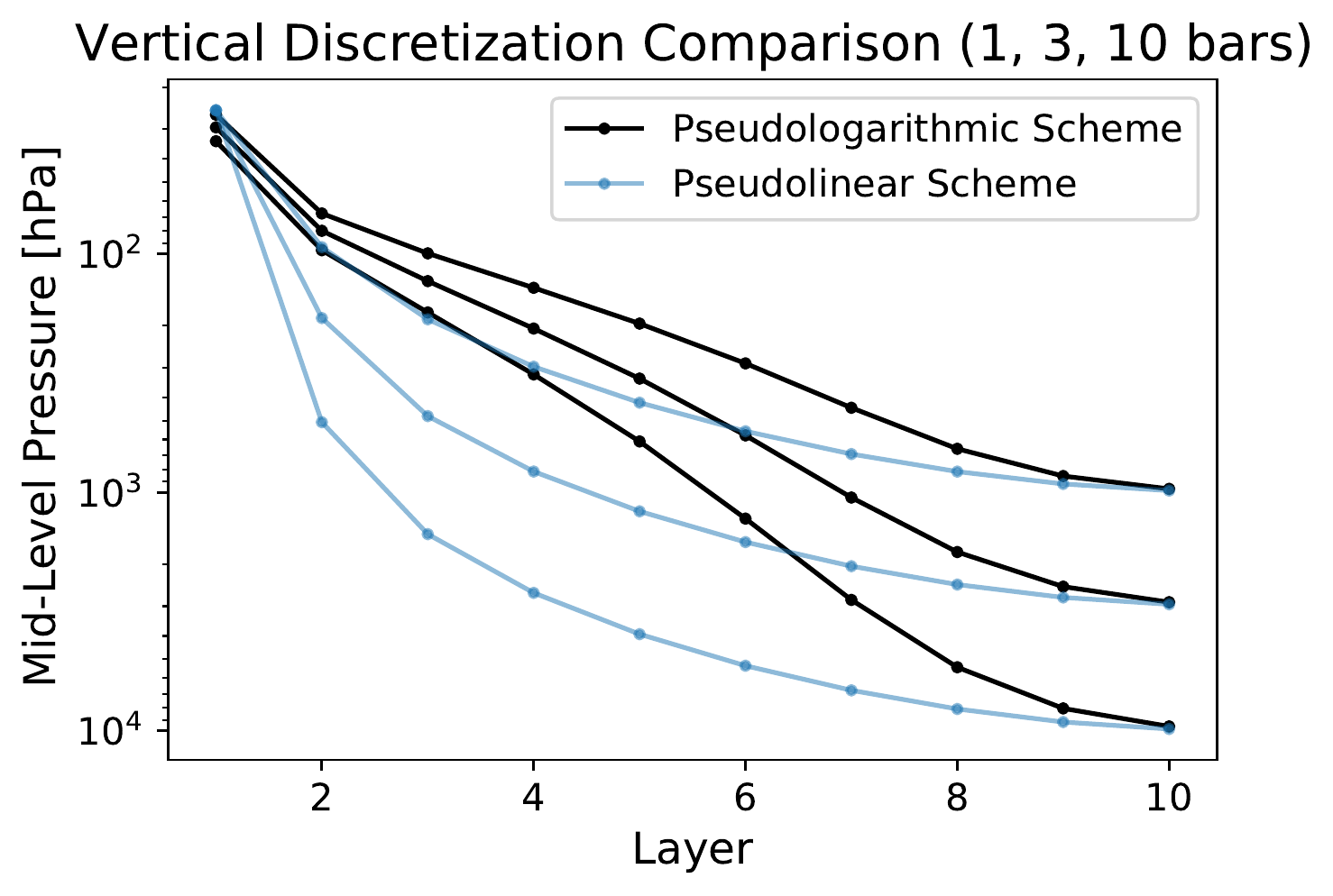}
\end{center}
\caption{Comparison of the pseudolinear and pseudologarithmic vertical discretizations, for 1 bar, 3 bars, and 10 bars, with 10 vertical layers. The pseudolinear scheme provides more resolution in the lower atmosphere, at the expense of resolution near the tropopause. The pseudologarithmic scheme provides more resolution near the tropopause. Both can be pinned to a specified model top to ensure that the entire troposphere is resolved, and both have heightened near-surface resolution compared to purely linear or purely logarithmic discretizations.}\label{pyfig:vertical}
\end{figure}

As described in \citep{Paradise2021}, we therefore introduce additional optional vertical discretizations, including one that mimics the default, but re-scaled so that the top layer's mid-level pressure is pinned to a user-prescribed pressure. This ``pseudolinear" discretization is our recommended mode, because it is nearly identical to PlaSim's default at 1 atm surface pressures. We have also included a pseudologarithmic discretization that provides more resolution in the upper layers, where the mid-layer pressures of all but the top layer are given by:
\begin{align}
\Delta z &= \frac{0.99-\log_{10}\frac{p_\text{top}}{p_s}-1}{N-1} \\
\zeta_n &= \exp\left[-20\left(1-\frac{n-1}{N-1}\right)^2\right] \\
\sigma_n &= (1-\zeta_n)10^{\log_{10}\frac{p_\text{top}}{p_s}+(n-1)\Delta z} + \zeta_n\left(\frac{n-1}{N}\right)^\frac{1}{4}
\end{align}
In this mode, the pressure of the top layer is set to half the pressure of the layer beneath it. PlaSim makes use of both interface pressures and mid-level pressures, and the mid-level pressure must always be halfway between the interfaces, so from $\sigma_n$, we derive $\sigma_{h,n}$, with $\sigma_{h,N}=1$, and the other interfaces halfway between the other mid-levels. We then re-derive the mid-level pressures at the true pressure midpoints of each layer, so that the top and bottom layers satisfy the requirement that mid-layers be halfway between layer interfaces. The rescaled pseudolinear scheme and pseudologarithmic scheme are shown in \autoref{pyfig:vertical}, for 1 bar, 3 bar, and 10 bar atmospheres, all pinned to 50 hPa at the top. The pseudologarithmic function we have chosen converges to the pseudolinear discretization near the surface, providing near-surface resolution, and also slightly increases the resolution near the model top, to better-resolve the tropopause. We have not validated the pseudologarithmic discretization, and find that for surface pressures between 0.1 and 10 bars, the scaled pseudolinear discretization returns comparable results. Furthermore, the pseudologarithmic mode has proven to require shorter timesteps in our testing, and runs into stability problems more often.

We have also added a hybrid mode for models with more than 10 layers, such that the bottom 10 layers are pseudolinear, pinned to a prescribed upper pressure, and all layers above those are logarithmically-spaced, up to a prescribed model top. This mode is intended to add the ability to model the lower atmosphere with a discretization that has been validated, while also modeling parts of the stratosphere with a logarithmic discretization. This mode has also not been validated, and more work is needed to study ExoPlaSim's performance in regimes that demand nonlinear discretizations.

\subsection{Radiation}

The majority of the modifications to PlaSim needed to model synchronously rotating planets around M dwarfs have to do with the radiation calculation in the model. This includes changes to the energy partitioning between shortwave bands, surface albedos, and the efficiency of Rayleigh scattering. Each of these changes is described in the sections that follow.

\subsubsection{Energy partitioning}

While many past studies exploring exoplanet climate with intermediate-complexity GCMs have used a Sun-like spectrum \citep[e.g.][]{Edson2011,Kaspi2015,Checlair2017,Abbot2018,Paradise2021}, in reality, the incident spectrum varies from system to system, and can have a significant effect on the climate \citep{Kasting1993,Kopparapu2013,Kopparapu2016}. Water absorption in temperate conditions happens primarily at wavelengths greater than 1 $\mu$m \citep{Kasting1993}, the albedos of surfaces such as ice and snow have strong spectral dependencies \citep{Joshi2012,vonParis2013,Shields2013}, and the efficiency of Rayleigh scattering has a $\lambda^{-4}$ dependence \citep{Kasting1993,Rybicki2004}. This is particularly important to account for in the near-term, as the vast majority of potentially-habitable Earth-sized planets that can be characterized with the next generation of instruments orbit around low-mass stars with much cooler effective temperatures \citep{Morley2017}---and as low-mass stars make up the vast majority of stars within our galaxy \citep{Bochanski2010}, these planets are likely to make up a large fraction of Earth-sized planets in habitable zones in our galaxy as well \citep{Cloutier2018,Cloutier2020}. Extending PlaSim to model the climates of these planets therefore requires a radiation model that can account for the effects of different stellar spectra. 

PlaSim's radiation scheme is partitioned into three bands: shortwave light between 316 nm and 0.75 $\mu$m (hereafter SW1), shortwave light at wavelengths longer than 0.75 $\mu$m (hereafter SW2), and longwave thermal radiation. It is assumed that there is a source of shortwave light above the atmosphere, and that all longwave radiation is emitted by the surface and the atmosphere, such that net longwave radiation at the top of the atmosphere is equivalent to outgoing longwave radiation. Wavelength-dependent processes such as Rayleigh scattering, absorption, and reflection are parameterized as gray processes within these wavelength bands. The partitioning of incident flux between SW1 and SW2 is tuned to match the incident solar spectrum, such that the energy fraction in SW1, $Z_1$, is 51.7\%, and the fraction in SW2, $Z_2$, is 48.3\%. These wavelength bands are too broad to appropriately capture the changes in reflectivity and Rayleigh scattering efficiency at lower effective stellar temperatures through simple energy re-partitioning alone---computation at higher spectral resolution is required. Extending PlaSim's radiation scheme for non-solar incident spectra therefore requires two modifications:
\begin{enumerate}
    \item $Z_1$ and $Z_2$, the energy partitioning between SW1 and SW2, need to be recomputed to match the input spectral energy distribution. 
    \item The coefficients governing the efficiency of gray scattering, absorption, and reflection processes may need to be recomputed to account for differences in the shape of the true energy distribution within a radiation band.
\end{enumerate}

We adjust the energy partitioning, $Z_1$ and $Z_2$, by simple trapezoidal integration of the incident spectrum, dividing the incident spectrum into two logarithmically-spaced bands of 1024 wavelengths each, denoted by $\lambda_{1,k}$ and $\lambda_{2,k}$, where for example $\lambda_{1,1024}$ is the 1024\textsuperscript{th} wavelength of the blue band ($\lambda_1$), and $\lambda_{2,1}$ is the first wavelength of the red band ($\lambda_2$):
\begin{linenomath*}
\begin{equation}
Z = \int_{0.316\, \mu\text{m}}^{100\,\mu\text{m}}F_\lambda(\lambda)\,d\lambda 
\end{equation}
\begin{equation}
\begin{alignedat}{2}
  &= \sum_{k=1}^{1023}\Biggl(&&\frac{F_\lambda(\lambda_{1,k})+F_\lambda(\lambda_{1,k+1})}{2}\Delta\lambda_{1,k} \\
  & &&+\frac{F_\lambda(\lambda_{2,k})+F_\lambda(\lambda_{2,k+1})}{2}\Delta\lambda_{2,k}\Biggr) \\
  & &&+\frac{F_\lambda(\lambda_{1,1024})+F_\lambda(\lambda_{2,1})}{2}(\lambda_{2,1}-\lambda_{1,1024})
\end{alignedat}
\end{equation}
\begin{equation}
Z_1 = \frac{1}{Z}\int_{0.316\,\mu\text{m}}^{0.75\,\mu\text{m}}F_\lambda(\lambda)\,d\lambda
\end{equation}
\begin{equation}
    \begin{alignedat}{2}
&= \frac{1}{Z}\Biggl(&&\sum_{k=1}^{1023}\left[\frac{F_\lambda(\lambda_{1,k})+F_\lambda(\lambda_{1,k+1})}{2}\Delta\lambda_{1,k}\right] \\
& &&+\frac{F_\lambda(\lambda_{1,1024})+F_\lambda(\lambda_{2,1})}{2}(\lambda_{2,1}-\lambda_{1,1024})\Biggr)
    \end{alignedat}
\end{equation}
\begin{align}
Z_2 &= \frac{1}{Z}\int_{0.75\,\mu\text{m}}^{100\,\mu\text{m}}F_\lambda(\lambda)\,d\lambda \\ \nonumber \\
 &= \frac{1}{Z}\sum_{k=1}^{1023}\frac{F_\lambda(\lambda_{2,k})+F_\lambda(\lambda_{2,k+1})}{2}\Delta\lambda_{2,k}
\end{align}
\end{linenomath*}

We have implemented two ways to specify the incident stellar spectrum. The simplest option is to specify a blackbody effective temperature for the star. A stellar blackbody effective temperature of 5772 K produces $Z_1$=0.517 and $Z_2$=0.483, such that specifying the Sun's blackbody temperature produces model behavior identical to PlaSim's default behavior. Another option is to provide an actual input spectrum, with incident flux as a function of wavelength. This is provided in the form of a text file, whose path is provided to the model in a namelist parameter. If an explicit spectrum is being used, then it must be provided in both a high-resolution version, with 1024 wavelengths in each band, and a lower-resolution version with 965 wavelengths in the 0.34--14.01 $\mu$m range, which is used to compute broadband surface albedos (this range and number of wavelengths is chosen to match the spectral resolution of the reference reflectance spectra chosen for ExoPlaSim, which are described in the next section).

\subsubsection{Surface albedos}

PlaSim includes a range of surface types, including water, land, snow, sea ice, and glacial ice. In the original version of the model, each is given a prescibed reflectivity, which is tuned to roughly match Earth behavior, with a temperature dependence for snow and ice. If non-solar stellar spectra are allowed, however, then the surface reflectivity has to change as well. Snow, for example, is highly-reflective at short wavelengths, but very dark in near-infrared wavelengths \citep{Joshi2012,vonParis2013,Shields2013}. Even within the near-infrared SW2 band, the reflectance spectrum of snow has a distinct shape. Therefore, as the stellar blackbody temperature changes, not only will near-infrared absorption be of greater importance to the overall energy budget, but the reflectivity of snow in that band will change as well. Furthermore, the stellar spectrum may not be a blackbody--low-mass stars often have molecular absorption in the photosphere, which might mean less incident light in the wavelengths where water is particularly absorptive. 

\begin{figure*}
\begin{center}
\includegraphics[width=6.5in]{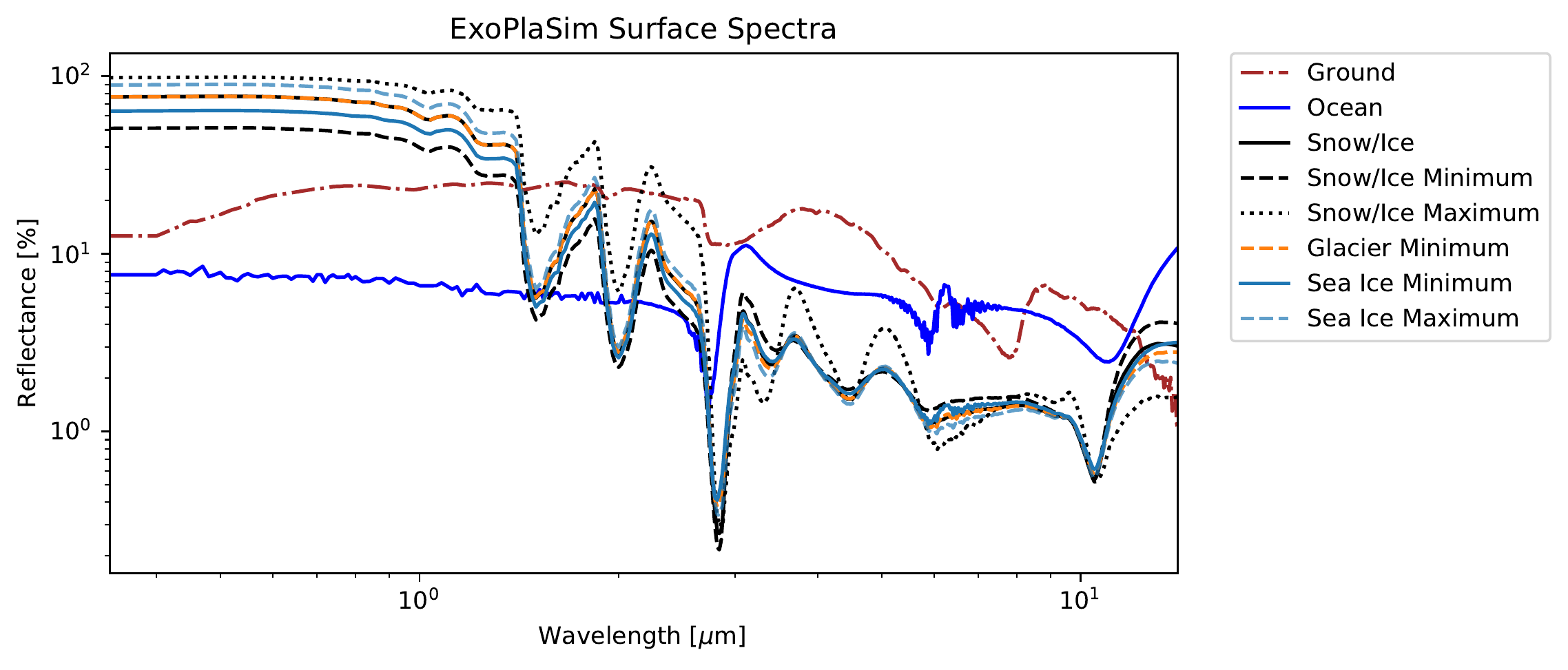}
\end{center}
\caption{Composite spectra for each of the surface types used in ExoPlaSim. Ground is a mix of fine-grained and solid andesite, basalt, and brown and yellow loam and sand, along with dune sand. Snow and ice types are mixes of snow grain sizes, frost, and clear ice, with the fraction represented by clear ice used as a free parameter to match PlaSim albedos for a 5772 K input spectrum. Spectra are taken from the JPL ECOSTRESS library \citep{aster,ecostress}.}\label{pyfig:exoplasimspectra}
\end{figure*}

To account for this, we define high-resolution spectra for each of the surface types, including the maximum and minimum reflectance scenarios for snow, sea ice, and glacial ice. These spectra are sourced from the JPL ECOSTRESS library \citep{aster,ecostress}, which provides both analytically-calculated and empirical spectra, defined out to 14 $\mu$m. For ground and surface ices, we use blends of ECOSTRESS spectra. Land is represented by a mix of fine-grained andesite (a gray mineral common in Earth's crust), solid andesite, solid basalt, brown sand, dune sand, yellow loam, and yellow sand, weighted to produce an albedo of 0.2 under a 5772 K spectrum. Most of the spectrum is comprised of fine-grained andesite, solid andesite, and basalt (43.2\%, 18\%, and 16.1\% respectively). ExoPlaSim stores multiple albedos for snow and ice, corresponding to generic snow and ice, snow and ice at very cold temperatures and at 0 $^\circ$C, glacial ice at cold and melting temperatures, and sea ice at cold and melting temperatures. The actual reflectance of the surface is allowed to be a combination of these surface types, such that snow can overlay sea ice. For each type of snow/ice reflectance, we compute a spectrum from a weighted average of coarse-grained snow, medium-grained snow, fine-grained snow, frost, and clear ice. We ignore darkening by soot, dust, and aerosols. We use the fraction of the albedo that is clear ice as a free parameter to match the spectrum to PlaSim's defaults for 5772 K. The remainder of the spectrum is comprised of 40\% medium-grained snow, and 20\% each of frost, fine-grained snow, and coarse-grained snow. Ocean reflectivity is derived by normalizing the spectrum for water to PlaSim's expected ocean albedo (the reflectivity of water in the ECOSTRESS library is only about 0.2). The resulting composite spectra used in ExoPlaSim are shown in \autoref{pyfig:exoplasimspectra}. The choices of weights and specific reference spectra used to generate these composite spectra are to a degree arbitrary, but are motivated by physical considerations--land surface is likely to represent a mix of compositions, and real snow and ice are likely to be represented by multiple grain sizes.

At the start of a model year, the stellar spectrum is convolved with the reflectance spectrum for each surface type and then integrated to produce a mean reflectivity for each of the two shortwave bands:
\begin{linenomath*}
\begin{align}
a_\text{SW1} &= \frac{\int_0^{0.75\mu\text{m}}F_\lambda(\lambda)a(\lambda)\,d\lambda}{\int_0^{0.75\mu\text{m}}F_\lambda(\lambda)\,d\lambda} \\ \nonumber \\
a_\text{SW2} &= \frac{\int_{0.75\mu\text{m}}^{2.5}F_\lambda(\lambda)a(\lambda)\,d\lambda}{\int_{0.75\mu\text{m}}^{2.5}F_\lambda(\lambda)\,d\lambda}
\end{align}
\end{linenomath*}
where $a(\lambda)$ is the reflectance spectrum of the surface in question, and $F_\lambda(\lambda)$ is the incident stellar spectrum. We do not account for Rayleigh scattering making the incident spectrum near the surface appear more blue, nor water vapor absorption aloft reducing water absorption at the surface. We include two operating modes: one in which separate albedos are used in SW1 and SW2, as defined above, and a simplified mode, in which a single bolometric albedo is computed for each surface by combining the two band albedos. Anecdotally, we have found that there is little quantitative difference in model output, and the simplified mode is less likely to crash.


\subsubsection{Rayleigh scattering}

We have made two substantial modifications to PlaSim's Rayleigh scattering parameterization. The unmodified version of PlaSim has Rayleigh scattering parameterized to match Earth's atmosphere and a solar spectrum \citep{Lacis1974,Fraedrich2005}. In ExoPlaSim, the Rayleigh scattering optical depth depends on both surface pressure and the input stellar spectrum. We described and explored the added surface pressure dependence in \citet{Paradise2021}, but revisit those modifications here. 

PlaSim's Rayleigh scattering is calculated at the model's bottom layer, just above the surface, as parameterized transmittances for the diffuse and direct beam ($T_\text{direct}$ and $T_\text{diffuse}$), the latter of which depends on the cosine of the solar zenith angle, $\mu_0$ \citep{Fraedrich2005}:
\begin{linenomath*}
\begin{align}
T_\text{direct} &= 1-\frac{0.219}{1+0.816\mu_0} \label{pyeq:plasimrayleighdir}\\ \nonumber \\
T_\text{diffuse} &= 0.856 \label{pyeq:plasimrayleighdiff}
\end{align}
\end{linenomath*}
The optical depth, $\tau=-\ln{T}$ \citep{beerlambert}, should be roughly proportional to the column mass of the scattering gas. As introduced in \citet{Paradise2021}, ExoPlaSim therefore scales the transmittance in each beam by a transmissivity scaling factor derived from the column mass, expressed here as $p_s/g$, where $p_s$ is the surface pressure and $g$ is the surface gravity:
\begin{linenomath*}
\begin{equation}\label{pyeq:exoplasimrayleigh1}
T = \exp\left[\left(\frac{p_s}{p_{s,\oplus}}\frac{g_\oplus}{g}\right)\ln{T_\oplus}\right]
\end{equation}
\end{linenomath*}

As shown in \citet{Paradise2021}, this modification produces relatively good agreement with more-sophisticated radiative codes such as SBDART \citep{Ricchiazzi1998} between 0.1--10 bars, with systematic biases that are of course larger at the ends of that range. We have not tested this modified scattering at pressures much higher than 10 bars, nor much lower than 0.1 bars, and therefore urge caution using ExoPlaSim in those parameter spaces. In particular for high-pressure models, we note that ExoPlaSim still computes all of its Rayleigh scattering in the bottom layer. In the models we tested in \citep{Paradise2021}, this shortcoming was not enough to cause qualitative deviation from more advanced codes, but this may be important especially for high-pressure hot models, where Rayleigh scattering aloft might limit the amount of light that can be absorbed by abundant water vapour in the lower troposphere, altering the energy budget of the lower troposphere. 

\begin{figure}
\begin{center}
\includegraphics[width=3in]{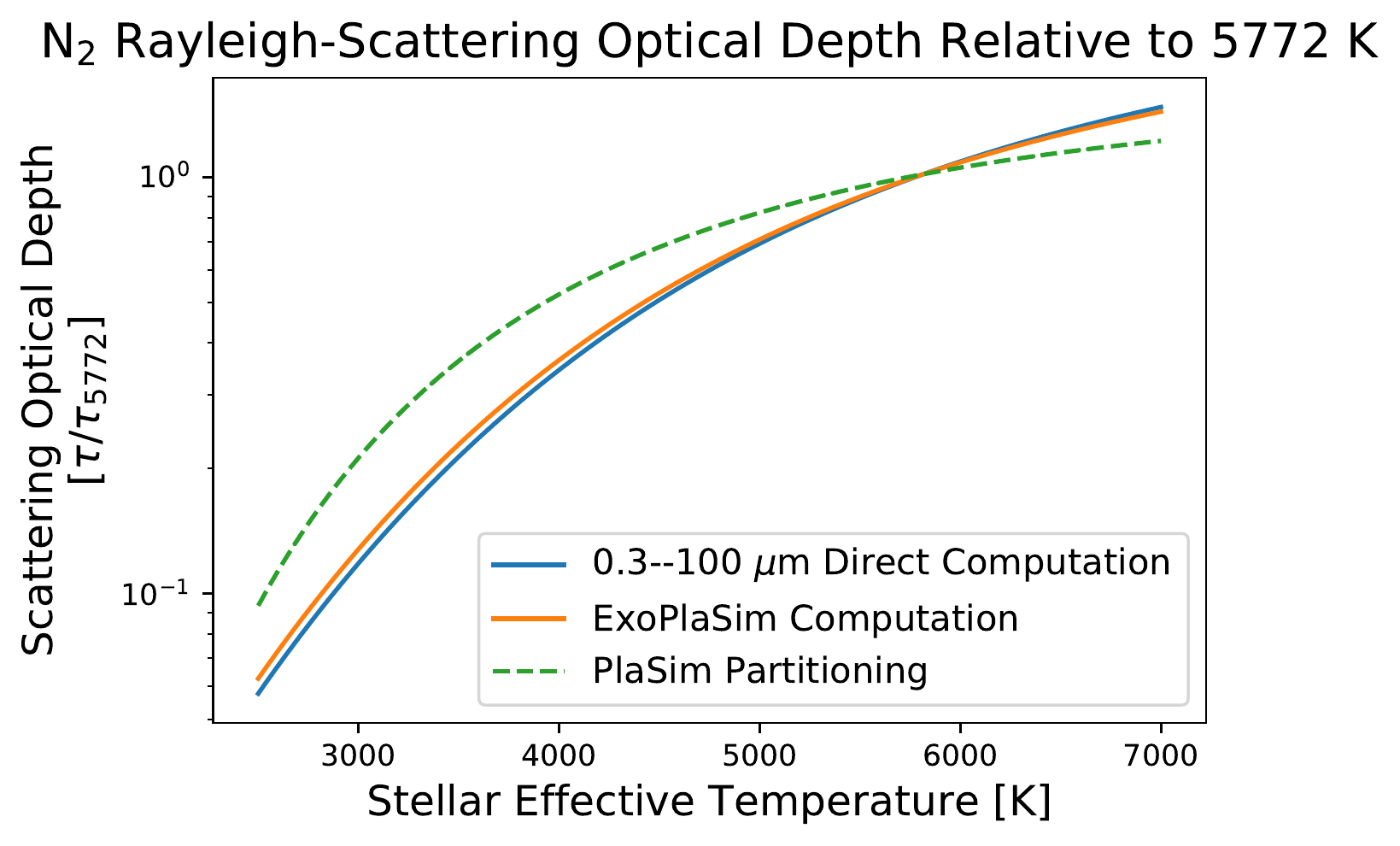}
\end{center}
\caption{Comparison of the Rayleigh scattering broadband optical depth for a 1 bar N$_2$ atmosphere as a function of stellar effective temperature, showing the true optical depth, the optical depth that is produced by simply relying on PlaSim's shortwave energy partitioning, and the optical depth produced by the implementation in ExoPlaSim. The ExoPlaSim implementation matches the analytical result closely, while the ``naive" approach of simply relying on PlaSim's energy partitioning shows significant error for cooler stars.}\label{pyfig:rayleighcomparison}
\end{figure}

We must also adjust the Rayleigh scattering optical depth to account for the stellar spectrum. PlaSim only computes Rayleigh scattering in SW1, the shortwave band blueward of 750 nm. Adjusting the energy partitioning between the two shortwave bands would therefore naturally lead to a reduction of Rayleigh scattering optical depth for redder spectra. However, the $\lambda^{-4}$ wavelength dependence of the Rayleigh scattering cross-section means that the amount of Rayleigh scattering within a band may change significantly between Sun-like stars and redder stars. The difference is shown in \autoref{pyfig:rayleighcomparison}. 

The correct optical depth is related to the scattering optical depth under a Sun-like spectrum by the ratio of mean scattering cross-sections:
\begin{linenomath*}
\begin{equation}
\tau = \frac{\mean{\sigma}}{\mean{\sigma_\oplus}}\tau_\oplus
\end{equation}
\end{linenomath*}
where $\tau_\oplus$ and $\mean{\sigma_\oplus}$ are the optical depth and mean scattering cross-section under a Sun-like spectrum, respectively, and $\mean{\sigma}$ is the mean scattering cross-section under any spectrum. With the cross-sections expanded, and only the $\lambda^{-4}$ dependence included, this yields
\begin{linenomath*}
\begin{equation}
\tau = \left(\frac{\int_0^\infty \lambda^{-4}F_\lambda(\lambda)\,d\lambda}{\int_0^\infty F_\lambda(\lambda)\,d\lambda}\right)\left(\frac{\int_0^\infty F_{\lambda,\odot}(\lambda)\,d\lambda}{\int_0^\infty \lambda^{-4}F_{\lambda,\odot}(\lambda)\,d\lambda}\right)\tau_\oplus
\end{equation}
\end{linenomath*}
where $F_\lambda(\lambda)$ is the spectrum of the star in question, and $F_{\lambda,\odot}(\lambda)$ is the Sun's spectrum. ExoPlaSim only computes Rayleigh scattering in the SW1 ($<0.75$ $\mu$m) band, however, so deriving a scaling coefficient requires that the two wavelength bands be taken into account. We can introduce a band-dependence by multiplying by 1, introducing $\lambda_2$, the wavelength at which the second band, SW2, begins (0.75 $\mu$m for ExoPlaSim):
\begin{linenomath*}
\begin{align}
\tau &= \left(\frac{\int_0^{\lambda_2} F_\lambda(\lambda)\,d\lambda}{\int_0^{\lambda_2} F_\lambda(\lambda)\,d\lambda}\right)\left(\frac{\int_0^\infty \lambda^{-4}F_\lambda(\lambda)\,d\lambda}{\int_0^\infty F_\lambda(\lambda)\,d\lambda}\right)\left(\frac{\int_0^\infty F_{\lambda,\odot}(\lambda)\,d\lambda}{\int_0^\infty \lambda^{-4}F_{\lambda,\odot}(\lambda)\,d\lambda}\right)\tau_\oplus \\ \nonumber \\
 &= \left(\frac{\int_0^{\lambda_2} F_\lambda(\lambda)\,d\lambda}{\int_0^\infty F_\lambda(\lambda)\,d\lambda}\right)\left(\frac{\int_0^\infty \lambda^{-4}F_\lambda(\lambda)\,d\lambda}{\int_0^{\lambda_2} F_\lambda(\lambda)\,d\lambda}\right)\left(\frac{\int_0^\infty F_{\lambda,\odot}(\lambda)\,d\lambda}{\int_0^\infty \lambda^{-4}F_{\lambda,\odot}(\lambda)\,d\lambda}\right)\tau_\oplus \label{pyeq:switcheroo} \\ \nonumber \\
 &= Z_1\left(\frac{\int_0^\infty \lambda^{-4}F_\lambda(\lambda)\,d\lambda}{\int_0^{\lambda_2} F_\lambda(\lambda)\,d\lambda}\right)\left(\frac{\int_0^\infty F_{\lambda,\odot}(\lambda)\,d\lambda}{\int_0^\infty \lambda^{-4}F_{\lambda,\odot}(\lambda)\,d\lambda}\right)\tau_\oplus \label{pyeq:z1sub}
\end{align}
\end{linenomath*}
Here, in \autoref{pyeq:z1sub}, we have recognized that the first term in \autoref{pyeq:switcheroo} is the definition of the SW1 energy partition, $Z_1$, or the fraction of incident flux that is within that band.

\begin{figure}
\begin{center}
\includegraphics[width=3in]{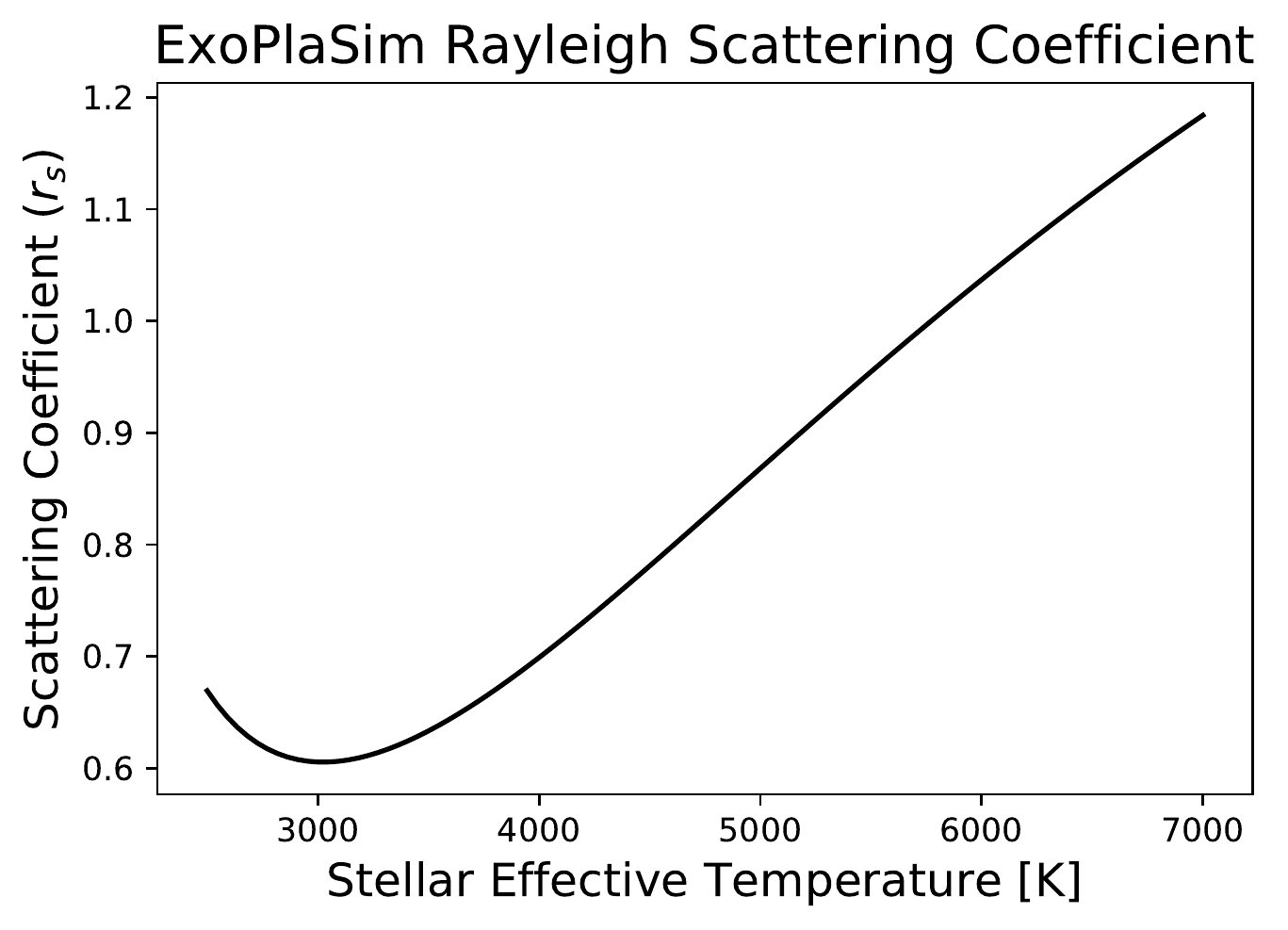}
\end{center}
\caption{The Rayleigh scattering correction coefficient used in ExoPlaSim, for a range of stellar blackbody spectra. This coefficient is needed to account for the fact that ExoPlaSim only computes Rayleigh scattering in one band, and the amount of scattering should have a spectral dependence. The largest corrections to the optical depth scaling introduced by PlaSim's energy partitioning scheme are needed around 3000 K, where PlaSim would overpredict scattering by 65\%. The scattering coefficient $r_s$ begins being less important below 3000 K because so little energy is left in the SW1 band that simply relying on energy partitioning begins to become more appropriate once again.}\label{pyfig:scatteringcoeff}
\end{figure}

The presence of PlaSim's two-band model poses an additional complication here, in that because Rayleigh scattering is only computed in the SW1 band, this introduces a $Z_1$ dependence on its own. Without the addition of a Rayleigh scattering coefficient, the scattering optical depth in PlaSim is thus
\begin{linenomath*}
\begin{equation}
\tau = \frac{Z_1}{Z_{1,\oplus}}\tau_\oplus
\end{equation}
\end{linenomath*}
We therefore want to compute a coefficient $r_s$ which corrects this optical depth:
\begin{linenomath*}
\begin{equation}\label{pyeq:exoplasimtau}
\tau = r_s\frac{Z_1}{Z_{1,\oplus}}\tau_\oplus
\end{equation}
\end{linenomath*}
We can therefore solve for $r_s$ by combining \autoref{pyeq:z1sub} and \autoref{pyeq:exoplasimtau}:
\begin{linenomath*}
\begin{equation} \label{pyeq:rcoeff}
r_s = Z_{1,\oplus}\left(\frac{\int_0^\infty \lambda^{-4}F_\lambda(\lambda)\,d\lambda}{\int_0^{\lambda_2} F_\lambda(\lambda)\,d\lambda}\right)\left(\frac{\int_0^\infty F_{\lambda,\odot}(\lambda)\,d\lambda}{\int_0^\infty \lambda^{-4}F_{\lambda,\odot}(\lambda)\,d\lambda}\right)
\end{equation}
\end{linenomath*}
This coefficient need only be computed once, when the model starts up, and then can be added to the expression for the Rayleigh scattering transmittance:
\begin{linenomath*}
\begin{equation}\label{pyeq:exoplasimrayleigh2}
T = \exp\left[\left(r_s\frac{p_s}{p_{s,\oplus}}\frac{g_\oplus}{g}\right)\ln{T_\oplus}\right]
\end{equation}
\end{linenomath*}
The scattering coefficient $r_s$ is shown in \autoref{pyfig:scatteringcoeff}, and the resulting optical depth is shown in \autoref{pyfig:rayleighcomparison}. The optical depth that results from this scaling is very close to that found by computing the optical depth directly using the Rayleigh scattering cross-section. The value of $r_s$ is 1 for a 5772 K blackbody spectrum, meaning this modification produces the same behavior for a Sun-like spectrum as the original version of PlaSim.

\subsection{Dynamical Core}

One of the challenges that all numerical models with discrete resolution face is how to handle features whose size scales are near or below the model resolution \citep{Lander1997}. This is often encountered in computational fluid dynamics in the context of modeling and resolving shocks, which need to be resolved correctly in order to satisfy conservation laws and to properly obey fluid transport equations \citep{Woodward1986}. An analogous challenge however is often encountered by GCMs, which must occasionally model aspects of the climate that are intrinsically sharp, such as mountain ranges, ice-lines, the terminator, land-ocean boundaries, etc. Just as with shocks, if not treated carefully, such features can either produce inaccurate results, or may produce noticeable artifacts in the model output. Often, in GCMs which use a spectral dynamical core, these artifacts take the form of waves or ripples, which are visible in the cloud field as bands of clouds whose length scale is roughly the model resolution \citep{Navarra1994, Lander1997}. Although they may appear small and localized in many GCMs, they can cause significant model biases \citep{Miyakoda1993}, necessitating careful attention from modellers. 

These ripples originate from the Gibbs effect, whereby a sharp feature (such as a step-function) that is represented in spectral space by a finite number of modes will produce undershoots and overshoots on either side of the feature when transformed back into physical space \citep{Wilbraham1848,Gibbs1899,Hewitt1979}. The Gibbs effect is well-studied in the context of spectral GCMs \citep[e.g.][]{Eliasen1970,Hoskins1980,Navarra1994}, but most applications relate to sharp topography in Earth models. In the following section, we describe the particular challenge the Gibbs effect poses in ExoPlaSim, and how we have modified the dynamical core in response.

\subsubsection{Gibbs oscillations and sharp discontinuities in GCM models of synchronously rotating planets}

\begin{figure*}
\begin{center}
\includegraphics[width=6in]{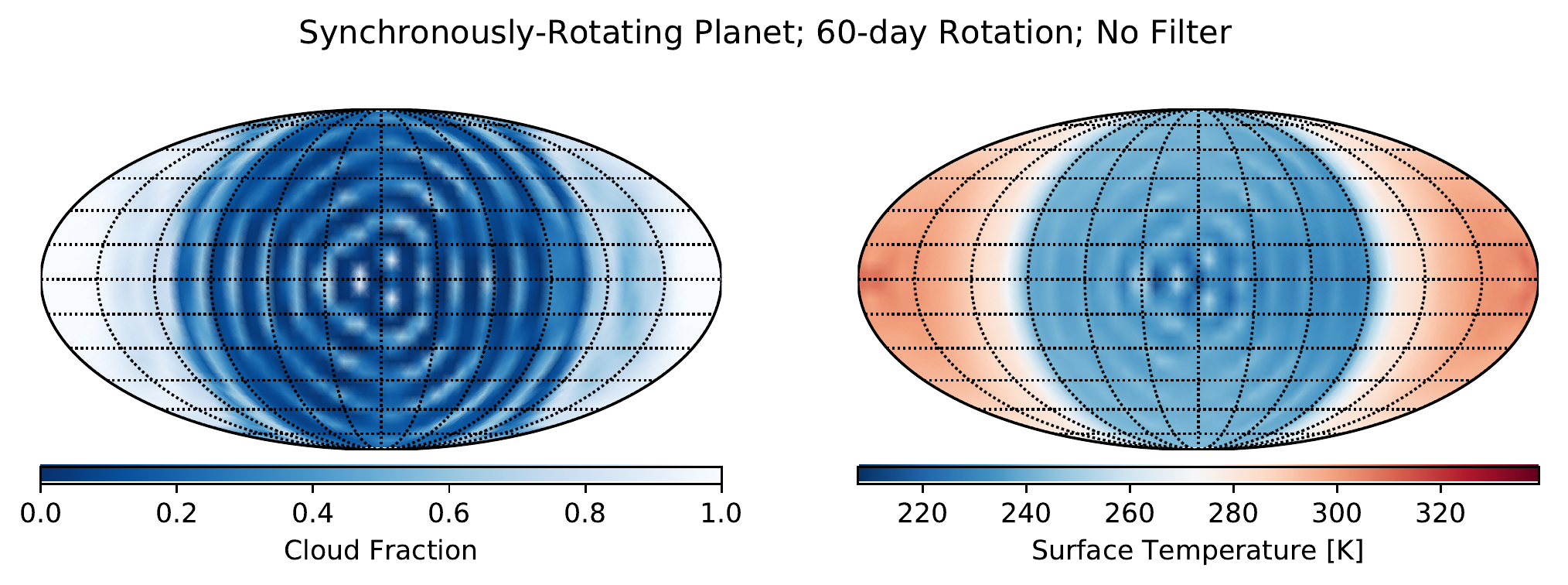}
\end{center}
\caption{The cloud field and surface temperature of a synchronously rotating aquaplanet with 60-day rotation, and no spectral filters used to damp any Gibbs phenomena. The planet is shown in a Mollweide projection, with the nightside and antistellar point centered. Sharp axisymmetric features resulting from the dipolar geometry of the synchronously rotating climate produce concentric rings, visible in most physical fields but especially in the cloud field. Moist processes serve to amplify the magnitude of this feature, such that surface temperature is significantly affected as well. The rings are slightly non-axisymmetric, likely because eastward heat transport means the sea ice boundary is also slightly non-axisymmetric.}\label{pyfig:gibbsunfiltered}
\end{figure*}

Tidally-locked planets pose a challenge other than sharp topography---in models that lack a dynamic ocean, the ice-line represents a sharp transition that is nearly axisymmetric \citep{Hu2014}. The terminator is also a relatively sharp transition that is nearly axisymmetric. The spectral core of a model such as PlaSim decomposes the model atmosphere into spherical harmonics, which means sharp axisymmetric features transfer directly to one of the global spherical modes, rather than remaining relatively local in their influence. Left unaddressed, this produces Gibbs ripples on a global scale, which are easily-visible as concentric rings of clouds on the nightside, as shown in \autoref{pyfig:gibbsunfiltered}. 

It is not clear in which physical variable the ripples first arise, or even if there is a single source variable. Because the atmosphere is a coupled system, the ripples show up in almost every variable in some way. Replacing the ocean with initially-saturated land does not remove the ripples, ruling out the ocean model, nor does removing horizontal advection, which rules out physical waves. Latent heat flux from evaporation and precipitation amplifies the impact of the ripples on surface temperature, but while removing water from the model reduces the impact of the ripples, they are still visible in dynamical fields such as vorticity and divergence, and in the surface temperature. Furthermore, the length scale of the ripples is set directly by the model resolution---doubling the horizontal resolution of the model halves the wavelength of the ripples. Making matters worse, the strength and prevalence of the ripples appears insensitive to the choice of horizontal hyperdiffusion parameters, which would damp out any features emerging from energy cascades piling up at the truncation scale. We therefore conclude that the ripples must have a numerical origin, and are fundamentally a consequence of PlaSim's spectral core encountering an inescapable feature of synchronously rotating planets. We note that this is particularly visible in PlaSim because of the spectral nature of the model's dynamical core, but non-spectral models are likely to also suffer from problems stemming from sharp axisymmetric features like a stationary terminator or the ice-line on a synchronously rotating planet, but the implications may not be as obvious (although they may still be severe). 

The effect is less-pronounced on the dayside in wet models, because moist atmospheric dynamics on local scales damp out the ripples. At extremely high resolution (T170, approximately 75-km resolution), the ripples are smoothed out entirely by local dynamics. However, simply running at extremely high resolution for all synchronously rotating models is unfeasible. Not only has PlaSim not been validated for the weather dynamics that could be resolved at those resolutions, but many of the advantages PlaSim has over other models in terms of computational speed are lost at such high resolution, with the model taking a day or more to simulate a year of climate, even with 32 cores. 

Instead, we attempt to mitigate the planetary Gibbs ripples numerically, by modifying PlaSim's dynamical core. The Gibbs phenomenon has been addressed in GCMs before \citep[e.g.][]{Eliasen1970,Hoskins1980,Navarra1994,Lander1997,Scinocca2008}, though not necessarily in response to features of such global scale. A common way to mitigate the Gibbs phenomenon is with the use of a `physics filter', whereby a mathematical filter is applied to the spectral fields, either before or after the primitive equations are solved, or both \citep{Lander1997}. 

\subsubsection{Physics filters in ExoPlaSim}

\begin{figure*}
\begin{center}
\includegraphics[width=6in]{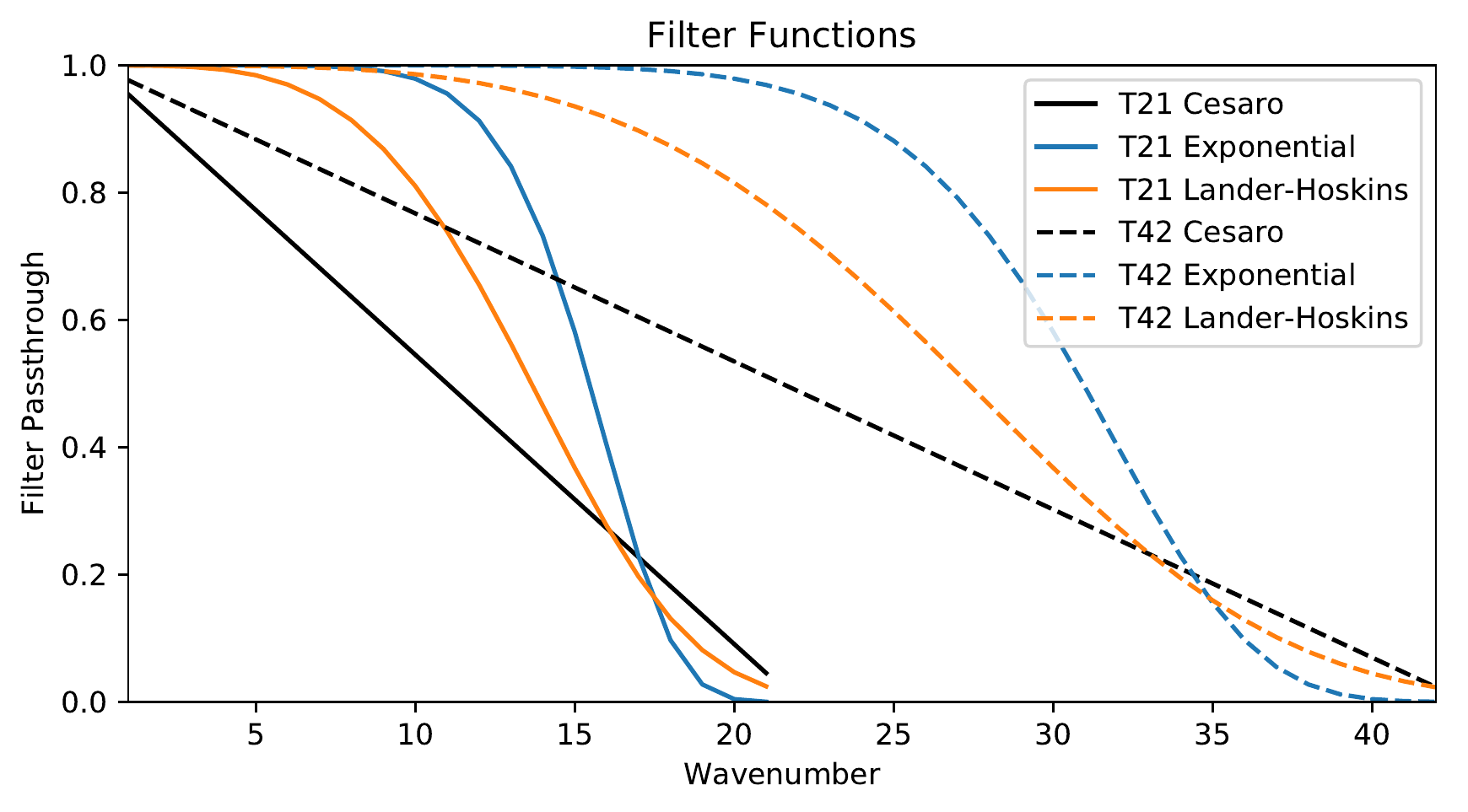}
\end{center}
\caption{Examples at T21 and T42 of the three filter functions available in ExoPlaSim. The Ces\`{a}ro function has the advantage that mathematically, it is more likely to preserve global quantities \citep{Navarra1994}, but the disadvantage that it damps lower wavenumbers. The exponential \citep{Navarra1994,Hoskins1980,Lander1997} and Lander-Hoskins \citep{Lander1997} filters, however, while being empirically derived, feature strong damping at large wavenumbers and low damping at small wavenumbers.}\label{pyfig:examplefilters}
\end{figure*}

\begin{figure*}
\begin{center}
\includegraphics[width=6in]{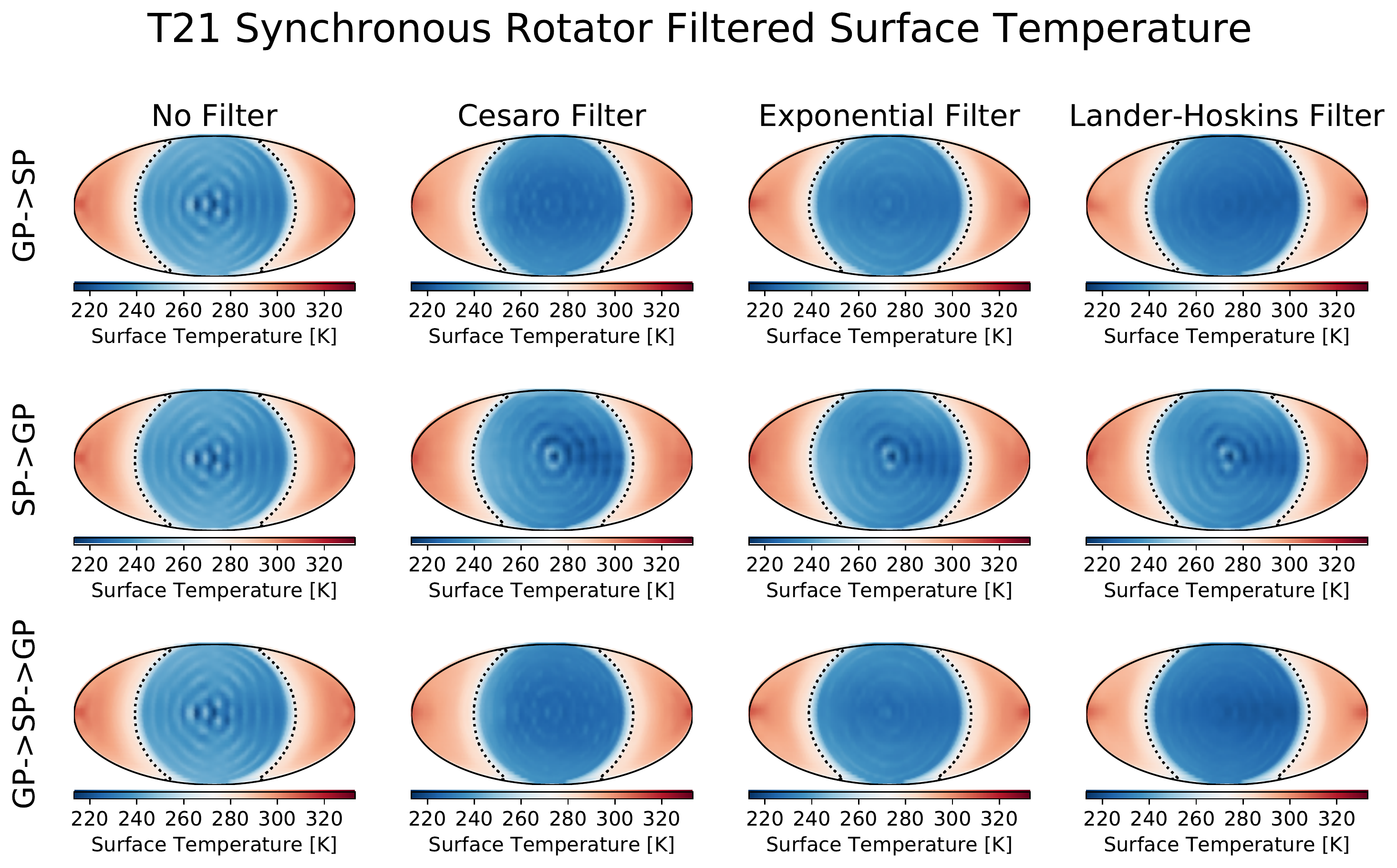}
\end{center}
\caption{Mean annual surface temperature of synchronously rotating aquaplanets with Earth bulk parameters, 60-day rotation, 3400 K incident spectrum, 1360 W/m$^2$, and no radiative effect from sea ice beyond latent heat flux. The antistellar point is centered. The models are at T21 resolution, or 32 latitudes and 64 longitudes, corresponding to approximately 5.6$^\circ$ horizontal resolution. Three different filters are shown, as well as the case with no filter, with 3 different filter configurations for each filter: `GP$\rightarrow$SP', corresponding to a filter during the transform from the gridpoint domain to the spectral domain, `SP$\rightarrow$GP', corresponding to a filter at the transform from the spectral domain to the gridpoint domain, and `GP$\rightarrow$SP$\rightarrow$GP', indicating a filter at both transforms. All three are of course identical for the case with no filter. At T21, the use of any of the three filters at both transforms significantly reduces the effect of Gibbs ripples on the surface temperature, despite the continued presence in the cloud field.}\label{pyfig:benchtl_tsT21}
\end{figure*}

\begin{figure*}
\begin{center}
\includegraphics[width=6in]{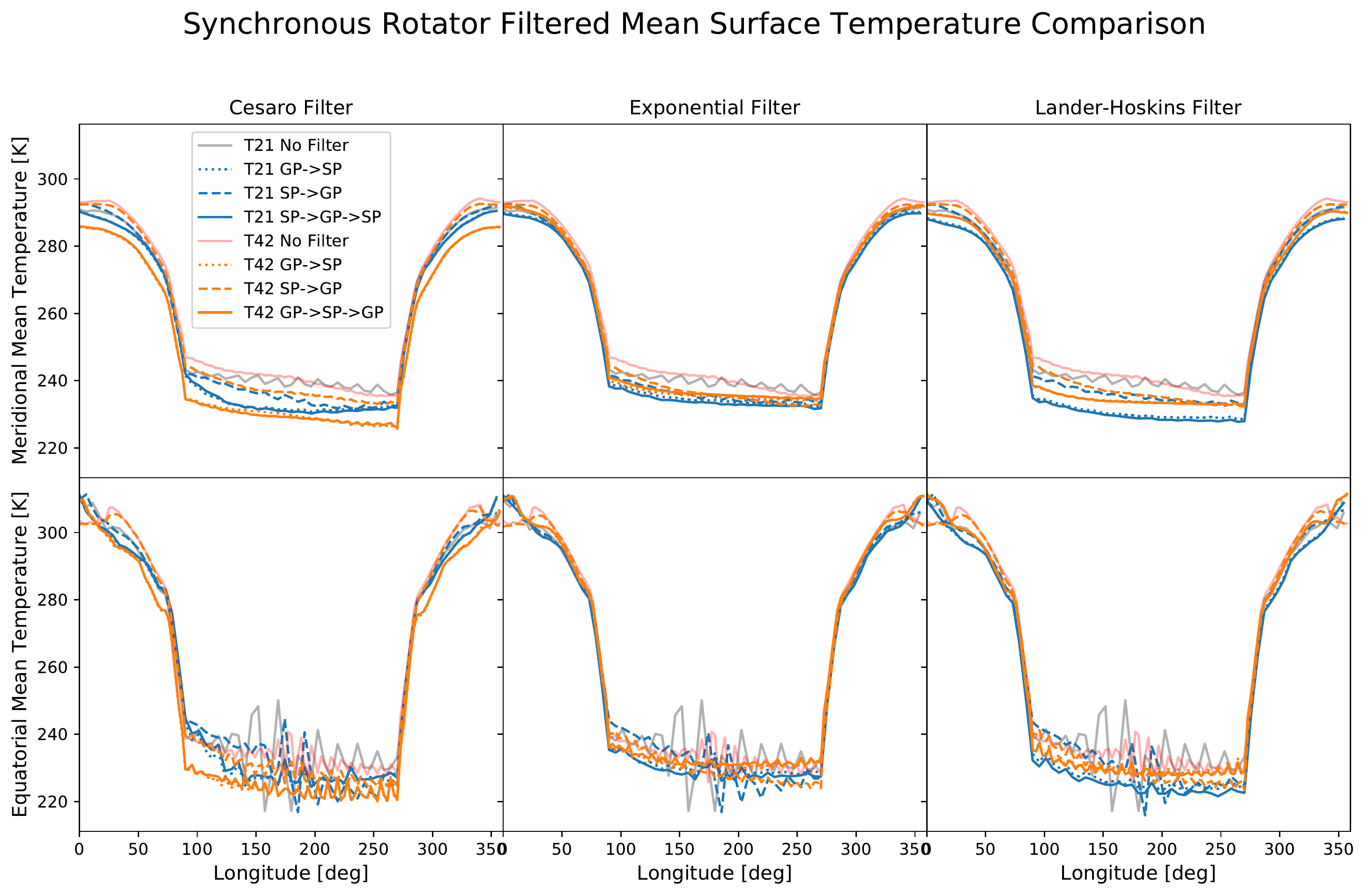}
\end{center}
\caption{Equatorial and mean meridional surface temperature for a synchronously rotating aquaplanet with 60-day rotation, Earth bulk and atmospheric parameters, and 3400 K incident spectrum, modeled at T21 and T42 with Ces\`{a}ro, exponential, and Lander-Hoskins \citep{Lander1997} filtration. The substellar point is at 0 degrees longitude. Models with no filtration are also shown for comparison. All three filters substantially reduce Gibbs-type ripples in surface temperature, with exponential and Lander-Hoskins filtration providing the most-robust results. In this test, the exponential filter appears to provide the best consistency between T21 and T42 resolutions.}\label{pyfig:benchtl_temps}
\end{figure*}

The PlaSim dynamical core uses a triangularly-truncated series of spherical harmonics to solve the primitive fluid equations, such that a variable $Q$ in latitude-longitude space at vertical level $\sigma$ and time $t$ is represented in spectral space by
\begin{linenomath*}
\begin{equation}
Q_n^m(\sigma,t) = \frac{1}{4\pi}\int_{-1}^1\int_0^{2\pi}Q(\lambda,\mu,\sigma,t)P_n^m(\mu)e^{-im\lambda}\,d\lambda\,d\mu
\end{equation}
\end{linenomath*}
where $\mu$ is the sine of the latitude, $\lambda$ is the longitude, and $P_n^m(\mu)$ is the associated Legendre polynomial of degree (total wavenumber) $n$ and order (zonal wavenumber) $m$ \citep{Eliasen1970,Orszag1970,Fraedrich2005}. Functionally, this transform is computed with a Fast Fourier Transform kernel. The inverse relationship is given by
\begin{linenomath*}
\begin{equation}
Q(\lambda,\mu,\sigma,t) = \sum_{m=-N}^N\sum_{n=|m|}^{N}Q_n^m(\sigma,t)P_n^m(\mu)e^{im\lambda}
\end{equation}
\end{linenomath*}
where $N$ is the truncation wavenumber (e.g. 21 for T21 resolution, or 32 latitudes and 64 longitudes, and 42 at T42 resolution, corresponding to 64 latitudes and 128 longitudes).

In order to avoid cumulative effects of any physics filter we apply, rather than modify temperature, vorticity, divergence, and humidity directly, we follow the approach described in \citet{Lander1997} and \citet{Scinocca2008}, and apply the filter as part of the spectral transform. We add a possible filter in each direction, such that
\begin{linenomath*}
\begin{align}
Q_n^m(\sigma,t) &= \frac{1}{4\pi}\int_{-1}^1\int_0^{2\pi}g(n)Q(\lambda,\mu,\sigma,t)P_n^m(\mu)e^{-im\lambda}\,d\lambda\,d\mu \\ \nonumber \\
Q(\lambda,\mu,\sigma,t) &= \sum_{m=-N}^N\sum_{n=|m|}^{N}h(n)Q_n^m(\sigma,t)P_n^m(\mu)e^{im\lambda}
\end{align}
\end{linenomath*}
where $g(n)$ is the filter for the transform from gridpoint to spectral space, and $h(n)$ is the filter for the reverse. We distinguish $g(n)$ from $h(n)$ because these filters serve different purposes, and it may be that in some cases, a filter is only desirable in one direction and not the other. $g(n)$ will help to remove sharp features that arise in physical tendencies (the time derivatives of the model's dynamical variables) computed through the model's physics modules in the latitude-longitude gridpoint domain. Conversely, $h(n)$ will help to remove small features in the spectral variables themselves that arise as a consequence of the model dynamics computed in the spectral core \citep{Lander1997}. In the case of the global Gibbs ripples observed in our models, we find both are necessary.

We consider three different kinds of filters, all of which are isotropic, meaning they depend only on total wavenumber $n$. \citet{Navarra1994} described an isotropic 2D Ces\`{a}ro filter, which makes use of the principle that certain infinite series may be approximated by the arithmetic mean of their partial sums:
\begin{linenomath*}
\begin{equation}
g(n) = 1-\frac{n}{N+1}
\end{equation}
\end{linenomath*}
The mathematical basis of Ces\`{a}ro summation makes this a potentially attractive choice, as the application of the filter is therefore unlikely to introduce significant global biases. In tests with a GCM at T30 resolution, \citet{Navarra1994} found such a filter dramatically reduced Gibbs effects from sharp orography. However, this filter suffers from the problem that significant power is also lost from spectral modes at lower wavenumbers, potentially reducing the resolvable scale of the model \citep{Lander1997}. \citet{Navarra1994} also described an empirical exponential filter,
\begin{linenomath*}
\begin{equation}
g(n) = e^{-\kappa\left(\frac{n}{N}\right)^\gamma}
\end{equation}
\end{linenomath*}
where $\kappa$ and $\gamma$ are constants, typically chosen from powers of 2. While this filter has no rigorous mathematical basis, as the Ces\`{a}ro filter does, it retains more power in lower wavenumbers, potentially improving model resolution \citep{Hoskins1980,Lander1997}. Finally, we also consider a filter form related to the exponential filter, described in \citet{Lander1997} and used in \citet{Scinocca2008}:
\begin{linenomath*}
\begin{equation}
g(n) = e^{-\left(\frac{n(n+1)}{n_0(n_0+1)}\right)^2}
\end{equation}
\end{linenomath*}
where $n_0$ must be determined experimentally such that the filter is equal to a desired minimum passthrough at the truncation wavenumber. \citet{Lander1997} suggest, based on the average wavenumber of a point-spread function represented in a spectral model, that the critical wavelength $n_0$ is likely approximately $N/\sqrt{2}$, which for T21 means $n_0\approx15$. We hereafter refer to this filter as the Lander-Hoskins filter. All three filters are available in ExoPlaSim (along with the option to not use a filter at all), and can be enabled for either the gridpoint-to-spectral transform ($g(n)$), the spectral-to-gridpoint transform ($h(n)$), or both. Examples of the filter functions are shown in \autoref{pyfig:examplefilters}.

\subsubsection{Physics filter performance}

We tested each of our three filters in two different sets of model configurations, corresponding to the synchronously rotating benchmark model in \citet{Yang2019} (60 days, 1360 W/m$^2$, 3400 K input spectrum) and the TRAPPIST-1e `Hab1' model described in \citet{Fauchez2020} as part of the THAI model intercomparison (6.1 days, 900 W/m$^2$, 2600 K input spectrum), at both T21 and T42. All models were aquaplanet models, and followed \citet{Fauchez2020} and \citet{Yang2019} for atmospheric pressure and composition. The TRAPPIST-1e models had the same bulk planetary parameters as in \citet{Fauchez2020}, and the \citet{Yang2019} models had Earth's bulk planetary parameters. As in \citet{Fauchez2020}, we prescribe a sea ice albedo of 0.25 and an ocean albedo of 0.06 for the TRAPPIST-1e models, but since in \citet{Yang2019} sea ice is turned off completely, which cannot be done in ExoPlaSim (due to its role in the computation as an interface between the ocean and atmosphere models), in those models we simply exclude ice and snow from the radiation component of the model. This means that latent heat fluxes from snowfall and snowmelt are still included. We use $\kappa=8$ and $\gamma=8$ for the exponential filters.

At T21 and T42, none of our filters are able to completely remove the Gibbs ripples, and the Gibbs effect remains apparent in the cloud field, although somewhat-reduced at T42. The effect on other fields such as surface temperature, however, is greatly-reduced, as shown for the YB models in \autoref{pyfig:benchtl_tsT21} and \autoref{pyfig:benchtl_temps}. The filter at the transform between the gridpoint and spectral domains seems most impactful, but using filters at both transforms results in the most robust behavior. The exponential and Lander-Hoskins filters seem ideal in these tests, removing more of the ripples than the Ces\`{a}ro filter and affecting spatial structures less. We also note that we do not observe any significant global biases with the exponential and Lander-Hoskins filters applied at both transforms.

\begin{figure*}
\begin{center}
\includegraphics[width=6in]{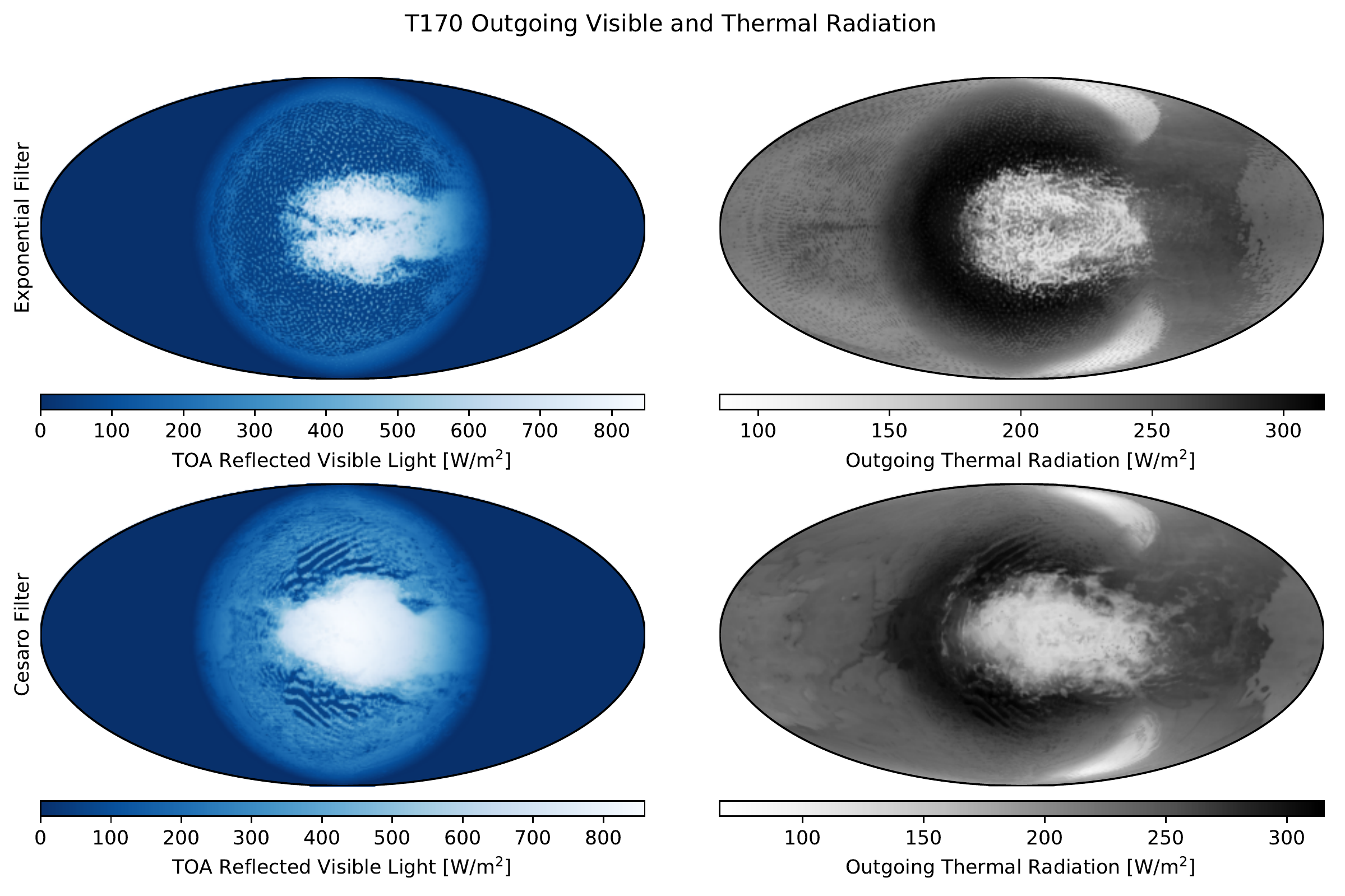}
\end{center}
\caption{Outgoing shortwave (visible) and longwave (thermal) radiation for a synchronously rotating aquaplanet with a rotation period of 9 days, 3000 K incident spectrum, and Earth bulk parameters and atmospheric pressure and composition, modelled at T170 resolution (256 latitudes and 512 longitudes; approximately 75-km resolution) with both an exponential filter (top) and a Ces\`{a}ro filter (bottom). In this case, the Ces\`{a}ro filter does a better job of treating Gibbs-type granularity, and potentially reveals physical behavior by allowing clouds in the substellar inflow region to aggregate into long bands that are advected towards the substellar upwelling region.}\label{pyfig:cesaroT170}
\end{figure*}

The choice of which physics filter is appropriate appears to be situational. At extremely high resolutions (T170), the Ces\`{a}ro filter appears to give the best performance, as shown in \autoref{pyfig:cesaroT170}. However, at lower resolutions, the exponential and Lander-Hoskins filters are ideal, as seen in the superior agreement between T21 and T42 resolutions in \autoref{pyfig:benchtl_temps}. For the YB models, an exponential filter with $\kappa=8$ and $\gamma=8$ yields the best performance, while for the THAI-B models, a Lander-Hoskins filter is best. A detailed description of the impact of the physics filters on other model fields such as clouds, humidity, wind speed, and relative vorticity, along with our assessment of the physics filters in different scenarios, including sensitivity tests to $\kappa$ and $\gamma$, is available in Appendix A of the Supplementary Materials (available online). For the sake of completeness, however, in our comparison between ExoPlaSim and other models later in this paper, we include multiple filter configurations, as well as comparison to the case with no filter.

\section{Model Performance}

\subsection{Model Speed}\label{pysec:speed}

We have tested ExoPlaSim's performance on a computing cluster with Intel Xeon E5-2650, Xeon Gold 6130, and Xeon Gold 6262V processors, using the Intel version 18.0 compiler and OpenMPI version 3.0.0. ExoPlaSim can run in parallel with as many MPI threads as there are latitudes in the model resolution, and if run with fewer threads, runs best if the number of latitudes is evenly divisible by the number of threads. We only run in configurations where all MPI threads can be localized to the same node; inter-node data transfer speeds can slow the model down to runtimes comparable to running in single-thread mode. We have run T21 models on these processors with 8 threads, 16 threads, and 32 threads. On these architectures, ExoPlaSim simulates one year of climate at T21 with a 45-minute timestep and 10 vertical layers in 30-60 seconds of walltime, depending on node and parallelization. With the 30-minute timestep we have found most-robust for synchronously rotating models, performance slows to 40-90 seconds of walltime per simulated year. When the timestep is reduced to 5 minutes, which we find can be necessary for models with high (10 bar) or low (0.1 bar) surface pressures, the walltime required per year rises to 6 minutes with 16 MPI threads. At T42, with 15-minute timesteps and 16 MPI threads, one year requires approximately 15-16 minutes of walltime on Xeon E5-2650 nodes. At the highest resolution we have tested, T170 (256 latitudes and 512 longitudes) with 30 layers, it takes approximately 60 hours of walltime at a timestep of 3.75 minutes to simulate one model year with 32 MPI threads on a Xeon Gold 6130 node. 

We have also tested ExoPlaSim's performance on consumer-grade computing architectures. AP's 2017 personal laptop contains an Intel i7-7700HQ quad-core processor, and has the GNU Fortran version 4.9.3 compiler installed, along with OpenMPI version 1.1.0. On this architecture, with 4 MPI threads and other programs running, ExoPlaSim runs a T21 model with 10 layers and a 45-minute timestep for one year in 3.75 minutes of walltime, and 5.5 minutes with a 30-minute timestep. On a per-cell, per-thread basis, we find that across the architectures tested, ExoPlaSim generally runs at a minimum of approximately $7.5\times10^4$ cell-timesteps per second per thread, with 2.5--4 times that more characteristic of typical performance. 

ExoPlaSim T21 aquaplanet models tend to require 50--200 years to run to energy balance equilibrium, which we define here as requiring mean top-of-atmosphere and surface net fluxes to be stable to less than 0.5 W/m$^2$ drift per decade. This means that a synchronously rotating aquaplanet model with an Earth-like atmosphere can be run nearly to energy balance equilibrium at T21 on a personal laptop over the course of an extended lunch break, and certainly far enough to provide a first-order overview of likely qualitative climate states. Crucially, this means that access to supercomputing infrastructure is not required to do productive GCM science with ExoPlaSim.

At the other end of the computing spectrum, even a modest allocation on a supercomputing cluster can enable large parameter sweeps with ExoPlaSim. At T21 with 45-minute timesteps, assuming minimal downtime and uninterrupted access to compute nodes, 40 simulations running simultaneously with 16 MPI threads for 150 years each permits on the order of 4000 simulations per week, corresponding to a 50x80 2D parameter grid, or a 20x20x10 3D parameter cube. The theoretical limit with a computing allocation of this size is therefore 1 million models run to equilibrium over the course of a 5-year PhD, representing over 100 million years of 3D climate and several petabytes of model output (or a few tens of terabytes if only the last model year is kept).

\subsection{Comparing ExoPlaSim to other models}\label{pysec:comparison}

One of the challenges inherent to developing climate models for exoplanets is that we have no real-world data against which to compare model predictions. We must therefore validate climate models for non-Earthlike habitable planets, such as synchronously rotating planets, through model intercomparisons \citep{Yang2019,Fauchez2020,THAI2021a,THAI2021b,THAI2021c}. Two such model intercomparison efforts have been undertaken for synchronously rotating planets. \citet{Yang2019} compared CAM3 \citep{Collins2004}, CAM4 \citep{Neal2010}, ExoCAM \citep{Wolf2015}, and two configurations of the LMD Generic model \citep[hereafter LMDG,][]{Wordsworth2010,Wordsworth2010a,Wordsworth2011,Forget2013}, for both an aquaplanet with Earth-like rotation and solar-like spectrum, and a synchronously rotating aquaplanet with a rotation period of 60 days, 1360 W/m$^2$, and a 3400 K blackbody spectrum. They also included the AM2 model \citep{Anderson2004}, but as it did not converge in the synchronously rotating case, we do not include it in ExoPlaSim's synchronously rotating comparisons. \citet{Yang2019} also compared the above models in a more Earth-like test case, involving an aquaplanet with a 24-hour rotation rate, solar-like spectrum, and zero obliquity or eccentricity. We do not directly validate ExoPlaSim against Earth climate, as PlaSim has been used extensively to model both modern Earth and paleo-Earth climates before \citep[e.g.][]{Garreaud2010,Lucarini2010,Haberkorn2012,Lucarini2013,Spiegl2015,Paradise2017,Andres2019,Duque2019,Holden2019,Paradise2019,Zuev2020}, and we have designed our modifications to the model such that in an Earth-like configuration, it converges to the original model. However, as a robustness test and to serve as a benchmark for the model's performance against other models, we do compare it to the fast-rotating test case in \citet{Yang2019} in addition to the synchronously rotating case. 

The TRAPPIST-1e Habitable Atmosphere Intercomparison \citep[hereafter THAI,][]{Fauchez2020,THAI2021a,THAI2021b,THAI2021c} similarly compared ExoCAM, ROCKE-3D \citep{Way2017}, LMDG, and the UK Met Office Unified Model \citep[hereafter UM,][]{Mayne2014,Boutle2017} to model several benchmark atmosphere and surface configurations for TRAPPIST-1e \citep[6.1-day rotation, 900 W/m$^2$, 2600 K spectrum;][]{Grimm2018}. To validate ExoPlaSim's performance in the context of synchronously rotating climates, we run ExoPlaSim with the same model configurations as each of these intercomparisons, and compare our results with the results of the various component models. We note however that due to the lack of empirical data on synchronously rotating climates, it is not possible to distinguish accurate ExoPlaSim predictions from collective inaccuracies shared by all current models, which stem from incorrect geophysical assumptions. 

For convenience, we will hereafter refer to the benchmark experiments from the \citet{Yang2019} intercomparison as YB experiments, and the moist benchmarks from the \citet{Fauchez2020} intercomparision (Hab1 and Hab2) as THAI-B experiments. We include a comparison to the CO$_2$-dominated THAI test case, but we note that prior work has shown clear radiative biases in PlaSim in high-CO$_2$ contexts \citep{Paradise2017}, and ExoPlaSim lacks crucial physics that would be relevant in high-CO$_2$ regimes, such as CO$_2$ Rayleigh scattering, CO$_2$ clouds, and CO$_2$ condensation. We therefore do not expect ExoPlaSim to be predictive of real climates in such regimes, and we expect larger disagreement with other models. For all ExoPlaSim models shown, we present only annual averages, where `annual' refers to approximately one Earth year. For the synchronously rotating YB models, this therefore means we show climate averaged over approximately 6 orbits, and for the THAI-B models, 59--60 orbits.  

\subsubsection{YB comparison}\label{pysec:yang}

We compare ExoPlaSim to the models in the \citet{Yang2019} intercomparison (the YB models) by running both the fast-rotator benchmark and the synchronously rotating benchmark described in that study. Both models have an instellation of 1360 W/m$^2$, zero obliquity, zero eccentricity, and Earth's radius, gravity, and surface pressure. Neither model configuration included ozone or aerosols, and both are aquaplanets, with no continents, and a prescibed ocean surface albedo of 0.5. The YB models assume uniform (Lambertian) reflectance of 0.05 from the ocean surface, but PlaSim's default is to use the ECHAM-3 zenith angle dependence for ocean surface reflectivity \citep{echam3} in the blue ($<0.75$ $\mu$m) shortwave band. We therefore replace the zenith angle dependence in PlaSim with an optional Lambertian reflectance model. We also include the zenith angle dependence used in ECHAM-6 \citep{echam6}, but we do not find major qualititative differences between the three, and all three are available in ExoPlaSim. The YB models had sea ice and snow turned off, but in PlaSim, it is not possible to fully-disable the sea ice module, as it functions as the interface between the ocean and air, and simply altering the thermodynamics of ice so that it doesn't form leads to errors elsewhere. We therefore disable the sea ice contribution to surface reflectivity and absorptivity, but leave its thermodynamic effects (latent heat flux and heat capacity) in place. The ocean is configured as a 50-meter mixed-layer slab ocean, with no dynamic heat transport. The fast-rotator model has a rotation period of 24 hours, and a solar-like incident spectrum, corresponding in ExoPlaSim to 51.7\% of the incident light in the blue shortwave band (SW1), and 48.3\% in the red shortwave band (SW2). The synchronously rotating model has a rotation period of 60 days, and an incident stellar spectrum corresponding to a 3400 K blackbody. We test ExoPlaSim in both T21 and T42 resolutions, with 10 vertical layers, and with ten different physics filter configurations: no filter, a Ces\`{a}ro filter, an exponential filter with $\kappa=8$ and $\gamma=8$, and a Lander-Hoskins filter. We used three different configurations of each of the three filter types: during the gridpoint-to-spectral transform, during the spectral-to-gridpoint transform, and during both transforms, for 9 different filtered models plus the unfiltered model. The other models in the intercomparison had horizontal solutions most similar to T42 (64 latitudes and 128 longitudes)---48 latitudes and 96 longitudes for CAM3, 96 latitudes and 144 longitudes for CAM4, 46 latitudes and 72 longitudes for ExoCAM, 90 latitudes and 144 longitudes for AM2, and 65 latitudes and 129 longitudes for the LMDG models. All of the models in the intercomparison however had substantially higher vertical resolution than our ExoPlaSim models, with 26 layers for CAM3 and CAM4, 45 layers for ExoCAM, 32 layers for AM2, and 30 layers for LMDG. Those models also all had model tops at pressures at least an order of magnitude lower than ExoPlaSim's model top. 

\begin{figure*}
\begin{center}
\includegraphics[width=6in]{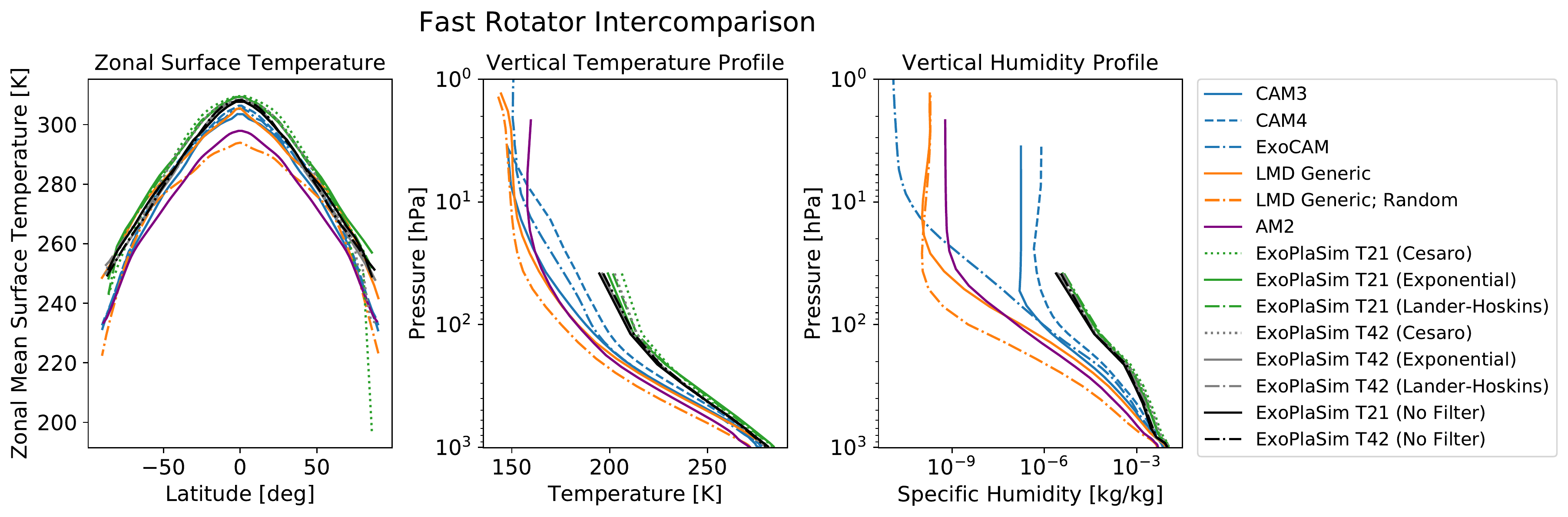}
\end{center}
\caption{Zonal mean surface temperature, mean air temperature, and mean specific humidity for the fast-rotator models in the \citet{Yang2019} model intercomparison, compared to ExoPlaSim models run with a series of physics filter configurations, at both T21 and T42 resolutions. The `LMD Generic; Random' model denotes the LMD configuration that uses random cloud overlap \citep{Yang2019}. ExoPlaSim models with filtration applied at both spectral transforms show good agreement with the unfiltered models, but the overall dependence on filter type is low. All of our ExoPlaSim models are warmer and wetter than the other models in the intercomparison, particularly in the upper atmosphere. Surface temperatures, however, while at the warm end of the intercomparison distribution, do not deviate significantly from the surface temperatures computed by other models. The shallower vertical profile may be a consequence of PlaSim's relatively simpler radiation and convection scheme.}\label{pyfig:yangAQbench1}
\end{figure*}

Temperature and humidity profiles for the results of our fast-rotator benchmark comparison are shown in \autoref{pyfig:yangAQbench1}. ExoPlaSim shows good agreement at the surface in this test case, although surface temperatures are near the warm end of the model distribution. The vertical temperature and humidity profiles, however, are shallower, resulting in a significantly warmer and wetter upper troposphere, by about 10 K and a factor of 10 in specific humidity at 50 hPa. We note however that there is a wide range of temperatures and humidities in the upper atmosphere among the other models in the intercomparison, and ExoPlaSim's model top is by far the lowest in the ensemble---the other models extend at least to 4 hPa, over a factor of 10 lower in pressure than ExoPlaSim's model top in this test. ExoPlaSim's upper atmosphere may therefore be affected by missing stratospheric physics and dynamics.

\begin{figure*}
\begin{center}
\includegraphics[width=6in]{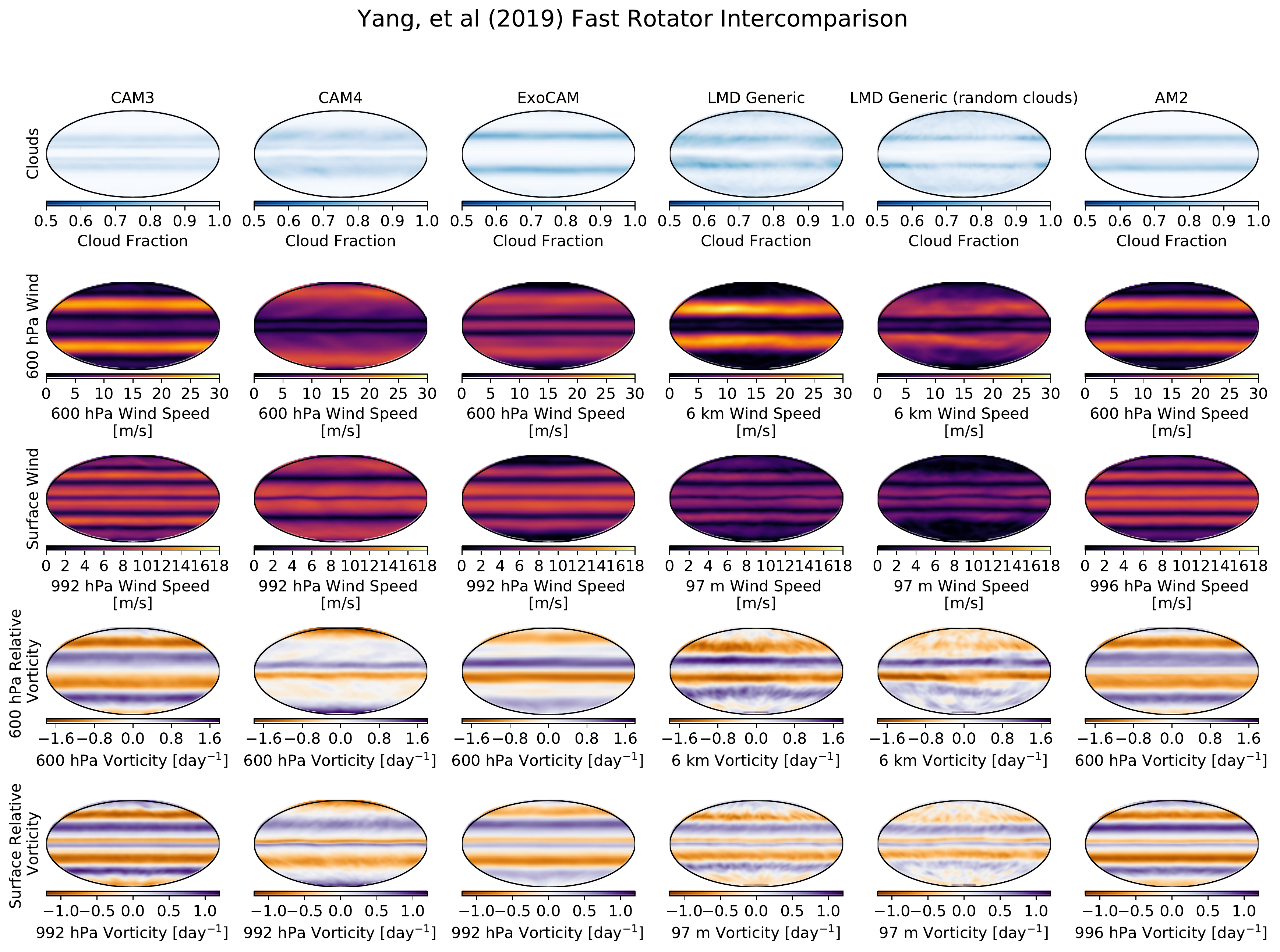}
\end{center}
\caption{Cloud fraction, wind speed and relative vorticity at or near 600 hPa, and wind speed and relative vorticity at or near the surface for each of the models included in the \citet{Yang2019} intercomparison fast-rotator benchmark. All models have the same qualitative structure, with three dominant cloud bands, two main zonal jets in the mid-troposphere, four dominant wind currents near the surface, and alternating bands of relative voriticity.}\label{pyfig:yangAQmaps}
\end{figure*}

\begin{figure*}
\begin{center}
\includegraphics[width=5in]{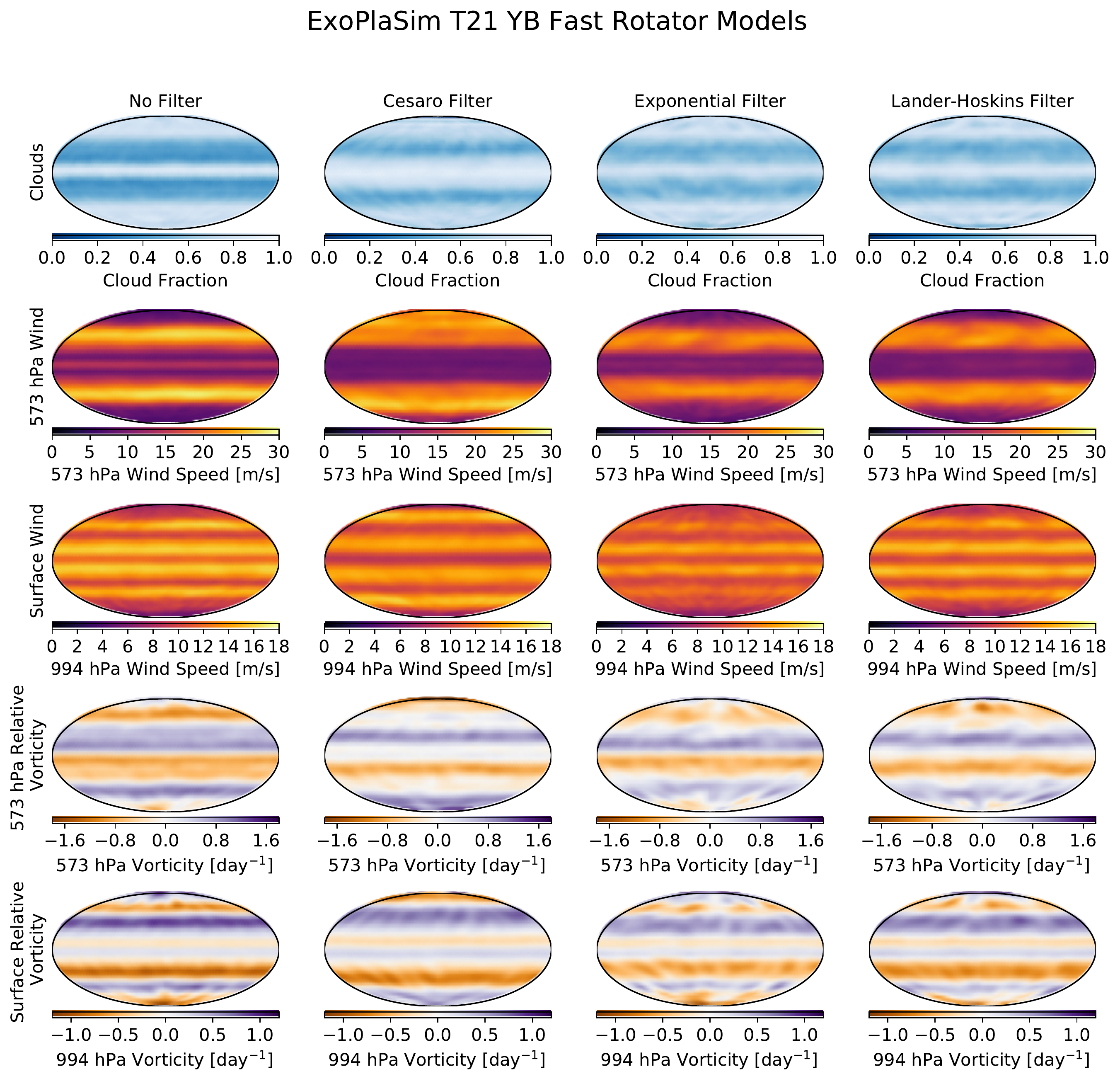}
\end{center}
\caption{Cloud fraction, wind speed and relative vorticity at 573 hPa (ExoPlaSim's closest layer to 600 hPa), and wind speed and relative vorticity at near the surface for T21 ExoPlaSim models configured similarly to the fast-rotator benchmark in the \citet{Yang2019} intercomparison. Four filter configurations are shown here, including a configuration with no physics filter (ExoPlaSim's default). Each filtered model shown here has the physics filter applied at both spectral transforms. At T21, ExoPlaSim shows good qualitative agreement with the other models, though wind speeds are generally higher, and cloud fraction outside the main cloud bands is lower. Use of the Ces\`{a}ro physics filter significantly broadens the tropical cloud belt.}\label{pyfig:exoplasimAQ_T21}
\end{figure*}

\begin{figure*}
\begin{center}
\includegraphics[width=5in]{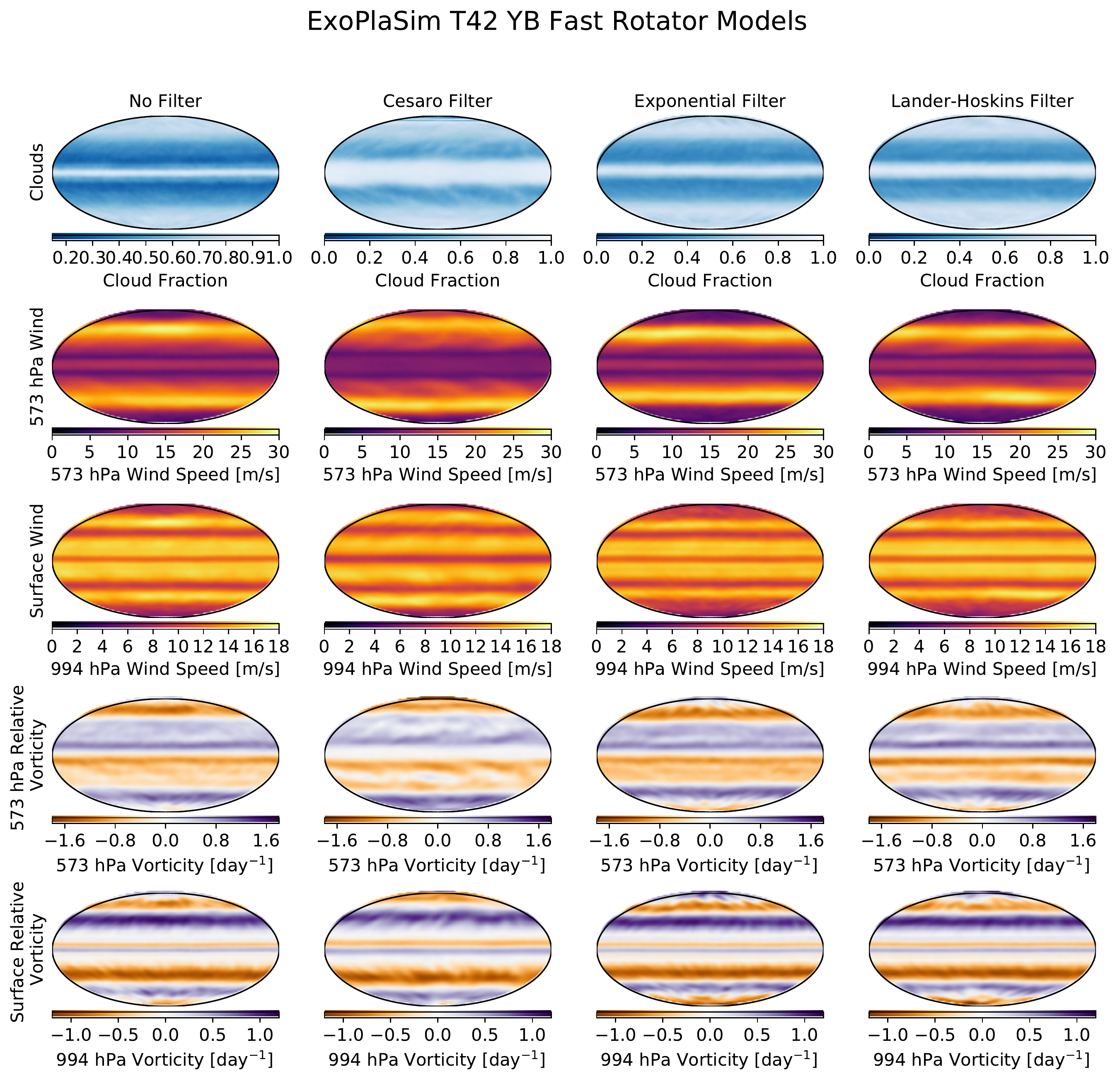}
\end{center}
\caption{Cloud fraction, wind speed and relative vorticity at 573 hPa (ExoPlaSim's closest layer to 600 hPa), and wind speed and relative vorticity at near the surface for T42 ExoPlaSim models configured similarly to the fast-rotator benchmark in the \citet{Yang2019} intercomparison. Four filter configurations are shown here, including a configuration with no physics filter (ExoPlaSim's default). Each filtered model shown here has the physics filter applied at both spectral transforms. At T42, ExoPlaSim shows the greatest similarity in atmospheric dynamics to the CAM3 and AM models.}\label{pyfig:exoplasimAQ_T42}
\end{figure*}

We also compare the morphology of the climate across the various models, through cloud fraction, wind speed, and relative vorticity. These quantities are shown for the models in \citet{Yang2019} in \autoref{pyfig:yangAQmaps}. The same quantities for the ExoPlaSim models are shown in \autoref{pyfig:exoplasimAQ_T21} (T21) and \autoref{pyfig:exoplasimAQ_T42} (T42), for each of the four filter types (showing only the configurations where filtration is applied during both spectral transforms). The cloud fractions shown are recomputed for each of the models, considering only clouds found below ExoPlaSim's model top, to facilitate direct comparison between models. We define the total cloud fraction as it is computed in ExoPlaSim, such that
\begin{linenomath}
\begin{equation}
\text{CLD}_\text{t}=1 - \prod_{k=0}^N(1-\text{CLD}(\sigma_k))
\end{equation}
\end{linenomath}
where $\sigma_k$ is a given vertical level, and $N$ is the number of vertical levels. This follows from the assumption of random cloud overlap used in ExoPlaSim's cloud transmissivity calculation.  We find that ExoPlaSim agrees qualitatively with the other models at both T21 and T42 resolutions. The Ces\`{a}ro filter results in a broader tropical cloud band, while the exponential and Lander-Hoskins filters do not significantly affect the dynamics of the atmosphere. ExoPlaSim has generally lower cloud fraction between the major cloud bands, and higher overall wind speeds, although the relative vorticity is approximately the same as the other models. Based on cloud fraction, wind speed, and vorticity, ExoPlaSim appears most similar to the CAM3 and AM models, and the surface temperature is most similar to the CAM4 model.

\begin{figure*}
\begin{center}
\includegraphics[width=6in]{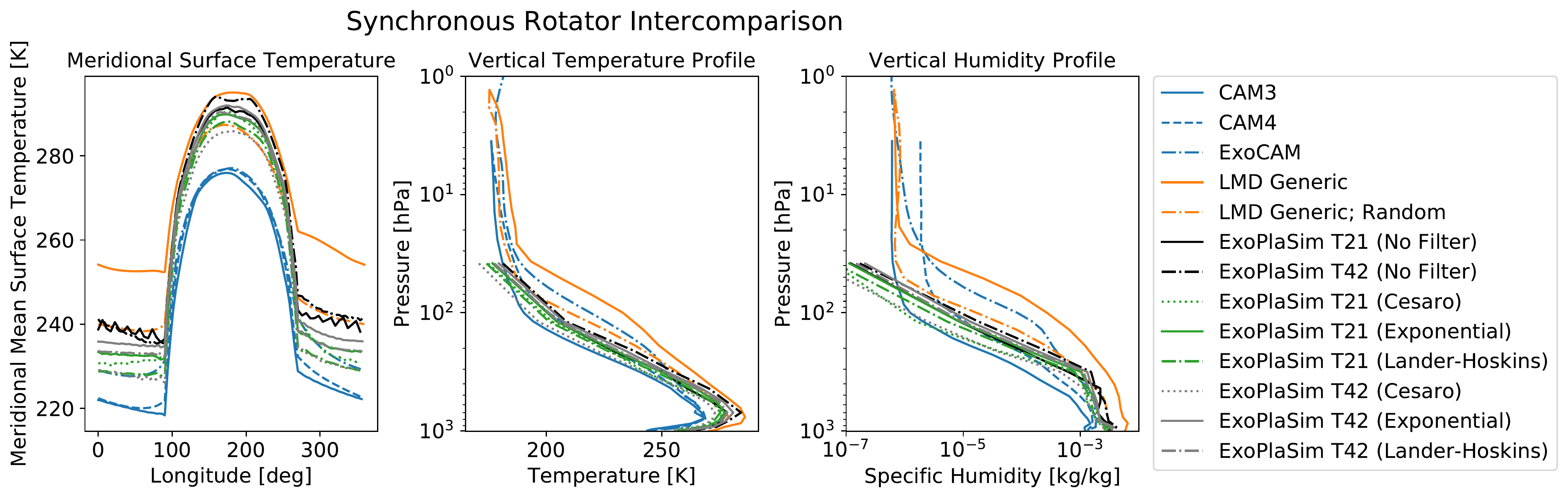}
\end{center}
\caption{Meridional mean surface temperature, mean air temperature, and mean specific humidity for the synchronously rotating models in the \citet{Yang2019} model intercomparison, compared to ExoPlaSim models run with a series of physics filter configurations, at both T21 and T42 resolutions. The `LMD Generic; Random' model denotes the LMD configuration that uses random cloud overlap \citep{Yang2019}. Unlike the fast-rotator benchmark, ExoPlaSim synchronously rotating models fall well within the ensemble distribution for surface temperature, vertical temperature profile, and vertical humidity profile. ExoPlaSim model tops are colder than the other models at the same pressure, but this may be a consequence of a lower model top and poor vertical resolution at that pressure. The choice of filter has a significant impact on nightside surface temperatures, although all ExoPlaSim variants are still within the model ensemble.}\label{pyfig:yangTLbench1}
\end{figure*}

Meridional mean surface temperatures, mean vertical temperature profiles, and mean vertical humidity profiles from the synchronously rotating benchmark comparison are shown in \autoref{pyfig:yangTLbench1}. Unlike in the fast-rotator case, we find that in the synchronously rotating benchmark, ExoPlaSim temperatures and humidity are well-within the model ensemble distribution, finding warmer temperatures than the CAM family of models, but cooler than the LMD models. Additionally, in the synchronously rotating case, the choice of filter can have a significant impact, particularly on nightside temperatures, where meridional mean temperatures can vary by 10 K based on the choice of filter. However, the nightside surface temperature is largely unconstrained by the model ensemble, with more than 30 K difference between the warmest LMD model and the coldest CAM model. 

\begin{figure*}
\begin{center}
\includegraphics[width=6in]{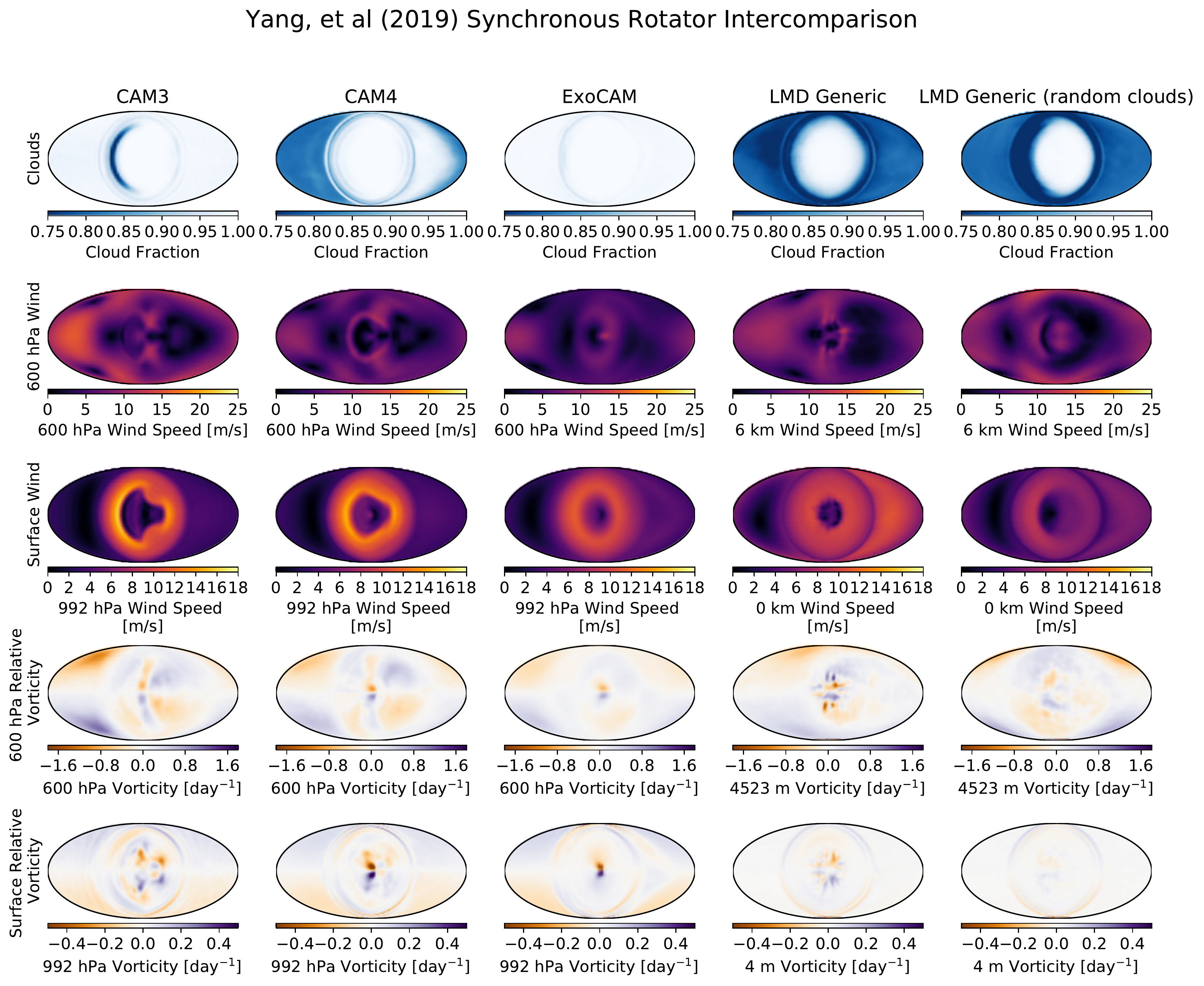}
\end{center}
\caption{Cloud fraction, wind speed and relative vorticity at or near 600 hPa, and wind speed and relative vorticity at or near the surface for each of the models included in the \citet{Yang2019} intercomparison synchronously rotating benchmark. All models have cloudy daysides, but differ on nightside cloud fraction, as well as on the strength of the nightside gyres and the structure of the substellar upwelling region.}\label{pyfig:yangTLmaps}
\end{figure*}

\begin{figure*}
\begin{center}
\includegraphics[width=5in]{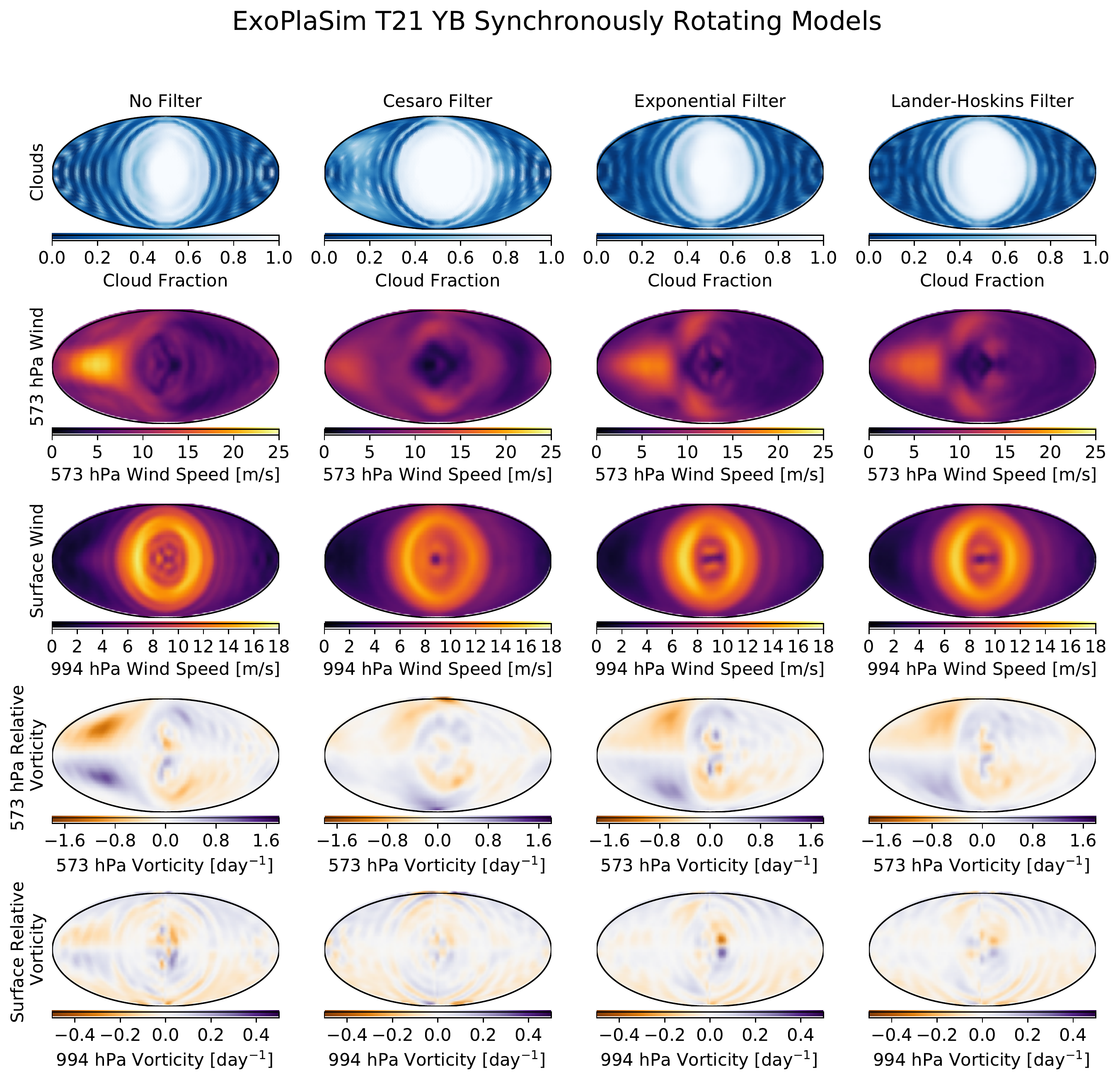}
\end{center}
\caption{Cloud fraction, wind speed and relative vorticity at 573 hPa (ExoPlaSim's closest layer to 600 hPa), and wind speed and relative vorticity at near the surface for T21 ExoPlaSim models configured similarly to the synchronously rotating benchmark in the \citet{Yang2019} intercomparison. Four filter configurations are shown here, including a configuration with no physics filter (ExoPlaSim's default). Each filtered model shown here has the physics filter applied at both spectral transforms. At T21, ExoPlaSim shows good qualitative agreement with the other models, though nightside clouds are affected by Gibbs ripples even with filtration, nightside cloud fractions are lower, and wind speeds are generally higher.}\label{pyfig:exoplasimTL_T21}
\end{figure*}

\begin{figure*}
\begin{center}
\includegraphics[width=5in]{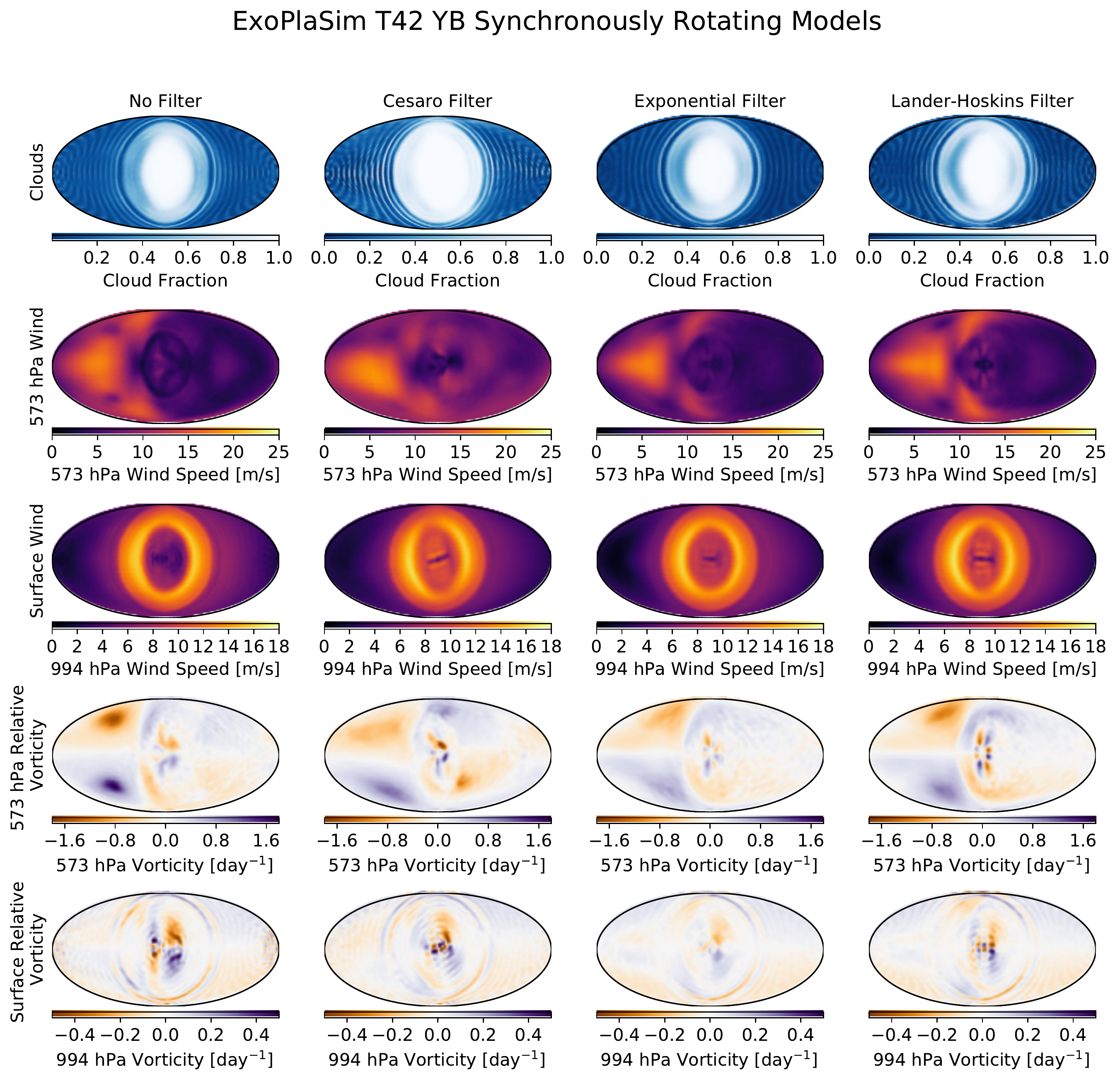}
\end{center}
\caption{Cloud fraction, wind speed and relative vorticity at 573 hPa (ExoPlaSim's closest layer to 600 hPa), and wind speed and relative vorticity at near the surface for T42 ExoPlaSim models configured similarly to the synchronously rotating benchmark in the \citet{Yang2019} intercomparison. Four filter configurations are shown here, including a configuration with no physics filter (ExoPlaSim's default). Each filtered model shown here has the physics filter applied at both spectral transforms. At T42, ExoPlaSim shows good qualitative agreement with the other models, though nightside clouds are affected by Gibbs ripples even with filtration, nightside cloud fractions are lower, and wind speeds are generally higher. ExoPlaSim's substellar upwelling is similar in structure to the CAM4, ExoCAM, and LMD Generic models.}\label{pyfig:exoplasimTL_T42}
\end{figure*}

As with the fast-rotating case, we show cloud fraction (recomputed), wind speed, and relative vorticity for the synchronously rotating models in \autoref{pyfig:yangTLmaps}, \autoref{pyfig:exoplasimTL_T21}, and \autoref{pyfig:exoplasimTL_T42}. All models produce a large mass of clouds on the dayside, with a slight reduction in cloud fraction on the morning limb of the dayside. Horizontal winds at the surface of the dayside are strongest away from the substellar point, forming an inflow region from which air is pulled into the upwelling region at the substellar point. In the mid-troposphere, the strongest winds are found near the equator on the nightside, where high-latitude gyres accelerate a zonal jet across the nightside. Beyond these features, however, models disagree on overall cloud fraction, especially on the nightside \citep{Yang2019}. ExoPlaSim models have the least nightside cloud coverage of all the models,  regardless of the choice of filter. ExoPlaSim models also have higher wind speeds, as in the fast-rotating case. The various models also show differences in atmospheric dynamics as shown through relative vorticity, both in the strength of the nightside gyres, and in the structure of the substellar upwelling region. ExoPlaSim shares features in this respect with the LMD Generic, CAM4, and ExoCAM models. 

To further explore ExoPlaSim's similarities and differences with the YB models, we consider their dynamical features when rotated into a coordinate system more appropriate for synchronous rotators. \citet{Koll2015} introduced a coordinate system in which the `North' pole is the substellar point, the `South' pole is the antistellar point, and the `equator' is the terminator. \citet{Hammond2021} showed that such a coordinate system could be useful for visualizing the divergent and rotational components of the circulation on a synchronously rotating planet. \citet{Hammond2021} further showed that such a coordinate system permitted the calculation of a `synchronously rotating streamfunction', which demonstrated the dipolar overturning circulation common to synchronously rotating planets better than the traditional streamfunction calculated in equatorial coordinates. 

We define a slightly different synchronous rotator coordinate system than \citet{Koll2015}, however. \citet{Koll2015} defined a left-handed coordinate system, with (0$^\circ$, 0$^\circ$) corresponding to the North equatorial pole, and 90$^\circ$ synchronous longitude corresponding to the point where the equator crosses the evening terminator. We define instead a right-handed coordinate system, such that the equator at the evening terminator is still 90$^\circ$ synchronous longitude, but (0$^\circ$, 0$^\circ$) is now the South equatorial pole. This has the desirable feature that in a latitude-longitude plot from 0$^\circ$ synchronous longitude to 360$^\circ$, the North equatorial pole is centered, and eastward flow (in the equatorial sense) revolves counter-clockwise about the center of the plot. This is analogous to looking down on the planet from above the North equatorial pole, and the coordinate transform itself is equivalent to a single 90$^\circ$ rotation from the North equatorial pole towards the substellar point. This coordinate system is also more consistent with the traditional equatorial coordinate system, which is also a right-handed coordinate system when viewed from above the atmosphere (it is a left-handed coordinate system  when looking outward from the ground). Because our modified coordinate system is a simple rotation, the rotational and divergent components of the horizontal circulation are unaffected, as they are invariant under rotation \citep{Hammond2021}. The synchronously rotating streamfunction is also unaffected. Routines to compute this coordinate transformation for both scalar and vector fields are included with ExoPlaSim's Python API (described in \autoref{pysec:Python}).

\begin{figure*}
\begin{center}
\includegraphics[width=6in]{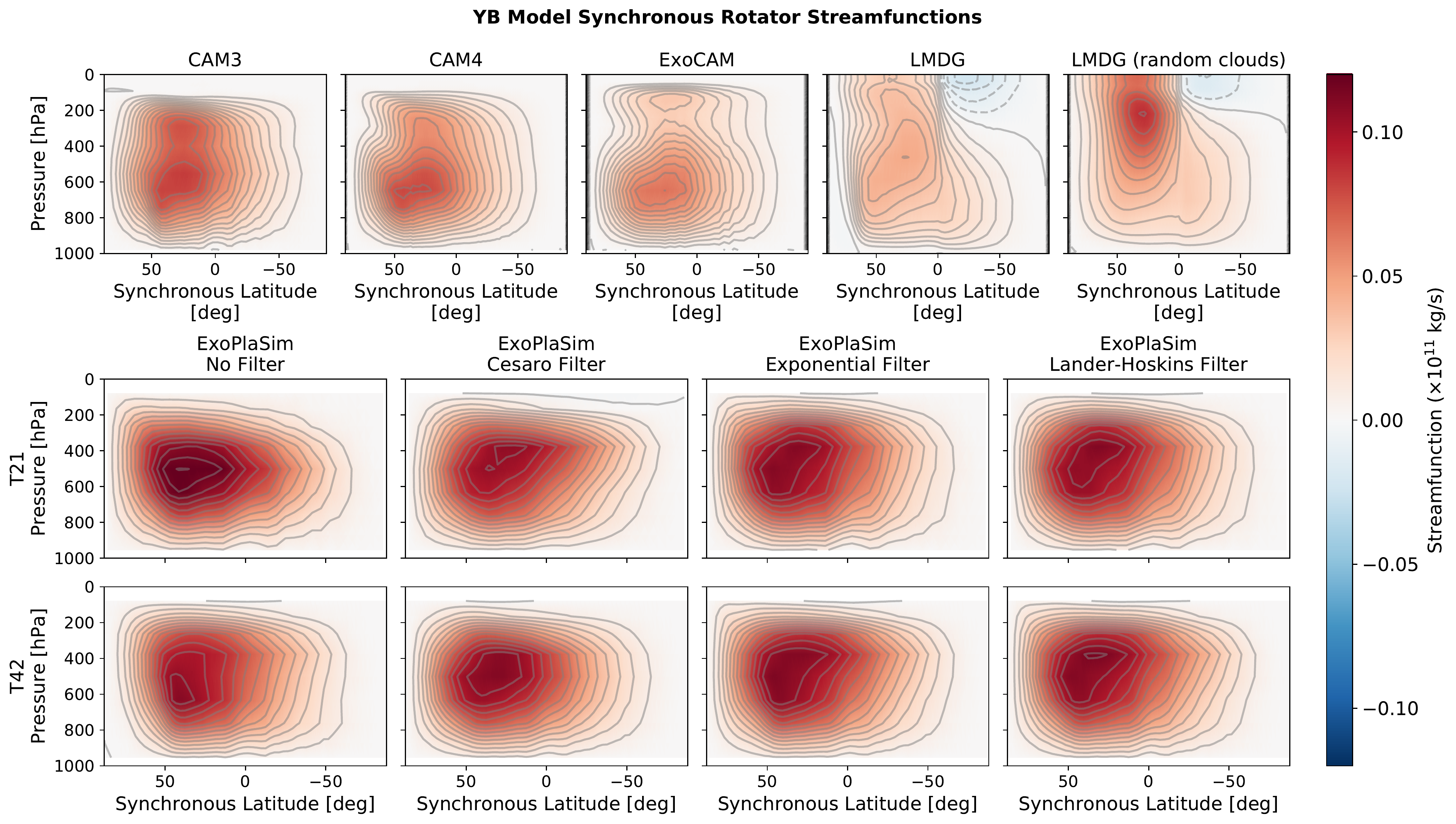}
\end{center}
\caption{The synchronously rotating streamfunction \citep{Hammond2021} of each of the models in the \citet{Yang2019} intercomparison, plus ExoPlaSim models at T21 and T42 resolutions, with four different physics filter types, each applied at both spectral transforms. This streamfunction is calculated in a rotated synchronous rotator coordinate system, such that $90^\circ$ latitude is the substellar point, $0^\circ$ is the terminator, and $-90^\circ$ is the antistellar point. Positive values indicate clockwise rotation. All models shown feature globally-dipolar overturning circulation, such that air rises at the substellar point, travels to the nightside, subsides at the antistellar point, and returns to the substellar point at the surface. The models disagree quantitatively on the shape and strength of the dipolar overturning circulation.}\label{pyfig:streamfxn}
\end{figure*}

\autoref{pyfig:streamfxn} shows the synchronously rotating streamfunctions of the \citet{Yang2019} intercomparison models, as well as for ExoPlaSim with four physics filter configurations each at T21 and T42 resolution, with the filter applied at both spectral transforms. All models shown are qualitatively similar, with one large dipolar overturning cell, such that air rises at the substellar point, is carried to the nightside, subsides near the antistellar point, and returns to the substellar point along the surface. The models disagree quantitatively on the strength of the overturning circulation, the vertical extent of the cell, its internal structure (ExoCAM shows evidence of a second dipolar cell atop the primary one), and whether there is a counterrotating cell aloft on the nightside (though this may also be a common feature of all models given sufficient vertical resolution, and is simply not resolved by models other than LMD in this test). ExoPlaSim generally shows stronger dipolar overturning circulation than any of the other models, but the morphology of the dipolar cell is most similar to CAM3.   

\begin{figure*}
\begin{center}
\includegraphics[width=6in]{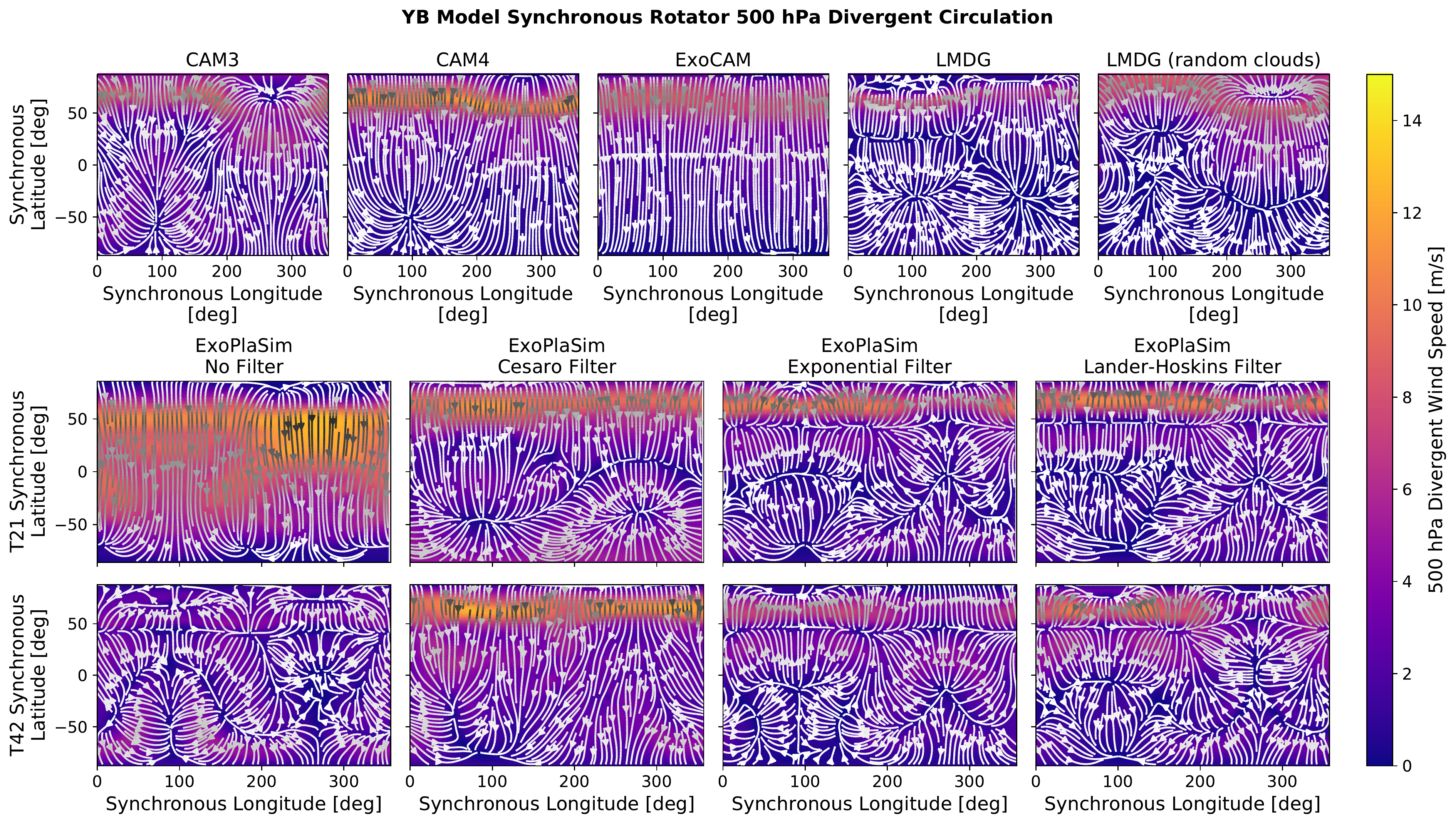}
\end{center}
\caption{The divergent (irrotational) circulation at 500 hPa of each of the models in the \citet{Yang2019} intercomparison, plus ExoPlaSim models at T21 and T42 resolutions, with four different physics filter types, each applied at both spectral transforms. $90^\circ$ synchronous latitude corresponds to the substellar point, $0^\circ$ latitude is the terminator, and $-90^\circ$ is the antistellar point. $0^\circ$ synchronous longitude is defined as the South equatorial pole, such that the North pole is centered in the plot, and eastward equatorial flows proceed counter-clockwise. The divergent flow at 500 hPa is largely disorganized, as this pressure corresponds to the center of the dipolar overturning cell. The exception is the unfiltered T21 ExoPlaSim model, suggesting that the absence of a physics filter can lead to qualitative changes in circulation. That model appears similar to ExoCAM, but since ExoCAM has two circulation cells, as shown in the previous figure, this may be a coincidence.}\label{pyfig:div500hpa}
\end{figure*}

\begin{figure*}
\begin{center}
\includegraphics[width=6in]{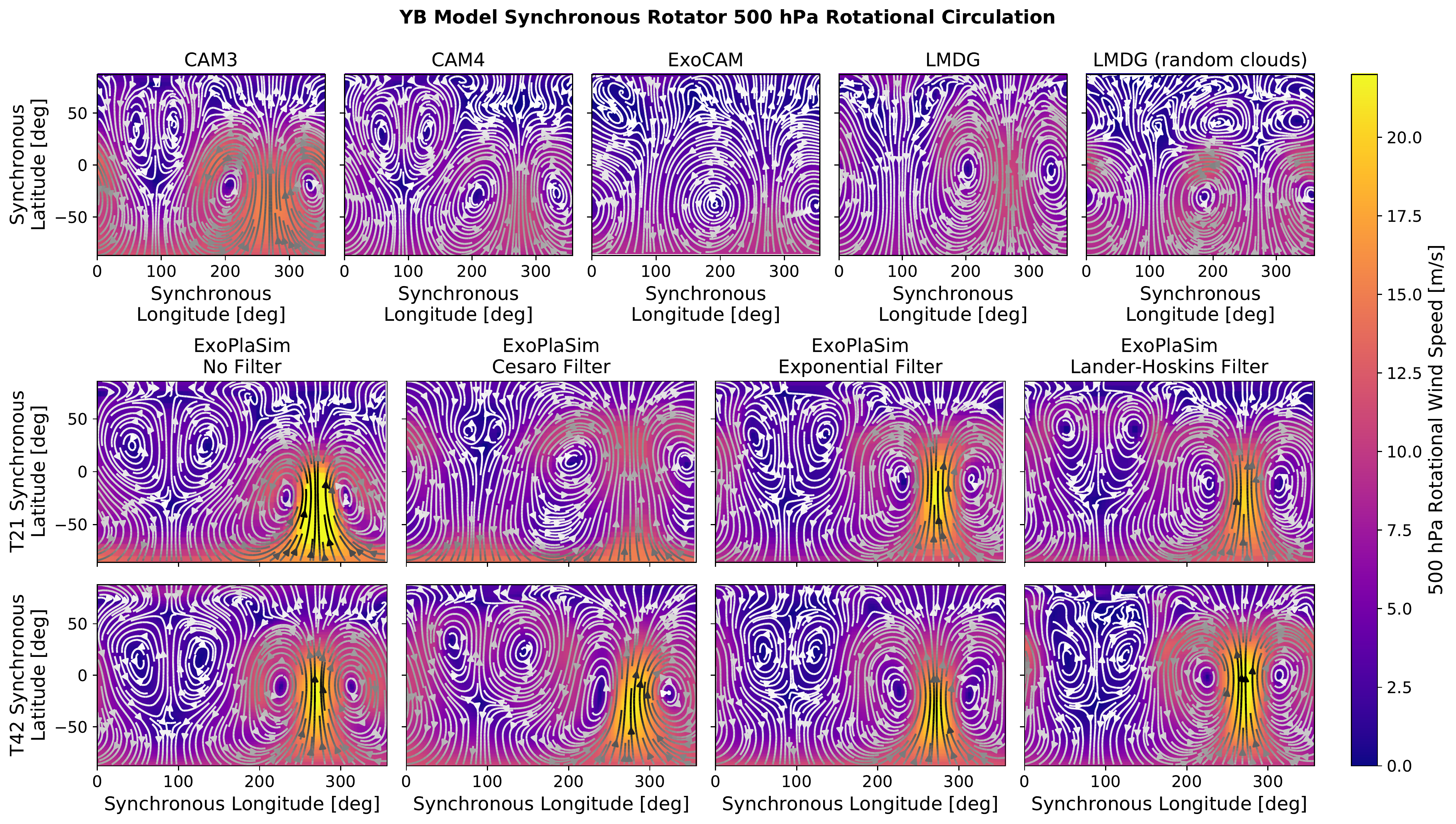}
\end{center}
\caption{The rotational (non-divergent) circulation at 500 hPa of each of the models in the \citet{Yang2019} intercomparison, plus ExoPlaSim models at T21 and T42 resolutions, with four different physics filter types, each applied at both spectral transforms. $90^\circ$ synchronous latitude corresponds to the substellar point, $0^\circ$ latitude is the terminator, and $-90^\circ$ is the antistellar point. $0^\circ$ synchronous longitude is defined as the South equatorial pole, such that the North pole is centered in the plot, and eastward equatorial flows proceed counter-clockwise. The rotational flow at 500 hPa is characterized by weak gyres immediately downstream of the substellar point, in the equatorial outflow region from the substellar upwelling region, where the zonal jet forms, and stronger gyres on the eastern limb of the nightside, driving a strong equatorial jet. This feature is shared by all the models, but is strongest in ExoPlaSim.}\label{pyfig:rot500hpa}
\end{figure*}

In \autoref{pyfig:div500hpa} and \autoref{pyfig:rot500hpa}, we show the divergent and rotational components of the circulation respectively at 500 hPa, in our synchronous coordinate system. The rotational and divergent components are computed through a Helmholtz decomposition following \citet{Hammond2021}, which we perform using the \texttt{windspharm} package \citep{Dawson2016}. This pressure level corresponds roughly to the center of the dipolar overturning cell, and therefore captures circulation not described by the bulk flows at the edges of the dipolar cell, which manifest as nearly-homogeneous divergent flows from one synchronously rotating pole to the other. At 500 hPa, the circulation is dominated by rotational flow, in the form of an equatorial jet that forms in the Eastern outflow of the substellar upwelling region amid weak gyres, and is accelerated across the nightside by much stronger high-latitude gyres. The divergent flow is thus largely disorganized in each of the models, except for the T21 ExoPlaSim model with no physics filter. This suggests that a physics filter is necessary to correctly resolve the atmospheric dynamics of synchronously rotating planets at low horizontal resolution in ExoPlaSim. We also note that the equatorial jet speed is higher in ExoPlaSim than in the other models, and the high-altitude divergent outflow and near-surface divergent return flow are also faster in ExoPlaSim than in the other models, which is consistent with the stronger synchronously rotating streamfunctions shown in \autoref{pyfig:streamfxn}. It is unknown whether this indicates a model limitation of ExoPlaSim, or if aspects of the flow are better-represented in ExoPlaSim than in the other models as a result of the spectral core. We recommend that other GCMs with spectral cores be tested on the \citet{Yang2019} synchronously rotating benchmark to determine if this is a desirable feature of spectral models, or a bias to account for.

In general, we find good qualitative agreement between ExoPlaSim and the models presented in \citet{Yang2019}, both in the Earth-like fast-rotator benchmark, and the synchronously rotating benchmark. ExoPlaSim shows a slight warm bias in the fast-rotating case relative to the other models, along with slightly higher wind speeds. Surface temperatures in the synchronously rotating benchmark however fall within the range produced by the \citet{Yang2019} ensemble. ExoPlaSim shows excellent agreement in the dynamical morphology of the atmosphere in the synchronously rotating case when either the exponential or Lander-Hoskins physics filters are used during both spectral transforms, although ExoPlaSim displays significantly less cloud cover on the nightside. The dipolar overturning circulation that spans from the substellar point to the antistellar point is stronger in ExoPlaSim than in the other models, as is the mid-troposphere equatorial jet, which together with the reduced nightside cloud cover may explain why ExoPlaSim is slightly warm in the fast-rotating case, but not the synchronously rotating case---the more-robust overturning circulation and equatorial jet transport more heat to the nightside, where it is lost more efficiently due to the lack of clouds, removing any warm bias that ExoPlaSim may have by virtue of its radiation scheme. We note as well that because the \citet{Yang2019} benchmarks use prescribed albedos and ignore sea ice, this test does not allow us to validate our surface spectral reflectivity parameterization against other models.

\subsubsection{THAI-B comparison}\label{pysec:thai}

While ExoPlaSim performs well when compared to other models on the \citet{Yang2019} synchronously rotating benchmark, that model configuration is not indicative of all synchronously rotating habitable planets. Its orbital (and rotation) period of 60 days, for example, is significantly longer than either TOI-700d \citep[37.4 days,][]{Gilbert2020,Rodriguez2020,Suissa2020} or Proxima Centauri b \citep[11.2 days,][]{Anglada-Escude2016}. These faster rotation rates can lead to qualitatively different circulation regimes \citep{Haqq-Misra2018}. The 3400 K blackbody spectrum assumed in the \citet{Yang2019} intercomparison is also on the warm and therefore blue end of the M dwarf distribution \citep{ReidHawley2005}. TRAPPIST-1e, as explored in the THAI model intercomparison framework \citep{Fauchez2020}, therefore serves as a useful second benchmark, as it lies near the opposite end of the distribution from the planet in \citet{Yang2019}, with a rotation period of 6.1 days and a 2600 K M8-type star \citep{Costa2006,Grimm2018}. Furthermore, TRAPPIST-1e also differs from Earth in its radius and surface gravity \citep{Grimm2018}, and is a real exoplanet that will be targeted by upcoming telescopes such as JWST. It is therefore additionally useful as a test case to measure how different climate models will perform relative to each other when informing climate retrievals of real exoplanets \citep{Fauchez2020,THAIreport}. The THAI intercomparison also includes a pure-CO$_2$ test case \citep{Fauchez2020}, which facilitates comparison in cases with non-Earthlike atmospheres.

We configure ExoPlaSim as described in \citet{Fauchez2020} for the `Hab1' and `Hab2' benchmarks, with a surface gravity of 9.12 m s$^{-2}$, a radius of 0.91 R$_\oplus$, and instellation of 900 W m$^{-2}$. Both Hab1 and Hab2 are aquaplanets, with 1 bar surface pressures. Hab 1 has 1 bar of N$_2$ and 400 ppm CO$_2$, while Hab2 simply has 1 bar of CO$_2$, and the mean molecular weight of each atmosphere is set to be consistent with those compositions. We remind the reader at this point that while we are including a comparison to the Hab2 benchmark, prior work has shown PlaSim to be unreliable at such high CO$_2$ \citep{Paradise2017}, and so we include Hab2 purely for the sake of comparison and completeness. The ocean is treated as a 100-meter-thick mixed-layer slab, with an ocean albedo fixed to 0.06, and snow and ice albedos fixed to 0.25. The THAI intercomparison protocol does not explicitly specify a direct-beam albedo zenith-angle dependence for the ocean surface \citep{Fauchez2020}, and some of the models in the intercomparison do include such a dependence in their benchmark simulations (Thomas Fauchez, private communication, June 27, 2020). We therefore use ExoPlaSim's default ECHAM-3 parameterization for the direct-beam ocean reflectivity.

While \citet{Fauchez2020} are careful to prescribe the same initial conditions across all models, we find that our equilibrium climate solutions for TRAPPIST-1e are not sensitive to our choice of initial temperature. This is likely due to the fact that the sea ice bistability present in fast-rotating Earth-like climates \citep{Budyko1969} likely does not occur on slow-rotating and synchronously rotating planets \citep{Checlair2017,Abbot2018}, meaning there is only one equilibrium climate state for the model to reach. We therefore use ExoPlaSim's default warm-start initial conditions, in which an isothermal 250 K atmosphere is allowed to adjust adiabatically to a fixed 288 K surface temperature prior to model integration. 

For the standard Hab1 and Hab2 tests, we use a 2600 K blackbody spectrum. We also present two additional sets of models, one denoted Hab1$^*$ and Hab2$^*$, following \citet{Fauchez2020}, in which surface albedos are computed in ExoPlaSim rather than prescribed, and one denoted Hab1$^*_\nu$ and Hab2$^*_\nu$, which is identical to Hab1$^*$ and Hab2$^*$ except that instead of a 2600 K blackbody, we use a 2600 K [Fe/H]=0 Bt-Settl spectrum from the PHOENIX CIFIST2011\_2015 precomputed stellar model grid \citep{Baraffe2015,Allard2016}. We re-sample this spectrum to the wavelengths needed by ExoPlaSim using linear interpolation via the \mintinline[fontsize=\footnotesize]{python}{exoplasim.makestellarspec} module. For the models where the albedo is calculated by ExoPlaSim, we use the simplified version of ExoPlaSim's albedo scheme, in which a single bolometric albedo is computed and used for both shortwave bands. As the version in which a different value is used for each band is likely to be more accurate, but the simplified version is more robust, this allows us to test ExoPlaSim's performance in its more-likely use cases, where it is likely to disagree with other models the most. 

We again test each of the three physics filters, as well as the unfiltered case, for both T21 and T42. Unlike the YB models, we find that the THAI-B models are less sensitive to Gibbs ripples, and both the exponential and Lander-Hoskins filters give qualitatively similar results to the case with no filter. A more-detailed discussion of the performance of individual filters in the THAI-B models is given in Appendix A (available in the online supplementary material). Because we know that Gibbs ripples can pose problems for synchronously rotating planets in ways that are not always obvious, as with the mid-atmosphere divergent flow in the YB T21 models, and the physics filters do nonetheless result in smoother nightside temperature distributions, we show here only the results of the models with Lander-Hoskins filtration.

\begin{figure*}
\begin{center}
\makebox[\textwidth][c]{\includegraphics[width=7in]{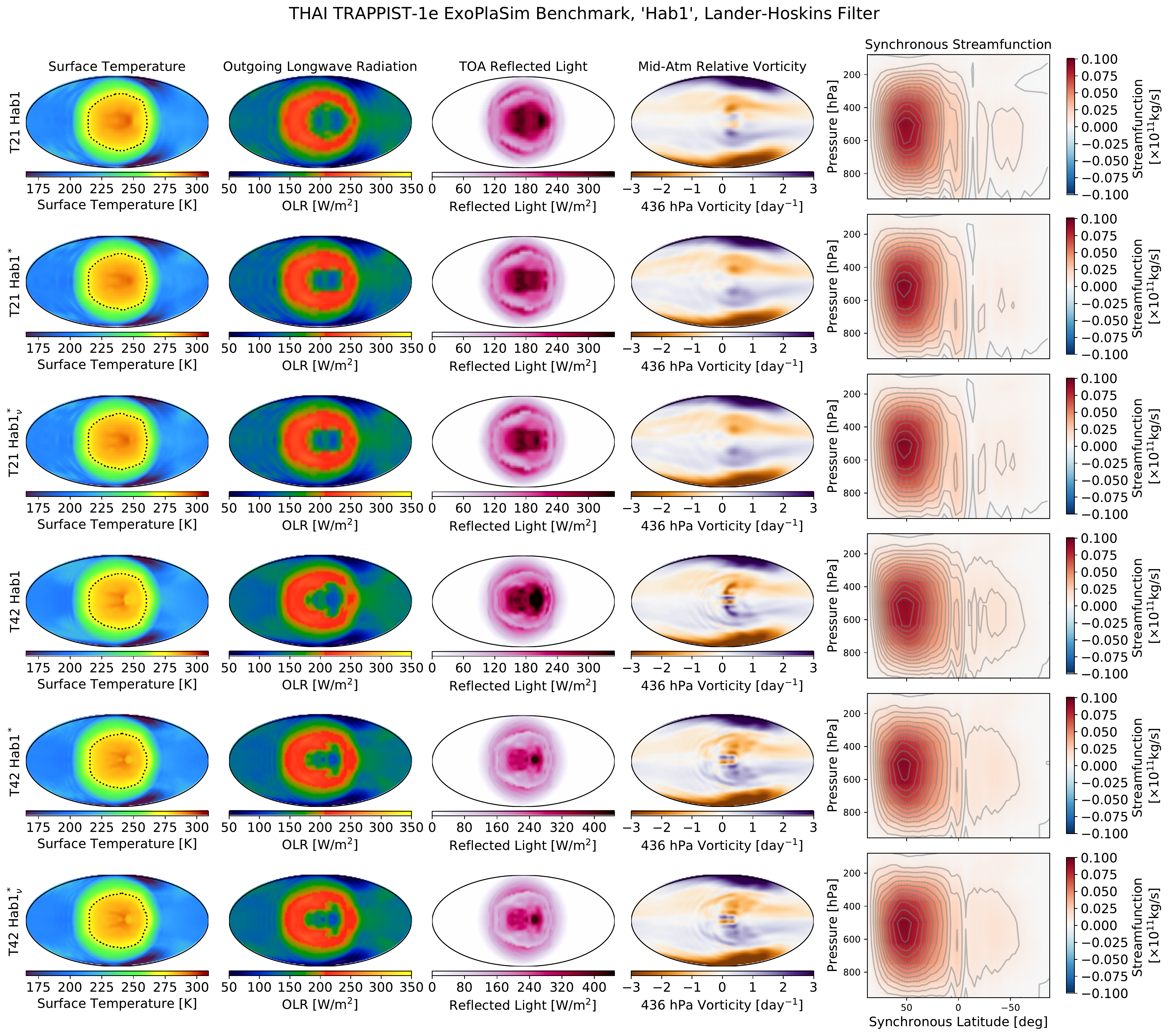}}
\end{center}
\caption{ExoPlaSim models of the `Hab1' scenario (1 bar N$_2$, 400 ppm CO$_2$) from the THAI intercomparison \citep{Fauchez2020}, at both T21 and T42 resolution, and with three different combinations of stellar spectrum and surface albedos. The Hab1 models use a 2600 K blackbody spectrum, and have prescribed ocean and ice albedos of 0.06 and 0.025, respectively. The Hab1$^*$ models replace the prescribed albedos with ExoPlaSim's internally-computed albedos. The Hab1$^*_\nu$ models further replace the blackbody spectrum with a 2600 K Bt-Settl spectrum with [Fe/H]=0, from the PHOENIX stellar model grid \citep{Baraffe2015,Allard2016}. Surface temperature, top-of-atmosphere (TOA) outgoing longwave (thermal) radiation, and TOA outgoing shortwave (reflected) light are shown, as well as the relative vorticity in the mid-troposphere, and the synchronously rotating streamfunction described in \citet{Hammond2021}. The colormaps used in the first three columns are chosen to facilitate comparison with \citet{Fauchez2020}, and the dotted black contour in the leftmost column indicates the 273.15 K isotherm. In this scenario, ExoPlaSim agrees well with the THAI models, and is relatively insensitive to albedo and spectrum parameterizations.}\label{pyfig:exoplasimthai_hab1}
\end{figure*}

\begin{figure*}
\begin{center}
\makebox[\textwidth][c]{\includegraphics[width=7in]{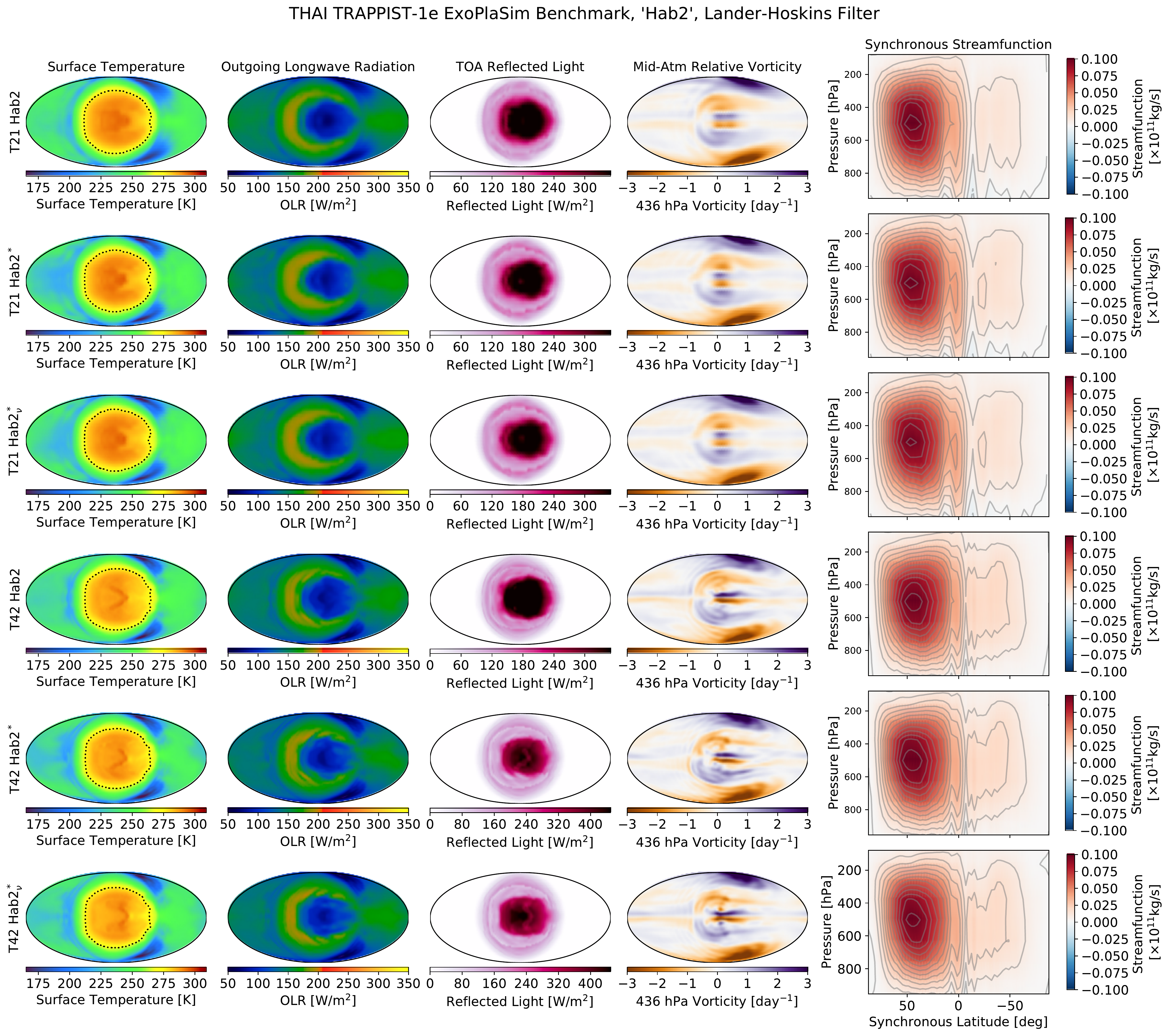}}
\end{center}
\caption{ExoPlaSim models of the `Hab2' scenario (1 bar CO$_2$) from the THAI intercomparison \citep{Fauchez2020}, at both T21 and T42 resolution, and with three different combinations of stellar spectrum and surface albedos. The Hab2 models use a 2600 K blackbody spectrum, and have prescribed ocean and ice albedos of 0.06 and 0.025, respectively. The Hab2$^*$ models replace the prescribed albedos with ExoPlaSim's internally-computed albedos. The Hab2$^*_\nu$ models further replace the blackbody spectrum with a 2600 K Bt-Settl spectrum with [Fe/H]=0, from the PHOENIX stellar model grid \citep{Baraffe2015,Allard2016}. Surface temperature, top-of-atmosphere (TOA) outgoing longwave (thermal) radiation, and TOA outgoing shortwave (reflected) light are shown, as well as the relative vorticity in the mid-troposphere, and the synchronously rotating streamfunction described in \citet{Hammond2021}. The colormaps used in the first three columns are chosen to facilitate comparison with \citet{Fauchez2020}, and the dotted black contour in the leftmost column indicates the 273.15 K isotherm. In this scenario, ExoPlaSim is significantly cooler than ExoCAM, likely due to the high-pCO$_2$ cooling biases identified in \citet{Paradise2017}.}\label{pyfig:exoplasimthai_hab2}
\end{figure*}

Surface temperature, outgoing longwave radiation (OLR), top-of-atmosphere net reflected light, mid-troposphere relative vorticity, and the synchronously rotating streamfunction as described in \citet{Hammond2021} for the Hab1 series of models are shown in \autoref{pyfig:exoplasimthai_hab1}. The Hab2 series of models are shown in \autoref{pyfig:exoplasimthai_hab2}. We find that ExoPlaSim performs well in the Hab1 case, with similar dayside maximum surface temperatures, outgoing longwave fluxes, and top-of-atmosphere reflection from clouds and sea ice as the other THAI-B models. We note that the THAI-B models show wide variation in terms of the regions on the dayside where thermal emission to space is maximized, and ExoPlaSim adds to that diversity---ExoPlaSim has a ring of maximum OLR on the dayside, with a larger central hole (due to clouds) than ROCKE-3D (the other model with a ring-like OLR structure) has. We also note that ExoPlaSim's nightside is on the cool end of the model ensemble, showing similar minimum temperatures to the UK Met Office's Unified Model, although most of the nightside is more similar in surface temperature to ExoCAM, LMDG, and ROCKE-3D. Notably, ExoPlaSim models run colder than all ensemble models in the cold traps on the night-side---approximately 40 K colder than the cold traps in the UM model, the coldest of the THAI ensemble \citep{THAI2021b}. It is unknown if this is due to inadequacies in ExoPlaSim's radiative transfer or if it is a consequence of the spectral core doing a better job of resolving the night-side high-latitude vortices. ExoPlaSim has significantly cooler global mean surface temperatures than the other GCMs in the Hab2 case, with dayside maximum temperatures approximately 10 K cooler than the coldest substellar temperatures in the ensemble (ROCKE-3D), and dayside and nightside mean temperatures 30--35 K colder than the overall coldest model in the ensemble (LMDG) \citep{THAI2021b}. The cold traps are again much colder in the Hab2 case, just over 80 K colder than the coldest cold traps in the ensemble (UM). This systematic cold bias in the Hab2 case indicates that ExoPlaSim is likely underperforming in the pure-CO$_2$ case, likely due to the cooling bias identifed at high pCO$_2$ in \citet{Paradise2017}.

Both our Hab1 and Hab2 models are characterized similar to the YB models, with nearly-circular ice-free regions on the dayside, a large mass of clouds at the substellar point, atmospheric circulation dominated by dipolar overturning circulation driven by divergent flow, two large high-latitude nightside gyres, a mid-troposphere equatorial jet, and elevated vorticity in the substellar upwelling region. Our THAI-B models differ from the YB models, however, in that more of the vorticity generated at the substellar point is advected eastward, and the equatorial jet has a larger contribution to overall circulation, as seen by the disruption and weakening of the dipolar overturning cell on the nightside of the planet in the rightmost columns of \autoref{pyfig:exoplasimthai_hab1} and \autoref{pyfig:exoplasimthai_hab2}. This is because the equatorial jet both brings air to and from the substellar point, so if it comes to dominate the nightside circulation, the dipolar overturning circulation seen in slower-rotating models will become less apparent. We also find that the differences between Hab1, Hab1$^*$, and Hab1$^*_\nu$ are minor, as with the Hab2 models---the versions with internally-computed albedos are slightly cooler, due to slightly higher albedos resulting from the internal calculation. This suggests that the climates of synchronously rotating planets in ExoPlaSim are not particularly sensitive to the details of the incident spectrum and albedo parameterization.

Taking the results of our THAI-B and YB comparisons together, we conclude that ExoPlaSim is able to reproduce to first-order the results of other, more-sophisticated GCMs when applied to synchronously rotating planets around M-type stars, which is comparable to the degree of variation between the more sophisticated models themselves. ExoPlaSim is not able to model the more-advanced physics often included in those models, so cannot be used for advanced modeling studies of certain phenomena on synchronously rotating planets, but for broad studies of general climate states, climate dynamics, and habitability, we conclude that ExoPlaSim is likely to produce acceptable results at relatively low computational cost. We also note that as neither the \citet{Yang2019} nor \citet{Fauchez2020} intercomparison included benchmarks with non-Earthlike surface pressures, we are unable to directly test how ExoPlaSim compares in cases of high or low surface pressure, in which changes to the Rayleigh scattering parameterization and vertical discretization may become important. We do note however that in \citet{Paradise2021}, we found that ExoPlaSim's Rayleigh scattering parameterization agrees well with SBDART for a solar-like spectrum. We note that \citet{Turbet2018} included a range of surface pressures in their simulations of the TRAPPIST-1 planets with LMDG, but as they included CO$_2$ condensation, we are unable to directly compare ExoPlaSim to their models. We encourage future studies of real exoplanets such as TRAPPIST-1e and TOI-700d that include benchmark-type models similar to those in the THAI intercomparison with low (0.1 bar) surface pressure and high (10 bar) surface pressure in addition to the more-typical Earth-like (1 bar) surface pressure.

\section{Demonstration: Confirming existing results, new results}\label{pysec:demonstration}

In this section, we demonstrate ExoPlaSim with two simple experiments. First, we use ExoPlaSim to replicate the findings in \citet{Kopparapu2013,Yang2014} and \citet{Kopparapu2016} that the habitable zone moves outward to lower fluxes as stellar mass decreases. Because ExoPlaSim has not been validated for the extremely hot and moist climates necessary to model the inner edge of the habitable zone, and has known radiative biases in the high-pCO$_2$ regimes necessary to model the outer edge \citep{Paradise2017}, rather than try to replicate the \citet{Kopparapu2013} results directly, we instead use temperature trends at temperate fluxes, and in particular the trend in the instellation at which the maximum surface temperature goes below freezing, as proxies for how the inner and outer edges of the habitable zone are likely to change with stellar effective temperature. Second, we use ExoPlaSim to test the prediction in \citet{Paradise2021} that accounting for the stellar spectrum will reduce the cooling efficiency of Rayleigh scattering at higher surface pressures on synchronously rotating planets than was found using a solar-like spectrum.

\subsection{Habitable zone trends}

The habitable zone is expected to move outward to lower fluxes around lower-mass stars primarily due to the lower Bond albedo that results with redder incident spectra (due to the lower reflectivity of water, snow, and ice at red and near-infrared wavelengths). This is complicated slightly by the trend towards albedo-raising dayside cloud coverage on synchronously rotating planets \citep{Yang2013}, but the overall trend is still expected to be one of higher mean and maximum surface temperatures at a given flux. As ExoPlaSim's surface albedos depend on the incident stellar spectrum and reproduces the expected day-side clouds, this habitable zone trend ought to emerge in habitable zone ExoPlaSim parameters sweeps in which the stellar effective temperature is varied. 

To test whether ExoPlaSim correctly finds that the habitable zone moves outward to lower fluxes around lower-mass stars, we run two grids of 100 aquaplanet models each, at T21 and T42 resolutions, varying both instellation and stellar effective temperature. We vary instellation from 500--1500 W m$^{-2}$, and stellar effective temperature from 2500--4000 K. Because synchronously rotating planets have shorter orbital periods and thus faster rotation rates for a given flux around lower-mass stars compared to higher-mass stars, we follow \citet{Yang2014} and \citet{Kopparapu2016} and allow the rotation rate to vary with stellar effective temperature, ranging from less than 5 days for the warmest planets around 2500 K stars, to greater than 150 days for the coolest planets around 4000 K stars. We run each model to energy balance equilibrium (typically 75--200 years), and use the average climate of the final year of output. We begin with a 30-minute timestep for T21 models and a 15-minute timestep for T42 models, but as we found that some of the warmer models crashed, we reduced their timesteps to 15 or 20 minutes for T21, and 5 or 10 minutes for T42, at which point they ran smoothly. We use an exponential physics filter, applied at both spectral transforms, and turn off ozone. We use 1 bar of N$_2$ and 400 $\mu$bar of CO$_2$ for each model. 

We calculate the orbital period, and thus the rotation rate, from an empirical mass-temperature relationship for K- and M-type stars above 3300 K \citep{Mann2013}. The luminosity is similarly taken from an empirical temperature-luminosity relationship \citep{Mann2013}:
\begin{linenomath}
\begin{multline}
M(T_\text{eff}\geq3300\text{ K}) = -22.297 + 1.544\times10^{-2}T_\text{eff} \\ -3.488\times10^{-2}T_\text{eff}^2+2.65\times10^{-10}T_\text{eff}^3
\end{multline}
\begin{multline}
L(T_\text{eff}\geq3300\text{ K}) = -0.781+7.4\times10^{-4}T_\text{eff} \\ -2.49\times10^{-7}T_\text{eff}^2+2.95\times10^{-11}T_\text{eff}^3
\end{multline}
\end{linenomath}
For stars below 3300 K, we derive a radius from an empirical temperature-radius relationship for very low-mass stars \citep{Cassisi2019}, and from there a luminosity. We then use a simple mass-luminosity relationship \citep{Duric2004} to estimate a stellar mass:
\begin{linenomath}
\begin{align}
R(T\text{eff}<3300\text{ K}) &= 0.763\left(\frac{T_\text{eff}}{5777.0\text{ K}}\right)-0.224 \\ \nonumber \\ 
L(T\text{eff}<3300\text{ K}) &= 4\pi{R}^2\sigma_{SB}T_\text{eff}^4 \\ \nonumber \\
M(T_\text{eff}<3300\text{ K}) &= \frac{1}{0.23}\left(\frac{L}{\text{ L}_\odot}\right)^\frac{1}{2.3} 
\end{align}
\end{linenomath}
The semimajor axis corresponding to a given instellation, and thus the orbital period, then follow from the stellar mass and luminosity via first principles and Kepler's Third Law.

\begin{figure*}
\begin{center}
\includegraphics[width=6in]{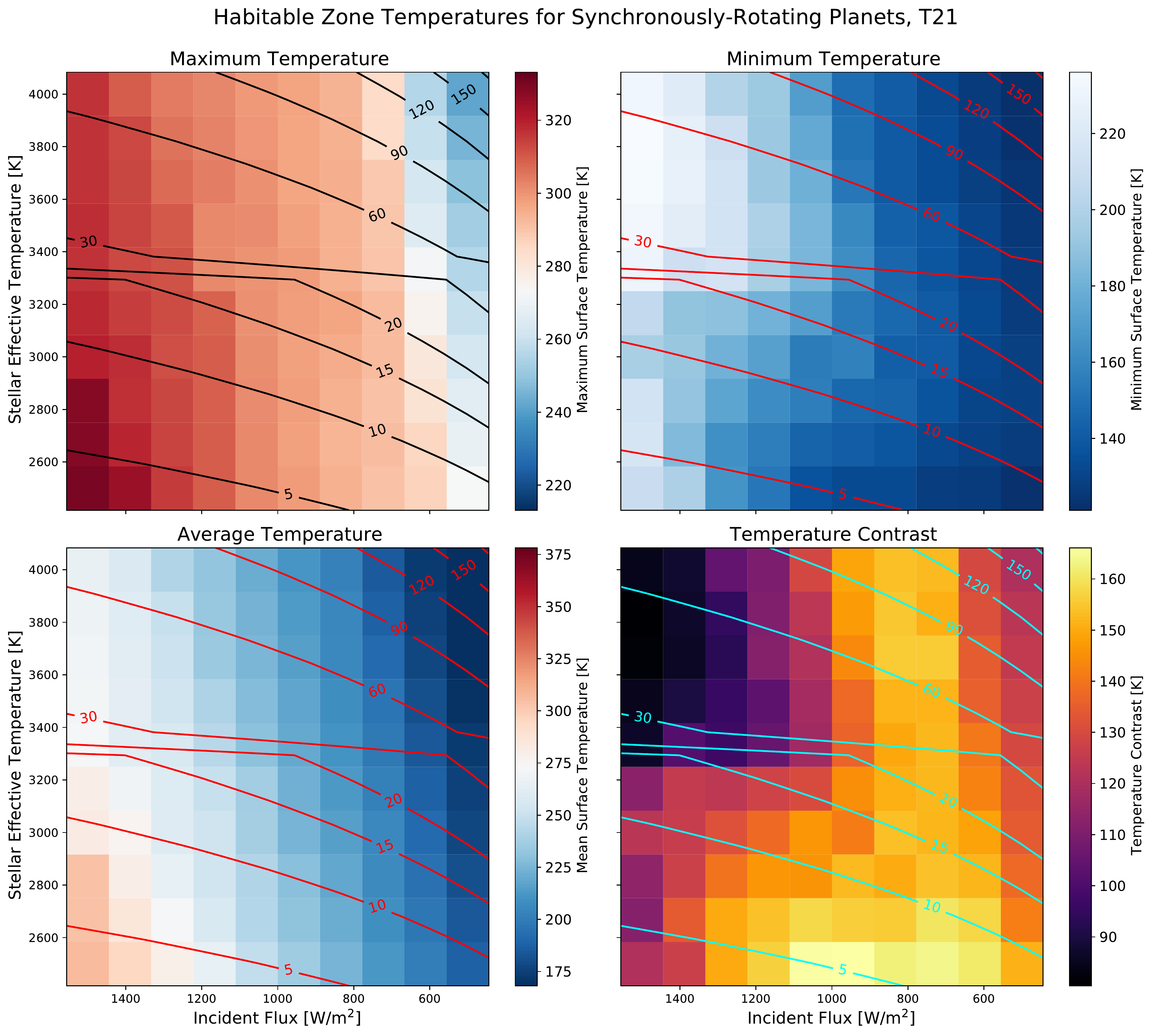}
\end{center}
\caption{Maximum, minimum, and mean annual surface temperatures, as well as temperature contrast, for 100 synchronously rotating models at varying instellations around stars of varying effective temperatures, at T21 resolution. The solid-line contours indicate rotation period contours (in days). Each model in the grid has a rotation rate that is consistent with the instellation it receives and the effective temperature of its parent star, assuming a main sequence star. Note that as stellar effective temperatures decrease, climates at a given flux warm up, as measured by maximum and mean temperatures, consistent with the habitable zone moving outward to lower fluxes around lower-mass stars.}\label{pyfig:colorgridT21}
\end{figure*}

\begin{figure*}
\begin{center}
\includegraphics[width=6in]{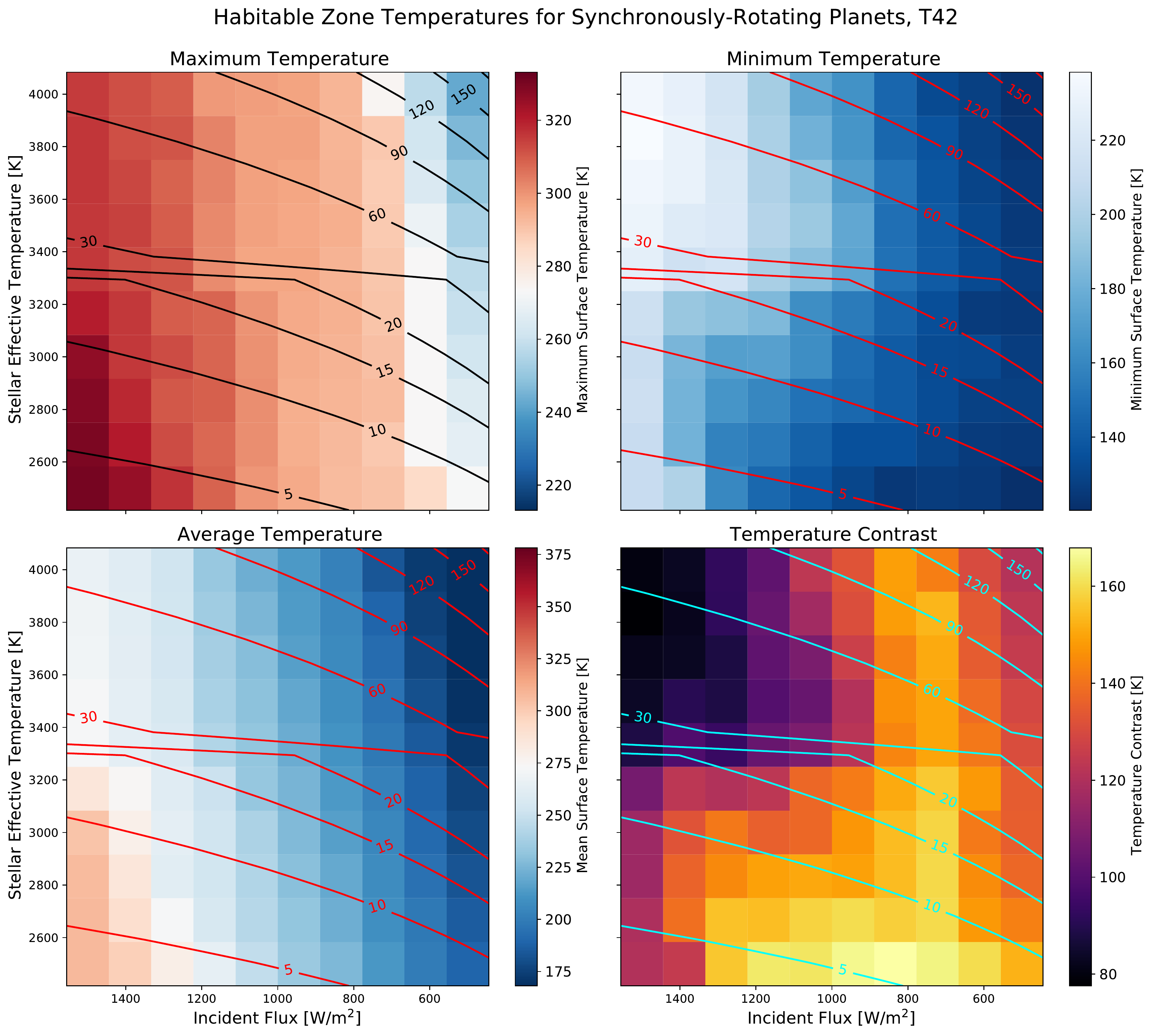}
\end{center}
\caption{Maximum, minimum, and mean annual surface temperatures, as well as temperature contrast, for 100 synchronously rotating models at varying instellations around stars of varying effective temperatures, at T42 resolution. The solid-line contours indicate rotation period contours. Each model in the grid has a rotation rate that is consistent with the instellation it receives and the effective temperature of its parent star, assuming a main sequence star. Note that as stellar effective temperatures decrease, climates at a given flux warm up, as measured by maximum and mean temperatures, consistent with the habitable zone moving outward to lower fluxes around lower-mass stars. These T42 results are also consistent with our T21 results \autoref{pyfig:colorgridT21}.}\label{pyfig:colorgridT42}
\end{figure*}

Our results are shown in \autoref{pyfig:colorgridT21} and \autoref{pyfig:colorgridT42}. We find that for a given incident flux, at both T21 and T42 synchronously rotating planets are warmer around cooler stars, which is consistent with the findings in \citet{Kopparapu2013,Yang2014}, and \citet{Kopparapu2016} that the habitable zone should move outward to lower fluxes around lower-mass stars.

\subsection{Effect of background gas pressure with an M dwarf spectrum}

In \citet{Paradise2021}, we included a grid of synchronously rotating models in our study of the role of background gas pressure (such as N$_2$) on the climate. We had found that background gas pressure, represented via pN$_2$ in our study, could have a large effect on the climate through three mechanisms: heating, via pressure-broadening of the CO$_2$ and H$_2$O greenhouse effects, and cooling, via increased heat transport efficiency and increased reflection via Rayleigh scattering \citep{Paradise2021}. In that study, however, we only used a Sun-like input spectrum. We found that on synchronously rotating planets, the heat transport mechanism became more important and more efficient, which meant that elevated pN$_2$ would cool a synchronously rotating planet at all but the highest incident fluxes we sampled. We predicted that with a realistic input spectrum, however, cooling via Rayleigh scattering would become less important, and this might affect our conclusions for synchronously rotating planets.

We therefore follow up on that work by running a new grid of 460 synchronously rotating models, spanning 0.1--10 bars and 400--2600 W m$^{-2}$. We use a uniform rotation period of 15 days, and a 3000 K blackbody stellar spectrum, with an exponential physics filter applied at both spectral transforms. We run at T21 with an aquaplanet configuration, ignore ozone, and add 400 $\mu$bar CO$_2$ to each model's pN$_2$. We use the rescaled PlaSim vertical discretization described in \autoref{pysec:vertical}, with the top layer pinned to 50 hPa. As in \citet{Paradise2021}, we use a default 30-minute timestep, but reduce the timestep for high-pressure and low-pressure models, as well as for the hottest models.

\begin{figure*}
\begin{center}
\includegraphics[width=6in]{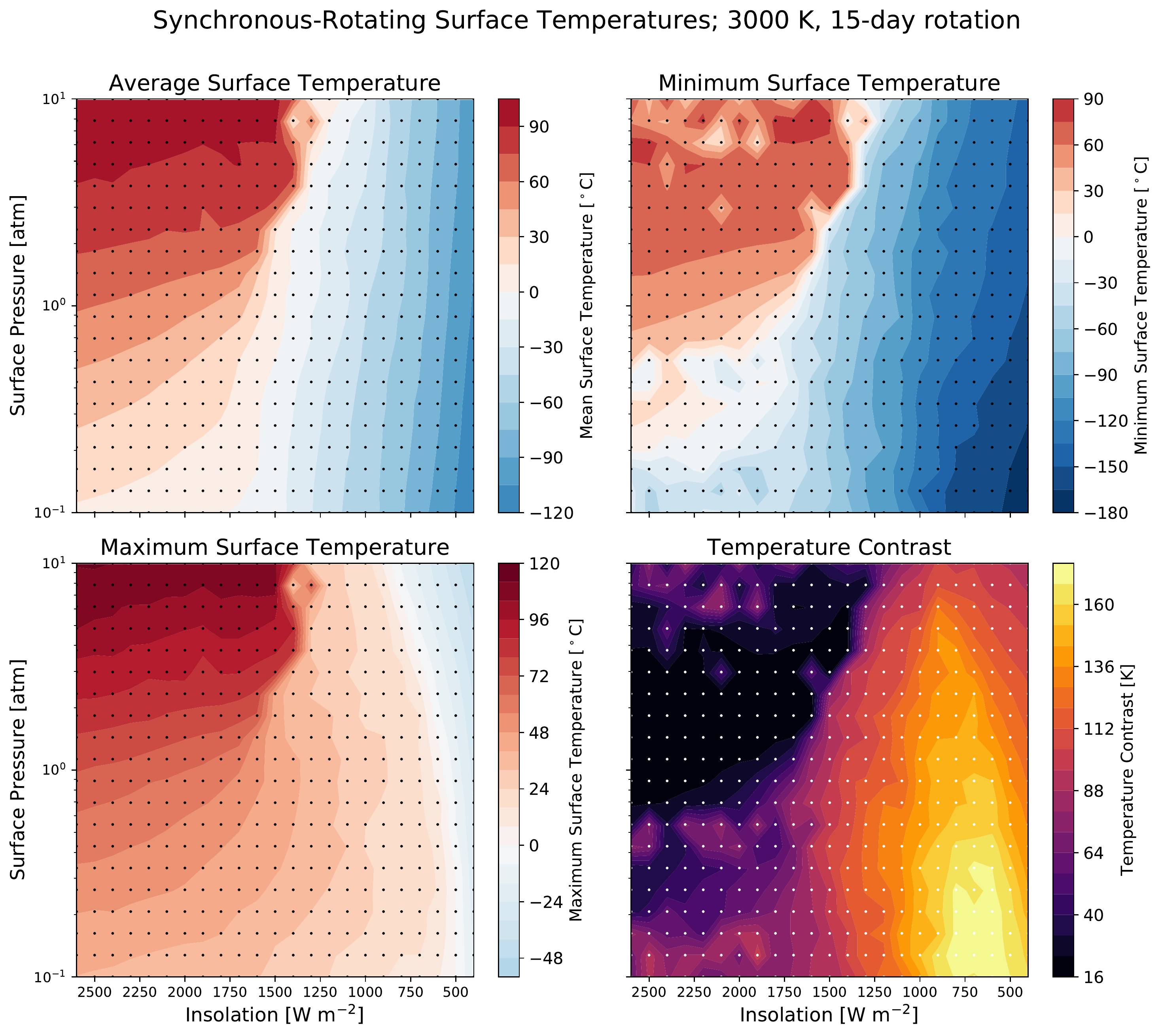}
\end{center}
\caption{Maximum, minimum, and mean annual surface temperatures, as well as temperature contrast, for 460 synchronously rotating models at varying instellations and N$_2$ partial pressures, at T21 resolution. Each dot indicates the location of a model in the parameter space. This experiment is similar to that described in \citet{Paradise2021}, except that a realistic 3000 K M dwarf spectrum is included. The redder spectrum dramatically reduces the cooling efficiency of Rayleigh scattering at high surface pressures, allowing heating through pressure broadening to dominate over much of the parameter space.}\label{pyfig:mdwarfpn2}
\end{figure*}

\begin{figure*}
\begin{center}
\includegraphics[width=6in]{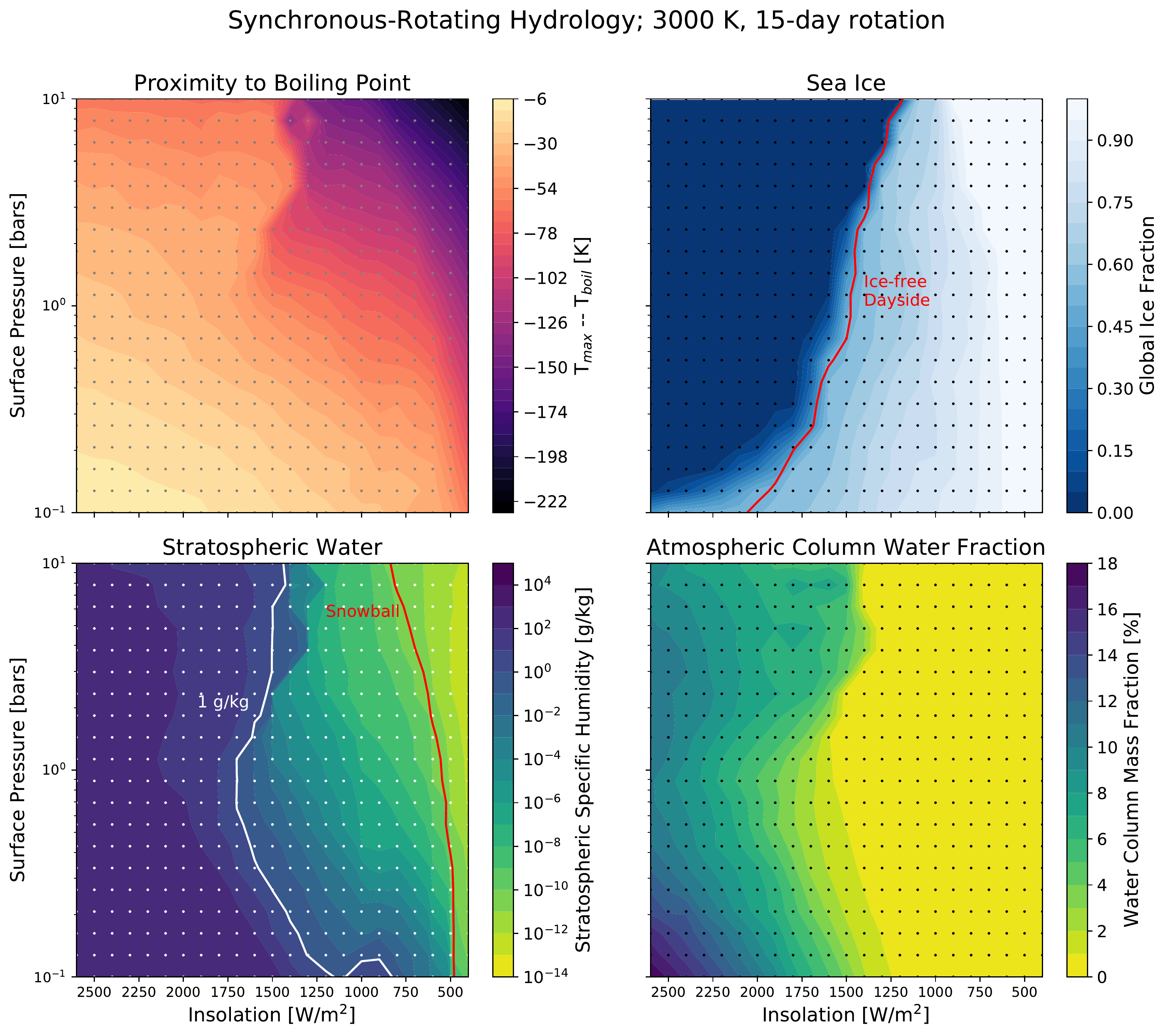}
\end{center}
\caption{Proximity of the maximum surface temperature to the boiling point, global sea ice fraction, water fraction in the top layer of the model, and the total water column mass fraction. We note that although some of our models at high instellations and high pressures get extremely warm, they are further from the boiling point of water than the low-pressure models. The proximity to boiling is used here as a sanity check for model stability. The 1 g/kg stratospheric limit for water loss, however, is reached well before those fluxes. Note that ExoPlaSim has not been designed nor validated for moist runaway or water loss regimes, but these results may help constrain the parameter space in which ExoPlaSim can be safely-used (i.e., not in these regimes). We note that the stratospheric limit coincides with the point where the column mass fraction begins to exceed 1--2\%.}\label{pyfig:mdwarfpn2_h2o}
\end{figure*}

Our results are shown in \autoref{pyfig:mdwarfpn2} and \autoref{pyfig:mdwarfpn2_h2o}. We find that with the redder 3000 K spectrum, Rayleigh scattering is far less effective at cooling the climate. Increased pN$_2$ can still be an overall cooling mechanism through increased heat transport efficiency, as seen in the snowball threshold in \autoref{pyfig:mdwarfpn2_h2o}, but at fluxes above approximately 1000 W m$^{-2}$ increasing pN$_2$ leads to increased surface temperatures and a reduction in sea ice. We note that at high fluxes, our high-pressure models reach very high temperatures, with mean temperatures in excess of 70 $^\circ$C and maximum temperatures of 100 $^\circ$C or higher. However, due to the higher pressure, these maximum temperatures are still 30 K or more below the boiling point of water at that pressure, as shown in \autoref{pyfig:mdwarfpn2_h2o}. The lowest-pressure models in the grid are actually the closest to boiling temperatures--maximum surface temperatures at 0.1 bar and 2600 W m$^{-2}$ are less than 10 K away from the boiling point. 

These models are unrealistic due to the fact that ExoPlaSim only treats H$_2$O as a minor constituent, and thus perturbations in H$_2$O vapor pressure are not expected to cause significant changes in surface pressure. At these temperatures, even though the boiling point has not been reached, evaporation should be so vigorous as to both change the surface pressure on short timescales and drive substantial water loss \citep{Wordsworth2013}. Because we have not validated ExoPlaSim in moist greenhouse and runaway greenhouse regimes, we cannot determine if the warm extremes in these results represent a realistic inner edge of the habitable zone. We include these warmer models for the sake of completeness. 

We also note in the lower-left panel of \autoref{pyfig:mdwarfpn2_h2o} the point at which the specific humidity in the model's top layer surpasses 1 g/kg, which has been identified as a relevant limit for water loss through atmospheric escape. This limit moves to lower fluxes at lower pressures, as at lower surface pressures, the ability of the tropopause to trap water in the lower atmosphere is weakened \citep{Wordsworth2013,Wordsworth2014}. We suggest that because this limit coincides with the point at which the total column mass fraction of water begins to exceed 1--2\%, this may set a useful limit on the regimes in which ExoPlaSim is likely to give reliable results. We note that there are low-pressure models in our sample that have high stratospheric humidity while still having sea ice on the dayside, suggesting that some temperate planets may require sophisticated GCMs to model appropriately, despite their Earth-like temperatures.

\section{Known limitations}\label{pysec:problems}

Because ExoPlaSim uses a simplified radiation scheme, there are several regimes in which ExoPlaSim may fall short and underperform compared to more-sophisticated GCMs. First, as shown in \citet{Paradise2017}, PlaSim has a significant cooling bias at high pCO$_2$ ($>$100 mbar), observed as excess outgoing longwave radiation compared to that predicted for the same temperature and pCO$_2$ by the \citet{wk97} polynomial fits. As we noted in \autoref{pysec:thai}, this might lead CO$_2$-dominated climates to be significantly colder in ExoPlaSim than in other models. Second, while ExoPlaSim nominally has the ability to respond to the dependence of shortwave atmospheric absorption on the incident stellar spectrum \citep{Kasting1993}, due to the fact that shortwave water absorption is only computed in the near-infrared band, and re-weighting the bands will correspondingly produce different amounts of water vapor absorption, the actual absorbed radiation is likely inaccurate for cool stars. This is because the absorptivity of water vapor in the near-infrared band is tuned to a solar spectrum, and the bolometric absorptivity across the band depends on the shape of the spectral energy distribution. An improvement to the model would involve re-tuning the water vapor absorptivity in each model run, by convolving a wavelength-dependent water vapor absorption spectrum with the incident spectral energy distribution to compute a new bolometric absorptivity. Due to the overlap between water vapor absorption and absorption by water, snow, and ice on the surface, an improved scheme would reduce the absorptivity of water surface types in columns with high amounts of water vapor. 

There are also likely to be biases that result from ExoPlaSim's simplified Rayleigh scattering scheme. As noted in \citet{Paradise2021}, ExoPlaSim currently computes all Rayleigh scattering in the bottom layer of the atmosphere. This is an acceptable approximation for Earth \citep{Lacis1974}, but for thicker atmospheres, this may lead to overestimates of shortwave absorption in the mid- and lower-troposphere, as light that should have been scattered back to space at higher altitudes is instead absorbed by tropospheric water vapor. This may result in a heating bias at high surface pressures. However, for the pN$_2$ ranges included in \citet{Paradise2021} and in the present work, the errors in TOA shortwave fluxes compared to those computed with SBDART \citep{Ricchiazzi1998} remain relatively small. This effect is also likely to be smaller around cooler stars, as Rayleigh scattering is less important at longer wavelengths.

ExoPlaSim also suffers from a lack of photochemistry and parameterizations for UV absorption. Ozone absorption is included in the model \citep{Fraedrich2005}, based off of the shortwave parameterization in \citet{Lacis1974}. However, the ozone distribution is prescribed as an idealized distribution described in \citet{Green1964} that is derived from measurements of Earth's ozone abundance, which may make it challenging to generalize to other planets. We have disabled ozone in all the experiments presented here, and recommend ozone be disabled for studies of planets other than Earth. Studies such as \citet{Chen2019} which investigate the atmospheric response to flares and UV fluxes from M dwarfs will therefore require additional modifications to ExoPlaSim, or must use other models.

ExoPlaSim's radiation scheme is also limited in that the range of surface types and albedos is limited, and not coupled to the climate beyond the presence or absence of snow and ice, and the temperature of the ice. There is a parameterization for wet soil described in \citet{Paradise2019}, which darkens the soil as soil moisture increases, but this has not been validated against other models nor spectroscopic observations of wet and dry soils and sands. ExoPlaSim does not, for example, consider whether high average wind stresses might affect the grain size of surface sediments, how snow grain size might interact with the climate, or how average precipitation patterns might affect the mineralogy of the surface, and therefore its reflectivity. Ocean reflectivity, similarly, has no dependence on salinity, suspended particulates, or wind stress, and an ocean glint parameterization is not included. Coupling a more-sophisticated land and ocean model might therefore improve ExoPlaSim's accuracy on small scales and its applicability to non-Earth-like planets.

Similarly, ExoPlaSim only has one tracer---water vapor. Other absorbing species whose spatial distribution may vary are either not included (CH$_4$, aerosols), or assumed to be uniformly distributed (CO$_2$). ExoPlaSim is therefore limited in its treatment of clouds, precipitation, and high-altitude absorption and reflectivity. The lack of other condensibles is particularly relevant in the case of CO$_2$. CO$_2$ condensation is likely to happen on most snowball planets \citep{Turbet2018,Paradise2019} and on many synchronously rotating planets in the habitable zone \citep{Ding2018,Turbet2018,Fauchez2020}. CO$_2$ condensation might affect the climate via cloud reflection, a cloud scattering greenhouse effect, direct removal of CO$_2$, and latent heat fluxes through evaporation and condensation \citep{Forget1997,Forget2013,Soto2015,Kitzmann2017}.

Finally, ExoPlaSim lacks a dynamic ocean model. A realistic treatment of ocean currents, their heat transport, and sea ice transport can lead to qualitative differences in the climates of synchronously rotating planets in particular \citep{Hu2014,Yang2014,DelGenio2018,Yang2019b}, as well as for studies of snowball climate \citep{Lewis2007,Checlair2019,Jansen2019}, and terrestrial climates in general \citep{Cullum2014}. In this work we have evaluated ExoPlaSim's atmosphere model, and compared it to two exoplanet GCM intercomparisons that also omitted dynamic oceans \citep{Yang2019,Fauchez2020}. PlaSim has been coupled with dynamic oceans in the past \citep[e.g.][]{Holden2016,Holden2019,Platov2017,Angeloni2019,Angeloni2020}, but the necessary work to couple ExoPlaSim to a dynamic ocean remains to be done. Until a dynamic ocean is coupled with ExoPlaSim, this necessarily limits ExoPlaSim to explorations of climate dynamics that do not rely on accurate ocean dynamics, and to exoplanet climate predictions that are relatively insensitive to the details of ocean heat transport. 

\section{Future work}\label{pysec:plans}

Further development of the ExoPlaSim climate model is ongoing. A feature that is present in the model but not yet validated is the calculation of storm climatology and a high-cadence output trigger to capture storm events in high temporal resolution. This module optionally computes at each timestep and in each grid cell the convective available potential energy, level of neutral buoyancy, entropy deficit, ventilation index, maximum potential intensity, ventilation-reduced maximum intensity, and the genesis potential index (GPI), following the approaches in \citet{Bin2018,Komacek2020}, and \citet{Yan2020}. The high-cadence output trigger can be set to begin writing output up to every timestep as soon as various climatological thresholds indicating storm activity are reached, and continue writing output either until storm activity falls below a certain threshold or a certain number of writes have been made, up to a set number of times per year. Testing of these features on high-resolution T170 models of synchronously rotating and Earth-like planets is ongoing.

It may also be possible to improve ExoPlaSim's radiation scheme in the near future by implementing layer-by-layer Rayleigh scattering, and potentially by adding additional absorbers, either as mixed species or as tracers. The latter could be done by duplicating water tracer transport with the properties of the new tracer. This could additionally enable inclusion of CO$_2$ condensation through both CO$_2$ clouds and CO$_2$ frost.

Finally, it may be possible to couple ExoPlaSim to other models to further its utility. PlaSim-ocean couplings have been explored in multiple studies \citep[e.g][]{Holden2016,Platov2017,Angeloni2019}, and many of these couplings could potentially be ported to the ExoPlaSim framework. PlaSim has already been coupled to geochemical and biochemical models, as with PLASIM-GENIE \citep{Holden2016} and PlaSim-ICMMG \citep{Platov2017}, and these couplings could inform future projects to expand ExoPlaSim's utility for exoplanets. It should additionally be possible to extend ExoPlaSim to a broader range of orbital scenarios, including circumbinary planets and habitable moons of giant planets, by coupling ExoPlaSim to an orbital integrator such as REBOUND \citep{REBOUND1,REBOUND2}. The orbital integrator would be used to compute the locations of multiple light sources in the sky as the system evolves, enabling studies of increasingly-exotic habitable worlds. A light-hearted proof-of-concept of this idea was demonstrated in \citet{Paradise2019b}, in which an Earth-like planet on a chaotic Sitnikov-type orbit \citep[][a special case of the three-body problem]{Sitnikov1961} with two parent stars was modeled with an earlier version of ExoPlaSim using a simple 4th-order Yoshida integrator \citep{Yoshida1990}

All of these model developments would require GCM intercomparisons to validate, as it is not yet possible to compare our model's performance to any real-world planets other than Earth. A wide range of habitable terrestrial exoplanet intercomparisons such as \citet{Yang2019} and \citet{Fauchez2020} are needed \citep{THAIreport}, including ones that test features already present in PlaSim, such as land models, biosphere models, climate sensitivity to continental distribution and topography, etc.

\section{Model Availability}
\subsection{Raw source code, model outputs}
The ExoPlaSim source code is available from \href{https://github.com/alphaparrot/ExoPlaSim}{AP's GitHub} \citep{exoplasim}. The model outputs we have presented here are available in an online \href{https://dataverse.scholarsportal.info/dataverse/kmenou}{Dataverse repository} \citep{Paradise2020}.

\subsection{Python API}\label{pysec:Python}

GCMs can often be complicated to configure and challenging to learn, and therefore only accessible to dedicated climate modellers. PlaSim, for example, requires that most model parameters be changed via namelist files \citep{Fraedrich2005}. Configuring a model is therefore a multi-step process of editing multiple namelist files. Not only is this time-consuming, but it poses a significant obstacle to running extremely large numbers of models, such as the parameter sweeps presented here. 

We therefore aim to improve ExoPlaSim's accessibility and utility by adding a Python abstraction layer. We have created a simple API wherein an ExoPlaSim model instance is defined and configured within a Python object, which has the added benefit of bundling several convenience routines, including commands to load output data, compute spatial and annual averages, run to equilibrium, and compute the coordinate transforms and synchronously rotating streamfunction described in \autoref{pysec:yang}. The abstraction layer also replaces the C++ postprocessor in PlaSim with a new postprocessor implemented in Python and Fortran, which supports multiple output formats, as well as additional derived variables such as streamfunctions, and both equatorial and synchronous coordinate systems. The model is initialized when the Python object is created, which sets the number of threads, the resolution, numerical precision, and optional optimization or debugging flags. If the model has never been compiled locally or a recompilation flag is set, the API will then compile ExoPlaSim. All of the namelist parameters can then be set in a single call to the Python object's \mintinline[fontsize=\footnotesize]{python}{configure()} function through the use of keyword parameters. This function call (or its sibling, \mintinline[fontsize=\footnotesize]{python}{modify()}) can be made multiple times throughout a simulation's run-time in order to cause a parameter such as instellation or pCO$_2$ to change over time. Consolidating all model configuration options to a single function call allows users to easily create large ensembles of models, by simply placing the model initialization and configuration function calls inside a loop or series of nested loops, which iterate over parameters of interest.

At this point, the model is ready to run, which can also be done with a single command. To improve the portability and reproducibility of studies that use ExoPlaSim, the model configuration can also be easily exported to a small text file, which can then be read by another ExoPlaSim user as a model input to configure their model in an identical fashion. We therefore encourage ExoPlaSim users to publish their configuration files alongside their model outputs. The Python object's state can also be saved to a local binary file in NumPy format \citep{numpy}, which is useful in supercomputing environments where a program is limited in wall-time, and must stop and restart itself multiple times. 

In addition to the Python API, we have uploaded ExoPlaSim to the \href{https://pypi.org/project/exoplasim/}{Python Package Index} (PyPI). This means that ExoPlaSim can be easily-installed via the popular \texttt{pip} command, which will install the program into the user's path, such that the Python module can be imported from any directory, without the user's knowledge of the source code's location. We have also documented the API, and a user manual and tutorial is available at \url{https://exoplasim.readthedocs.io/en/latest/}.

Our hope in adding these utilities and abstractions is that ExoPlaSim will be sufficiently easy to install, configure, and use that not only can it be easily used to model large parameter sweeps, but that observers may be able to use a GCM to probe the possible climates of their newly-discovered planets, and gain intuition for the climate states and dynamics those planets are likely to have. The accessible Python modular interface may also make it possible to incorporate ExoPlaSim into observational retrieval pipelines that would otherwise be forced to use lower-dimensional climate models. Finally, the relative ease and installation of use means that ExoPlaSim may continue the legacy of the original model and prove useful as an educational tool \citep{Fraedrich2005}, both for the interested public and in the classroom, similar to the CLIMLAB modeling platform \citep{Climlab}.

To illustrate the comparative ease of installing and using ExoPlaSim, the process of installing and configurating ExoPlaSim to model the THAI Hab1$^*$ benchmark \citep{Fauchez2020} is shown below. First, to install ExoPlaSim:
\begin{verbatim}
pip install exoplasim
\end{verbatim}
And then, in a Python interpreter, such as either the Python shell or a Jupyter notebook \citep{jupyter}, to initialize, compile, configure, and run the model:
\begin{minted}[
frame=lines,
framesep=2mm,
fontsize=\footnotesize
]{python}
import exoplasim as exo 

thaiHab1s = exo.Model(resolution='T42',layers=10,ncpus=32,
                      precision=8,optimization='mavx',
                      modelname="thai_hab1s",
                      workdir="thai_hab1s_run")
thaiHab1s.configure(startemp=2600.0,flux=900.0,radius=0.91,
                    gravity=9.12,fixedorbit=True,
                    synchronous=True,rotationperiod=6.1,
                    eccentricity=0.0,obliquity=0.0,
                    aquaplanet=True,ozone=False,
                    physicsfilter="gp|lh|sp",timestep=15.0,
                    snapshots=1440,pN2=1.0,pCO2=400.0e-6)
thaiHab1s.exportcfg()
thaiHab1s.runtobalance(postprocess=True)
\end{minted}
In this example, the model is initialized with a T42 resolution (64 latitudes and 128 longitudes), 32 MPI threads, double precision, an optimization flag useful for the Intel compiler on our cluster, a descriptive name that will be used to name the model outputs, and the name of a directory in which to run. This command creates that directory, compiles the model, and places the files ExoPlaSim needs to run in the working directory. The second command configures the model, specifying the TRAPPIST-1e planetary parameters given in \citet{Grimm2018} and \citet{Fauchez2020}. A Lander-Hoskins filter (``lh") is used, at both spectral transforms---both from gridpoint (``gp") and from spectral (``sp") spaces. We use a 15-minute timestep, and in addition to the usual time-averaged data, write instantaneous snapshots every 1440 timesteps, or once ever 15 days. The model's configuration file is exported, and then the model runs to energy balance equilibrium, postprocessing the raw output into formatted output files along the way (by default compressed NumPy archives, but NetCDF, HDF5, and uncompressed or compressed CSV files are also supported). Running the Hab2$^*$ benchmark involves a simple and obvious change; the \texttt{pN2} keyword should be set to 0, and \texttt{pCO2} set to 1.0, indicating 1 bar of CO$_2$ and no N$_2$.

\section{Conclusion}

We have modified PlaSim for use beyond Earth and paleo-Earth, by making modifications to the radiation scheme and dynamics that are necessary to accurately model synchronously rotating planets around low-mass stars. This included addressing severe Gibbs phenomena that emerge as an unavoidable consequence of the synchronously rotating geometry when modeled with a spectral core. We note that problems are likely to arise in non-spectral models as well, but may not be as obvious as the global axisymmetric rings observed in our model prior to addition of physics filters. We compared our modified model, which we call ExoPlaSim, to ExoCAM, CAM3, CAM4, the LMD Generic model, AM2, ROCKE-3D, and the UK Met Office's Unified Model through two synchronously rotating intercomparisons presented in \citet{Yang2019} and \citet{Fauchez2020}. We find that while ExoPlaSim has a slight warm bias for fast-rotating Earth-like planets, it agrees to a large degree with the model ensemble when applied to two different synchronously rotating test cases. ExoPlaSim differs in that the strength of the dipolar overturning circulation is greater, and wind speeds are higher. We confirm that in distinctly non-Earth-like regimes, such as CO$_2$-dominated atmospheres, ExoPlaSim is likely to deviate more from the other models, due to known radiative biases and missing or simplified physics parameterizations. We then demonstrated ExoPlaSim's potential for survey science by running 200 models to confirm that the habitable zone moves outward to lower fluxes around cooler stars, and by running 460 models to confirm our prediction in \citet{Paradise2021} that Rayleigh scattering would cease to be a dominant cooling mechanism for high-pN$_2$ synchronously rotating planets around cool stars. Finally, to make the model accessible to a broad range of scientists and educators, and to facilitate easy configuration of the types of large model ensembles we have presented, we have created an easy-to-use and fully-documented Python API and created a simplified installation procedure through \texttt{pip}.

Due to the lack of available observational data and comprehensive intercomparisons, exoplanet GCMs cannot yet be truly predictive, and are restricted to studying the behavior and sensitivity of various climate processes in exoplanetary contexts. While the more-sophisticated GCMs available are able to probe complicated questions related to photochemical processes, volatile cycling, and more on a wide variety of terrestrial exoplanets, a need nonetheless remains for GCMs that can quickly, cheaply, and accurately model the overall climate states of a wide diversity of terrestrial habitable zone planets, capturing the main distinctive features, such as the temperature distribution, circulation regimes, and cloud patterns. ExoPlaSim is fast enough and flexible enough to meet that standard, and with its simplified installation and the Python API, can also be used with little hassle by observers, theorists not used to working with 3D simulations, educators, and students. More work is needed to expand the range of climates ExoPlaSim can model, and to develop other models of comparable speed and flexibility that can be used to complement and test ExoPlaSim. The advent of such models, however, will promote greater intuition for exoplanet climate within the broader community, and enable new kinds of exoplanet climate studies that would be unfeasibly expensive with more-sophisticated GCMs.

\section*{Acknowledgments}

AP is supported by the Ontario Graduate Scholarship, and by the David A. Dunlap Department of Astronomy \& Astrophysics at the University of Toronto. EM is supported by the Department of Physics at the University of Toronto, as well as by the University of Toronto Faculty of Arts \& Science. CL is supported by the Department of Physics at the University of Toronto. KM is supported by the Discovery program of the Natural Sciences and Engineering Research Council of Canada. BF is also supported by the David A. Dunlap Department of Astronomy \& Astrophysics. The authors would like to extend deep gratitude to John Scinocca for invaluable contributions and discussion on the topic of ExoPlaSim's Gibbs ripples. We would also like to thank Jun Yang, J\'{e}r\'{e}my Leconte, Eric Wolf, and Tim Merlis for sharing model outputs from the \citet{Yang2019} intercomparison, and enabling us to compare ExoPlaSim to the other models. We would like to thank Martin Turbet and Thomas Fauchez for helpful discussions about the THAI intercomparison project, and about ExoPlaSim's capabilities. We would further like to thank Thaddeus Komacek and Mark Hammond for their helpful clarifications regarding their papers' methods. Finally, we would like to thank Nikolai Hersfeldt for catching several bugs in ExoPlaSim, and raising use cases we had not considered.

We would like to acknowledge that this work was performed on land which for thousands of years has been the traditional land of the Huron-Wendat, the Seneca, and the Mississaugas of the Credit. Today, this meeting place is still the home to many Indigenous people from across Turtle Island and we are grateful to have the opportunity to work on this land and benefit from its natural resources. Finally, we acknowledge the use of computing resources from the Canadian Institute for Theoretical Astrophysics (CITA). CITA's advanced computing infrastructure in Ontario uses renewable and low-carbon energy \citep{CER}, but we recognize low-carbon computing may not be possible everywhere. We therefore acknowledge that creating a more-accessible model may lead to increased demand on supercomputing resources, and increased carbon emissions as well as natural resource extraction on and near lands traditionally inhabited by Indigenous peoples.

\section*{Data Availability}

The version of PlaSim used in this study is available on AP's GitHub \citep{gplasim}, as well as through the Python Package Index via \texttt{pip install exoplasim}. Documentation is available online at \url{https://exoplasim.readthedocs.io/en/latest/}. The models used in our sample, as well as their postprocessed spectra, are available through an online \href{https://dataverse.scholarsportal.info/dataverse/kmenou}{Dataverse repository} \citep{Paradise2020}.



\bibliographystyle{mnras}

\appendix

\section{Choosing Appropriate Physics Filters to Mitigate Gibbs Ripples}\label{pysec:appendix}

As discussed in the main text, physics filters are necessary to remove Gibbs `ripples' in ExoPlaSim models of planets with dipolar forcing, such as synchronous rotators. As noted in Section 2.4.3, the best filter performance seems to require the use of a physics filter at both spectral transforms, going both from gridpoint to spectral space and vice versa. For most use cases in which a filter is warranted, exponential or Lander-Hoskins filter types appear to be work better than Ces\`{a}ro filters, although the latter appears to perform better in the limit of very high horizontal resolution (see Figure 9 of the main text). In this appendix, we describe in greater detail our tests to evaluate the different filters, and the tuning of $\kappa$ and $\gamma$ for the exponential filter $\left(f(n,N)=\exp\left[-\kappa\left(\frac{n}{N}\right)^\gamma\right]\right)$.

\subsection{Testing filter performance}

To evaluate filter performance, we first performed three different tests with four basic filter configurations. The three tests consisted of both the fast-rotator and synchronous-rotator benchmark scenarios from \citet{Yang2019}, and the `Hab1' scenario from the THAI intercomparison framework \citep{Fauchez2020}, at both T21 and T42 resolution ($32\times64$ and $64\times128$ horizontal resolution, respectively). The four filter configurations were (1) no filter, (2) a Ces\`{a}ro filter, (3) an exponential filter with $\kappa=8$ and $\gamma=8$, and (4) a Lander-Hoskins filter. For each test, each filter (except for the case with no filter), and each resolution, we ran 3 models, with the filter applied at the transform from gridpoint to spectral space (`GP$\rightarrow$SP'), at the transform from spectral to gridpoint space (`SP$\rightarrow$GP'), and at both transforms (`GP$\rightarrow$SP$\rightarrow$GP'). The models were configured as described in \citet{Yang2019} and \citet{Fauchez2020}, and as described in the main text of this paper. T21 models were run with 30-minute timesteps, while the T42 models were run with 15-minute timesteps. 

When evaluating filter performance, we looked for five criteria: First, the filter should not cause significant global biases in the fast-rotating control test relative to the unfiltered model. Second, the filter should have a noticeable impact on Gibbs features in model fields, especially in non-cloud fields such as temperature, and in the cloud field if possible. Third, in test cases with North-South symmetry in forcing (in terms of orbital parameters and boundary conditions such as land distribution), the filter should not result in North-South asymmetry in climate. Fourth, the ideal filter will preserve non-Gibbs small-scale structures. Some loss of fidelity here is inevitable \citep{Lander1997}, but a filter is only likely to broaden features, not introduce new small-scale features. It is however possible for organized small-scale structure to emerge in the presence of a filter that would otherwise have been disrupted and damped by Gibbs ripples of similar or slightly smaller size scales. This emergent structure however is more likely to be physical than numerical, as it represents the atmosphere's behavior when the Gibbs effect has been reduced. Finally, the model outcome at T21 and T42 should be qualitatively similar, with few systematic biases.

In order to tune the exponential filter, we ran $5\times6$ grids of models at T21 and T42 resolution, for both the YB scenario and the THAI-B Hab1 scenario. These grids varied $\kappa$ and $\gamma$, with $\gamma=4$, 8, 12, 16, and 32, and $\kappa=4$, 6, 8, 10, 12, and 16. We found that the THAI-B tests were relatively insensitive to tuning, likely due to the planet's relatively faster rotation. The YB models are much more sensitive to the exponential filter's tuning parameters, making it possible to identify preferred tuning parameters for the YB scenario. As the Lander-Hoskins filter also seems to produce better mitigation in the THAI-B tests, we focus our tuning tests on the YB models, and use the Lander-Hoskins filter for all other THAI-B tests. 

\subsection{THAI-B filter performance}

\begin{figure*}
\begin{center}
\includegraphics[width=6.5in]{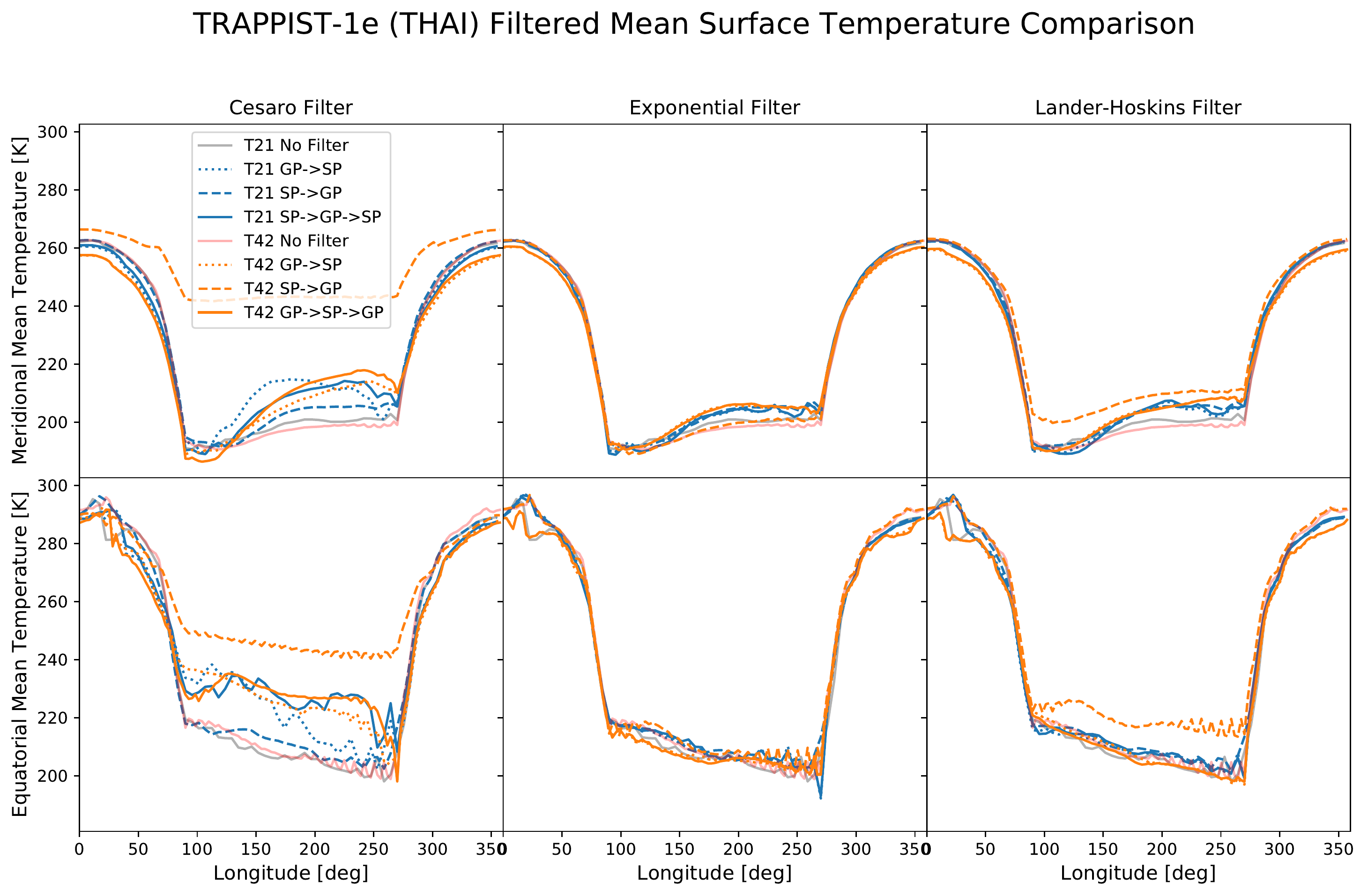}
\end{center}
\caption{Equatorial and mean meridional surface temperature for a synchronously-rotating aquaplanet with 6.1-day rotation, TRAPPIST-1e bulk and atmospheric parameters, and 2600 K incident spectrum, modeled at T21 and T42 with Ces\`{a}ro, exponential, and Lander-Hoskins \citep{Lander1997} filtration. Models with no filtration are also shown for comparison. The model configuration is chosen to match the `Hab1' benchmark described in the THAI intercomparison project \citep{Fauchez2020}. All three filters substantially reduce Gibbs-type ripples in surface temperature, with exponential and Lander-Hoskins filtration providing the most-robust results. In this test, both filters show good agreement between resolutions, but the Lander-Hoskins filter performs better at actually removing the Gibbs-type ripples.}\label{pyfig:benchthai_temps}
\end{figure*}

\begin{figure*}
\begin{center}
\includegraphics[width=6.5in]{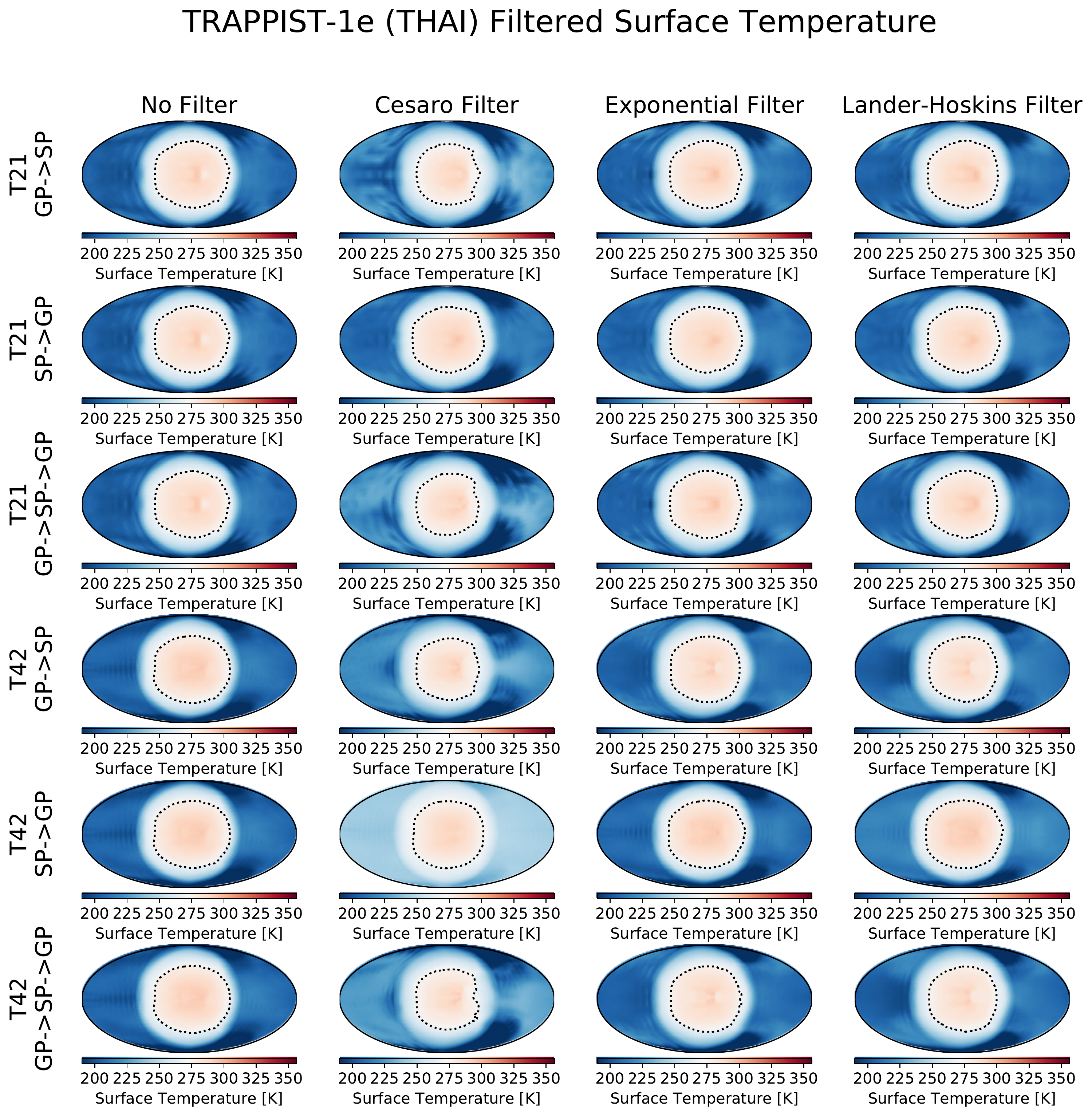}
\end{center}
\caption{Mean annual surface temperature of synchronously-rotating aquaplanets with TRAPPIST-1e bulk parameters, 6.1-day rotation, 2600 K incident spectrum, 900 W/m$^2$, and a sea ice albedo of 0.25. The substellar point is centered. Both T21 and T42 horizontal resolutions are shown. Three different filters are shown, as well as the case with no filter, with 3 different filter configurations for each filter: `GP$\rightarrow$SP', corresponding to a filter during the transform from the gridpoint domain to the spectral domain, `SP$\rightarrow$GP', corresponding to a filter at the transform from the spectral domain to the gridpoint domain, and `GP$\rightarrow$SP$\rightarrow$GP', indicating a filter at both transforms. All three are of course identical for the case with no filter. The use of either the exponential or Lande-Hoskins filter at both transforms almost completely removes the effect of Gibbs ripples on the surface temperature, despite the continued presence in the cloud field. The Ces\`{a}ro filter also removes the ripples, but produces a significantly warmer nightside.}\label{pyfig:benchthaiT21_ts}
\end{figure*}

\begin{figure*}
\begin{center}
\includegraphics[width=6.5in]{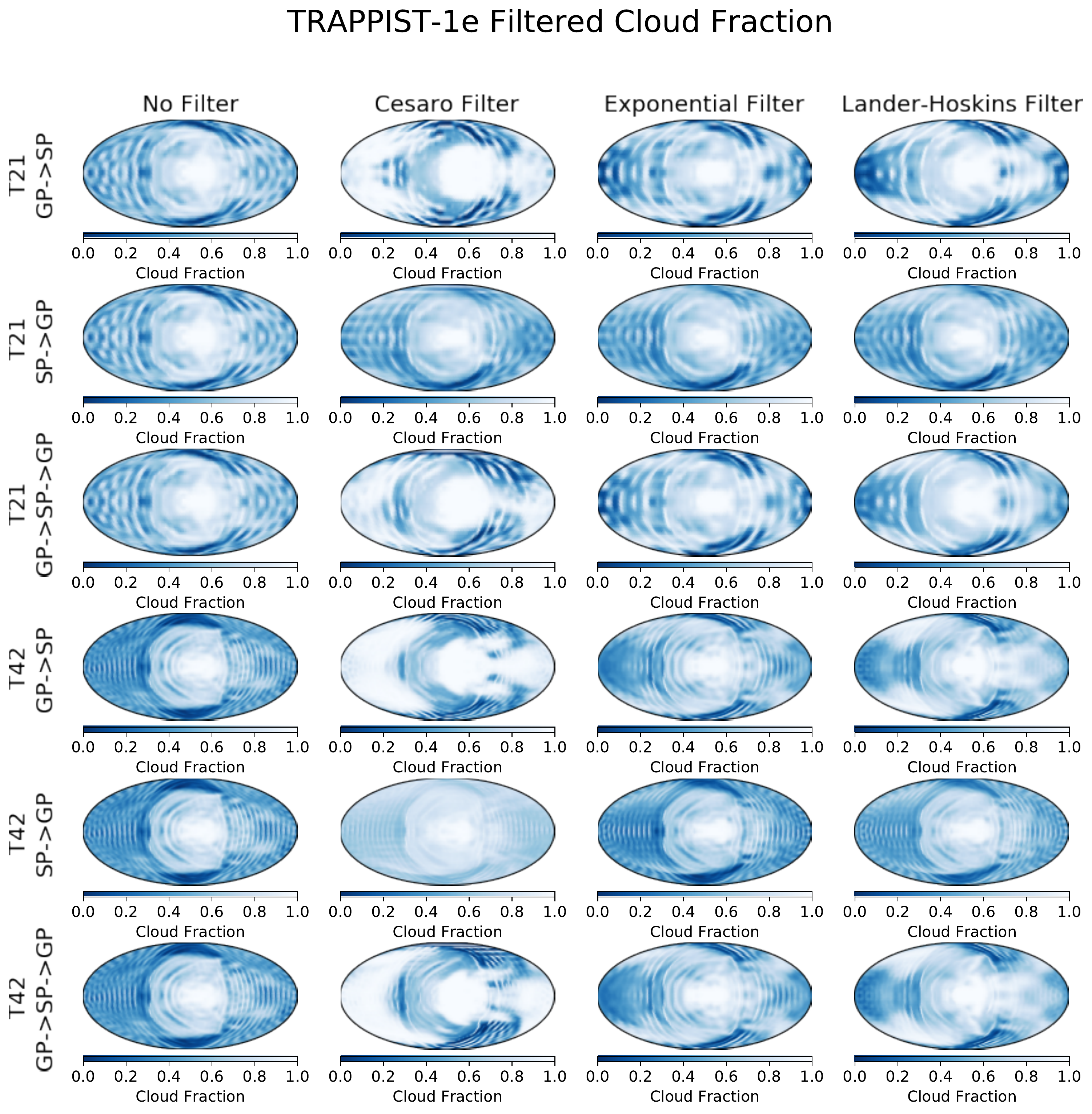}
\end{center}
\caption{Mean annual cloud fraction of synchronously-rotating aquaplanets with TRAPPIST-1e bulk parameters, 6.1-day rotation, 2600 K incident spectrum, 900 W/m$^2$, and a sea ice albedo of 0.25. The substellar point is centered. Both T21 and T42 horizontal resolutions are shown. Three different filters are shown, as well as the case with no filter, with 3 different filter configurations for each filter: `GP$\rightarrow$SP', corresponding to a filter during the transform from the gridpoint domain to the spectral domain, `SP$\rightarrow$GP', corresponding to a filter at the transform from the spectral domain to the gridpoint domain, and `GP$\rightarrow$SP$\rightarrow$GP', indicating a filter at both transforms. At both T21 and T42, it is difficult to fully-remove ripples with any filter, but applying the Lander-Hoskins filter at both transforms appears to give the best results.}\label{pyfig:benchthaiT21_clt}
\end{figure*}

Meridional mean and equatorial surface temperatures for the THAI-B tests are shown in \autoref{pyfig:benchthai_temps}. Maps of the annual average surface temperature are shown in \autoref{pyfig:benchthaiT21_ts}. We find that both the exponential and Lander-Hoskins filters are acceptable, as in the YB models. The Lander-Hoskins filter does a slightly better job however of reducing the ripples, as seen in the horizontal temperature profiles in \autoref{pyfig:benchthai_temps}. Similarly, the cloud field, shown in \autoref{pyfig:benchthaiT21_clt}, appears smoother with the Lander-Hoskins filter. We therefore conclude that for this particular test case, the Lander-Hoskins filter may be preferable to the exponential filter.  

\subsection{YB filter performance}

\begin{figure*}
\begin{center}
\includegraphics[width=6.5in]{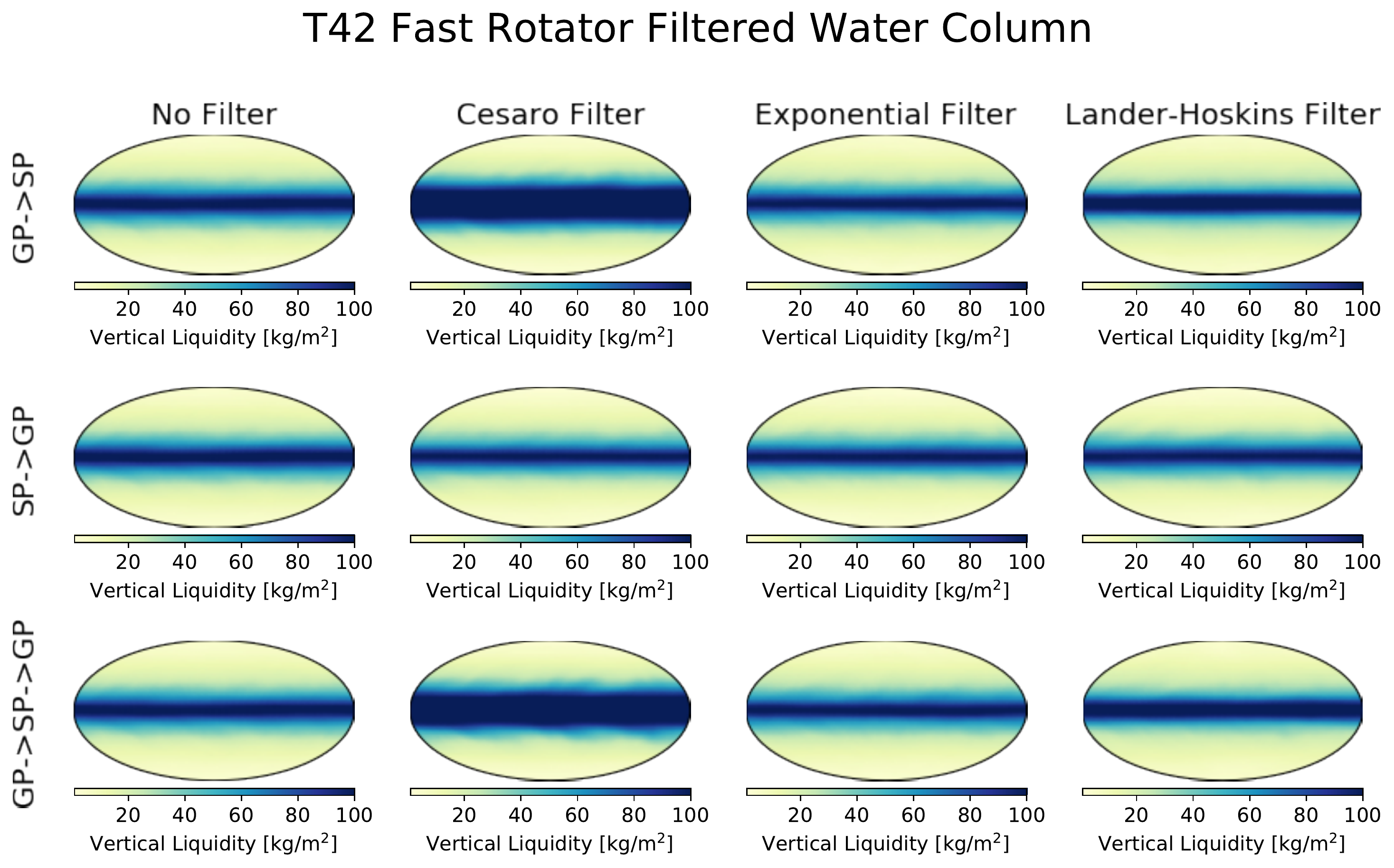}
\end{center}
\caption{Mean annual vertically-integrated water vapor column mass of fast-rotating aquaplanets with Earth bulk parameters, a solar-like spectrum, 1360 W/m$^2$, and no radiative effect from sea ice beyond latent heat flux. The models are at T42 resolution, or 64 latitudes and 128 longitudes, corresponding to approximately 2.8$^\circ$ horizontal resolution. Three different filters are shown, as well as the case with no filter, with 3 different filter configurations for each filter: `GP$\rightarrow$SP', corresponding to a filter during the transform from the gridpoint domain to the spectral domain, `SP$\rightarrow$GP', corresponding to a filter at the transform from the spectral domain to the gridpoint domain, and `GP$\rightarrow$SP$\rightarrow$GP', indicating a filter at both transforms. Fast-rotating planets do not particularly struggle with Gibbs ripples in ExoPlaSim, so testing the models here serves to test for global biases introduced by the filter in a case where the solution is well-constrained. It is apparent from the water vapor distribution that the Ces\`{a}ro filter can cause significant broadening of the tropics relative to the control case. }\label{pyfig:benchaq_prw}
\end{figure*}

We tested our filters on both an Earth-like fast-rotating aquaplanet, as a control comparison, and on the synchronous-rotating YB models. ExoPlaSim does not suffer from Gibbs ripples in fast-rotating aquaplanet models, so ideally, the resulting climate is very similar to the unfiltered case, which represents the unmodified PlaSim solution. We find however that the Ces\`{a}ro filter results in broader tropics in the control case, as shown at T42 in \autoref{pyfig:benchaq_prw}. At lower resolutions, therefore, it is likely that the Ces\`{a}ro filter will introduce errors when applied to both spectral transforms.

\begin{figure*}
\begin{center}
\includegraphics[width=6.5in]{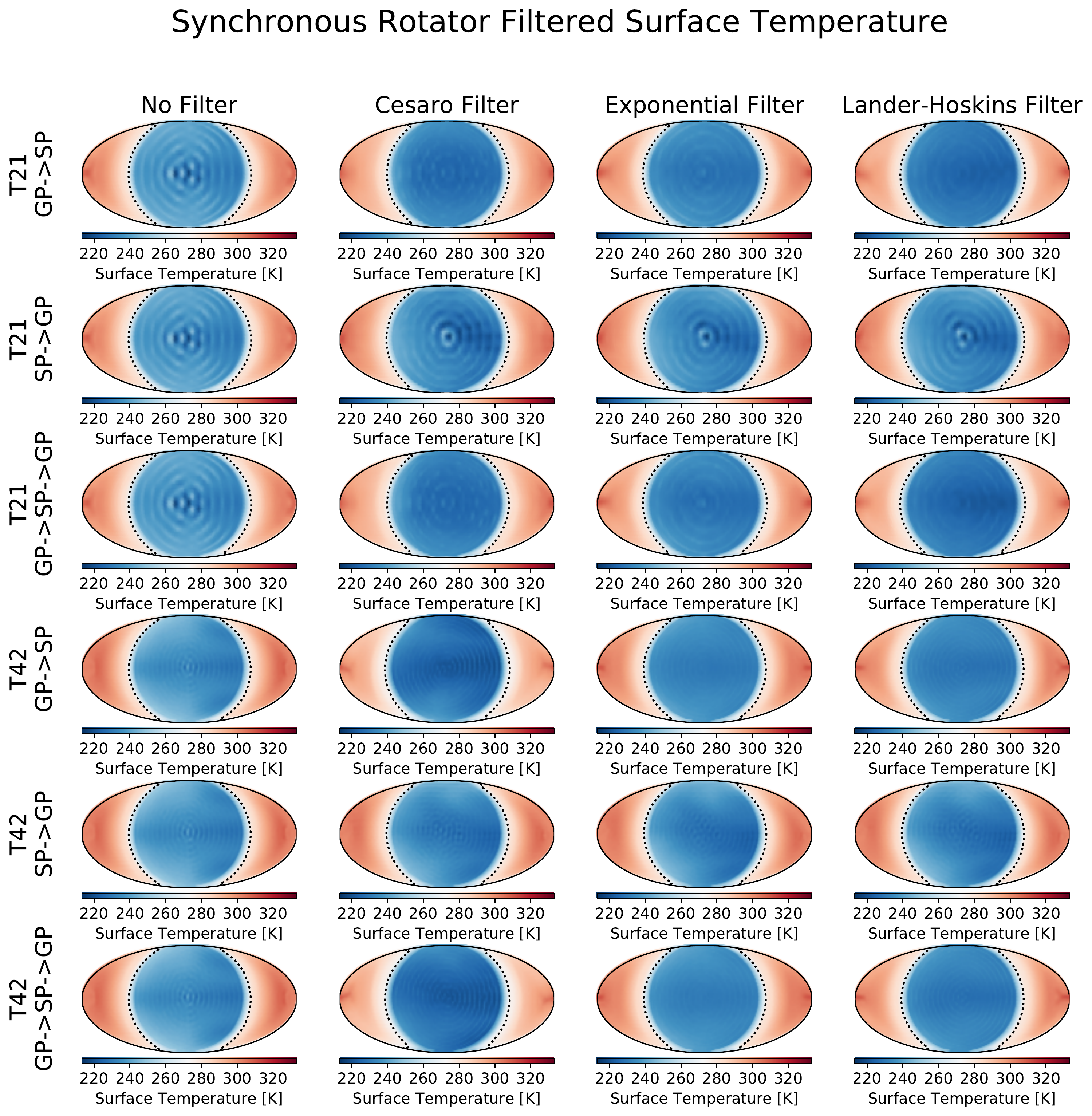}
\end{center}
\caption{Mean annual surface temperature of synchronously-rotating aquaplanets with Earth bulk parameters, 60-day rotation, 3400 K incident spectrum, 1360 W/m$^2$, and no radiative effect from sea ice beyond latent heat flux. The antistellar point is centered. Both T21 and T42 resolutions are shown. Three different filters are shown, as well as the case with no filter, with 3 different filter configurations for each filter: `GP$\rightarrow$SP', corresponding to a filter during the transform from the gridpoint domain to the spectral domain, `SP$\rightarrow$GP', corresponding to a filter at the transform from the spectral domain to the gridpoint domain, and `GP$\rightarrow$SP$\rightarrow$GP', indicating a filter at both transforms. All three are of course identical for the case with no filter. At both resolutions, the use of either the exponential or Lande-Hoskins filter at both transforms almost completely removes the effect of Gibbs ripples on the surface temperature, despite the continued presence in the cloud field.}\label{pyfig:benchtl_ts}
\end{figure*}

\begin{figure*}
\begin{center}
\includegraphics[width=6.5in]{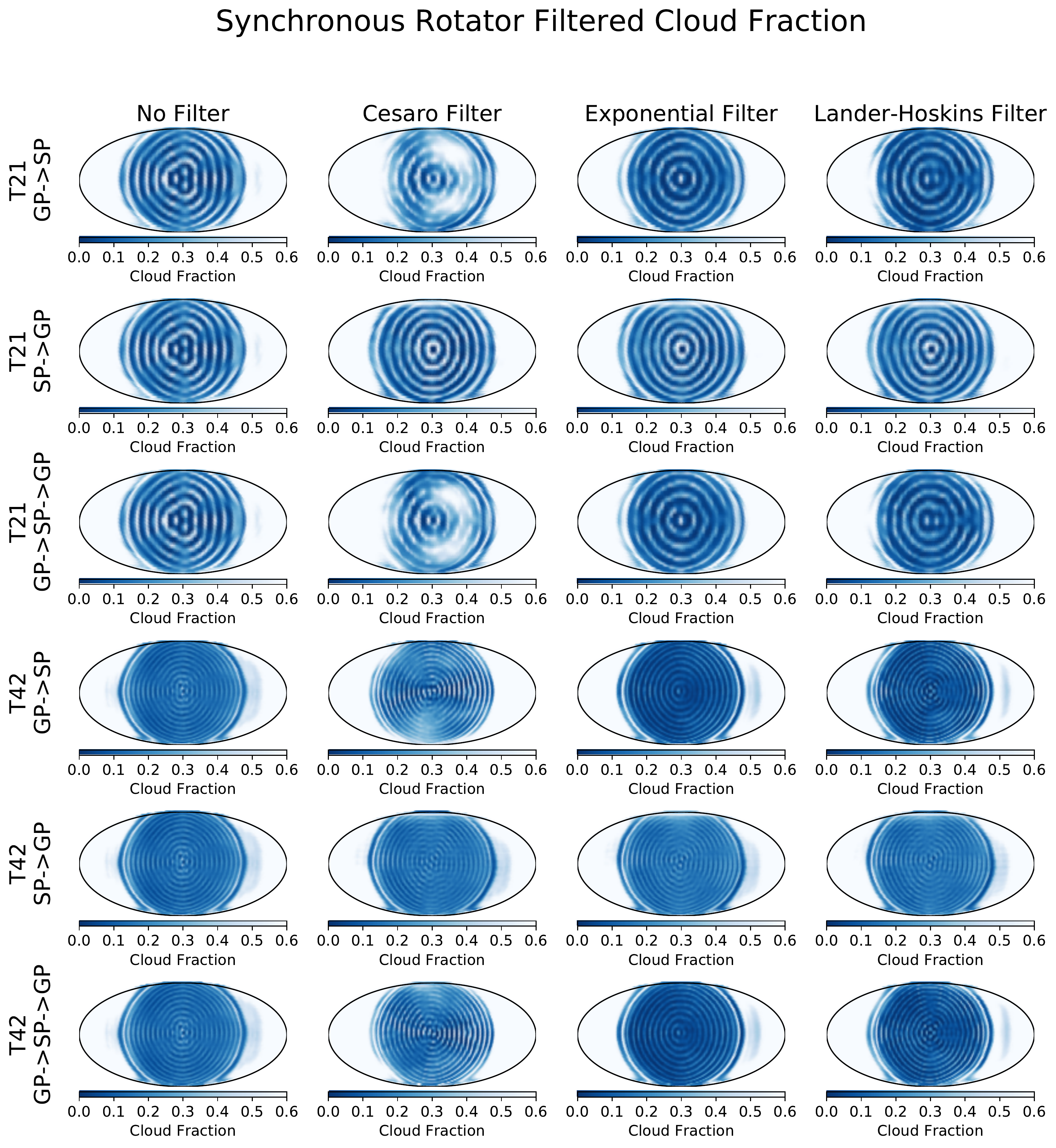}
\end{center}
\caption{Mean annual cloud fraction of synchronously-rotating aquaplanets with Earth bulk parameters, 60-day rotation, 3400 K incident spectrum, 1360 W/m$^2$, and no radiative effect from sea ice beyond latent heat flux. The antistellar point is centered. Both T21 and T42 resolutions are shown. Three different filters are shown, as well as the case with no filter, with 3 different filter configurations for each filter: `GP$\rightarrow$SP', corresponding to a filter during the transform from the gridpoint domain to the spectral domain, `SP$\rightarrow$GP', corresponding to a filter at the transform from the spectral domain to the gridpoint domain, and `GP$\rightarrow$SP$\rightarrow$GP', indicating a filter at both transforms. All three are of course identical for the case with no filter. At these resolutions, concentric cloud rings remain, but are damped somewhat with either an exponential or Lander-Hoskins filter at both transforms.}\label{pyfig:benchtl_clt}
\end{figure*}

\begin{figure*}
\begin{center}
\includegraphics[width=6.5in]{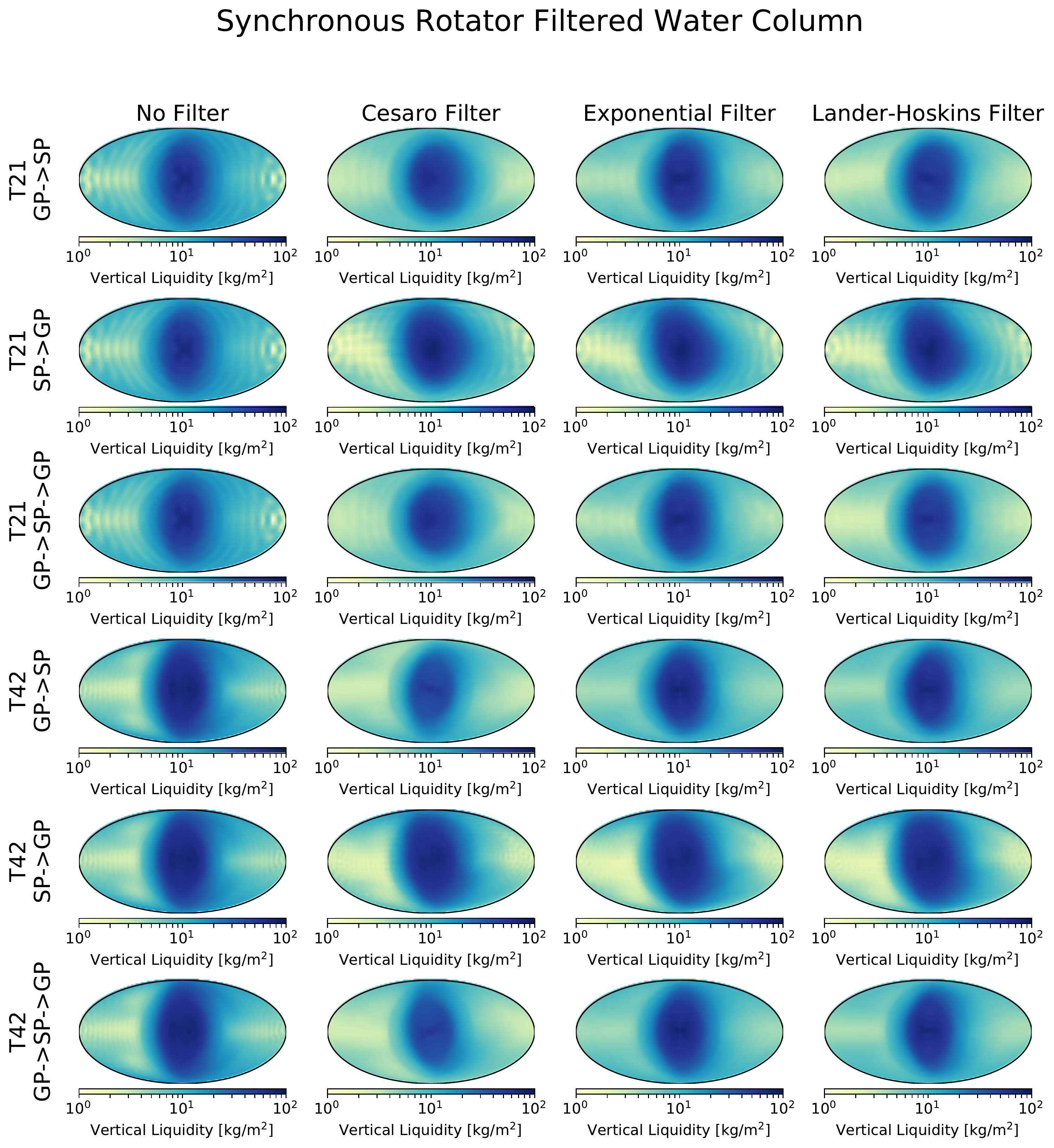}
\end{center}
\caption{Mean annual liquid water column mass of synchronously-rotating aquaplanets with Earth bulk parameters, 60-day rotation, 3400 K incident spectrum, 1360 W/m$^2$, and no radiative effect from sea ice beyond latent heat flux. The substellar point is centered. Both T21 and T42 resolutions are shown. Three different filters are shown, as well as the case with no filter, with 3 different filter configurations for each filter: `GP$\rightarrow$SP', corresponding to a filter during the transform from the gridpoint domain to the spectral domain, `SP$\rightarrow$GP', corresponding to a filter at the transform from the spectral domain to the gridpoint domain, and `GP$\rightarrow$SP$\rightarrow$GP', indicating a filter at both transforms. All three are identical for the cases with no filter. The use of any of the three filters at both transforms significantly reduces the effect of Gibbs ripples in atmospheric water, despite the continued presence in the cloud field. The Ces\`{a}ro filter however displays some hemispheric asymmetry.}\label{pyfig:benchtl_prw}
\end{figure*}

\begin{figure*}
\begin{center}
\includegraphics[width=6.5in]{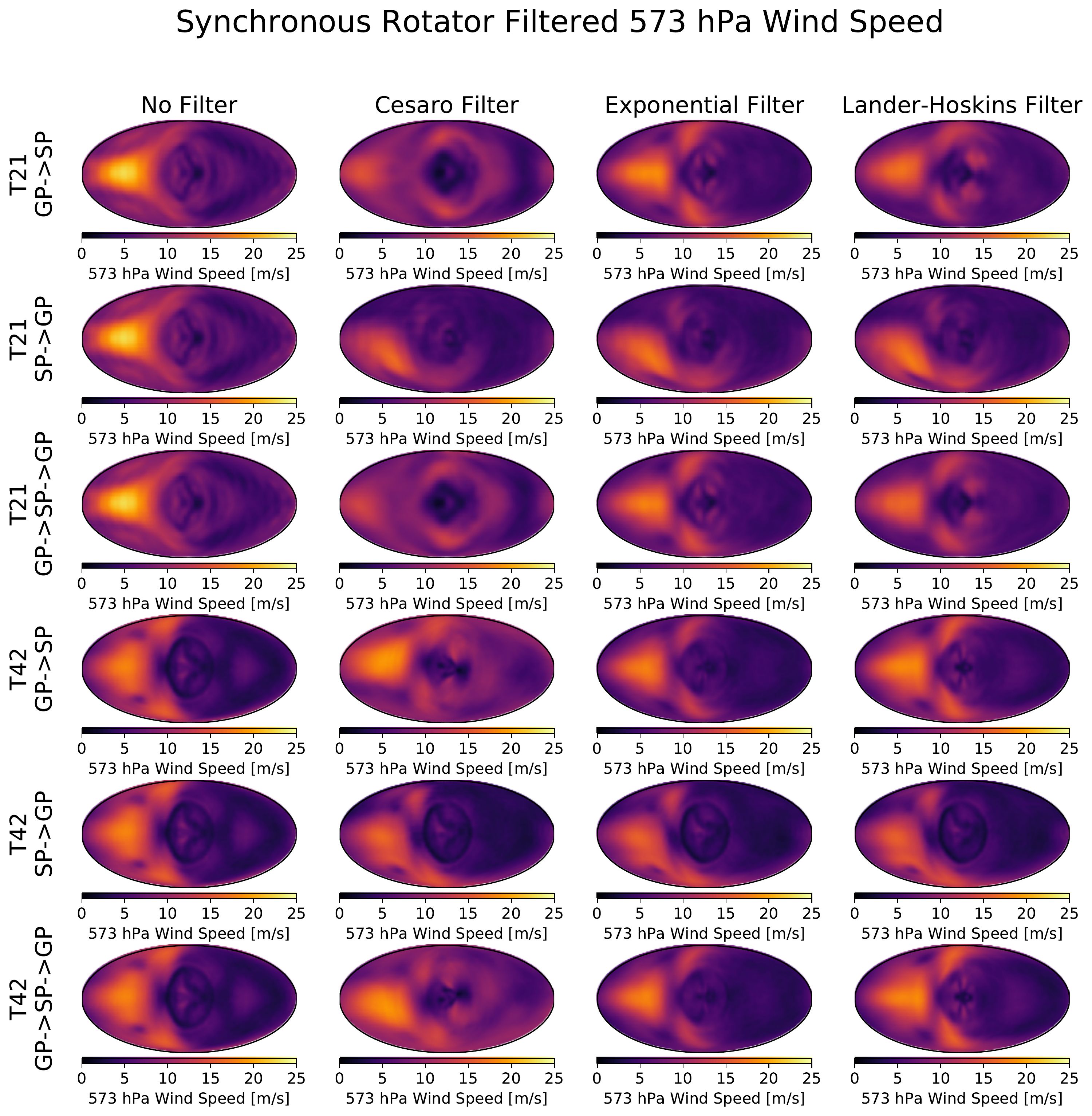}
\end{center}
\caption{Mean annual mid-atmosphere (573 hPa) wind speeds of synchronously-rotating aquaplanets with Earth bulk parameters, 60-day rotation, 3400 K incident spectrum, 1360 W/m$^2$, and no radiative effect from sea ice beyond latent heat flux. The substellar point is centered. Both T21 and T42 resolutions are shown. Three different filters are shown, as well as the case with no filter, with 3 different filter configurations for each filter: `GP$\rightarrow$SP', corresponding to a filter during the transform from the gridpoint domain to the spectral domain, `SP$\rightarrow$GP', corresponding to a filter at the transform from the spectral domain to the gridpoint domain, and `GP$\rightarrow$SP$\rightarrow$GP', indicating a filter at both transforms. Applying either an exponential or Lander-Hoskins filter at both transforms produces minimal North-South asymmetry, and retains small-scale wind structures.}\label{pyfig:benchtl_spd5}
\end{figure*}

\begin{figure*}
\begin{center}
\includegraphics[width=6.5in]{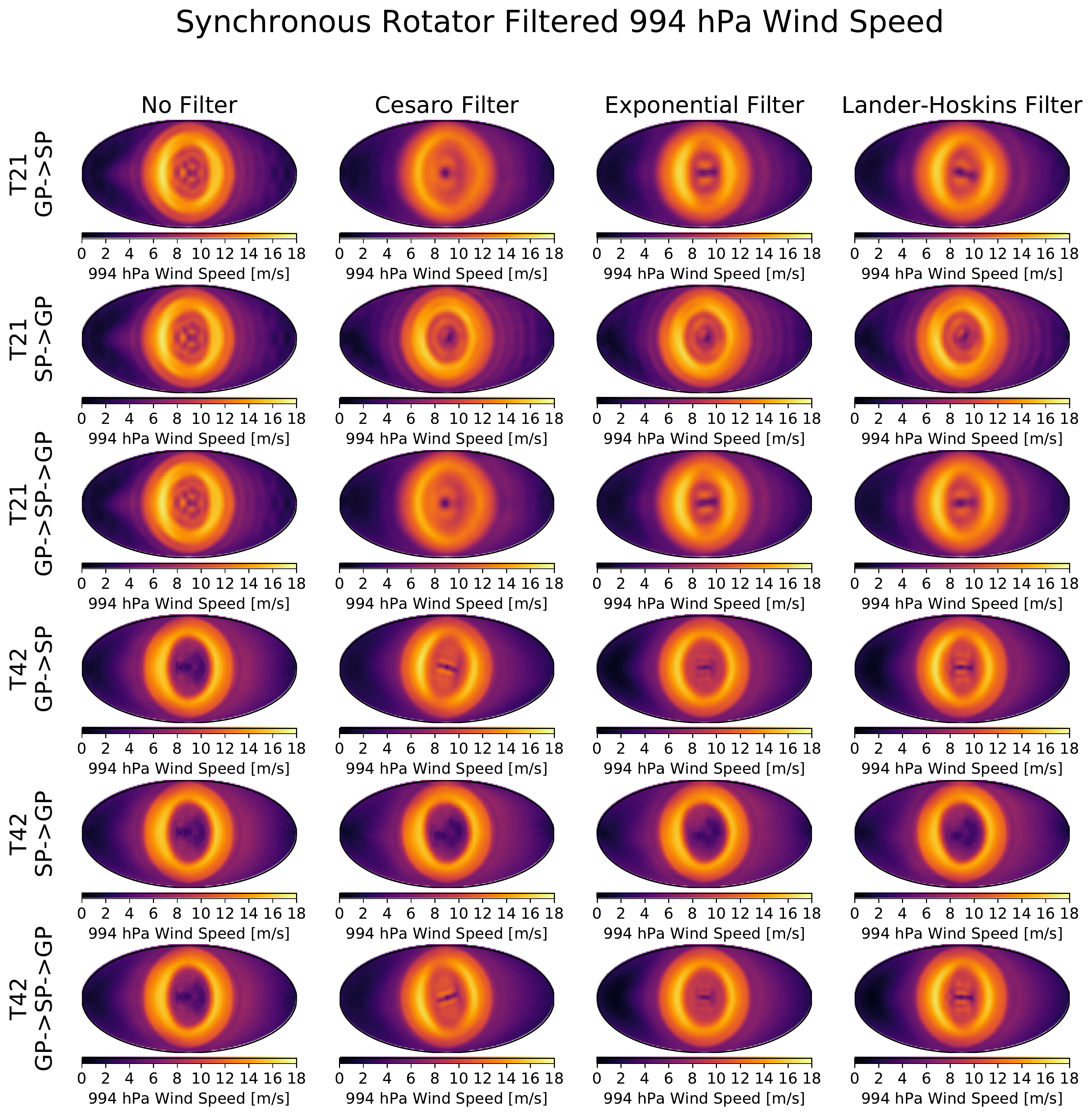}
\end{center}
\caption{Mean annual near-surface (994 hPa) wind speeds of synchronously-rotating aquaplanets with Earth bulk parameters, 60-day rotation, 3400 K incident spectrum, 1360 W/m$^2$, and no radiative effect from sea ice beyond latent heat flux. The substellar point is centered. Both T21 and T42 resolutions are shown. Three different filters are shown, as well as the case with no filter, with 3 different filter configurations for each filter: `GP$\rightarrow$SP', corresponding to a filter during the transform from the gridpoint domain to the spectral domain, `SP$\rightarrow$GP', corresponding to a filter at the transform from the spectral domain to the gridpoint domain, and `GP$\rightarrow$SP$\rightarrow$GP', indicating a filter at both transforms. Applying either an exponential or Lander-Hoskins filter at both transforms produces minimal North-South asymmetry, and retains small-scale wind structures. Notably, surface winds in the substellar upwelling region appear to be higher with a filter than without.}\label{pyfig:benchtl_spd9}
\end{figure*}

\begin{figure*}
\begin{center}
\includegraphics[width=6.5in]{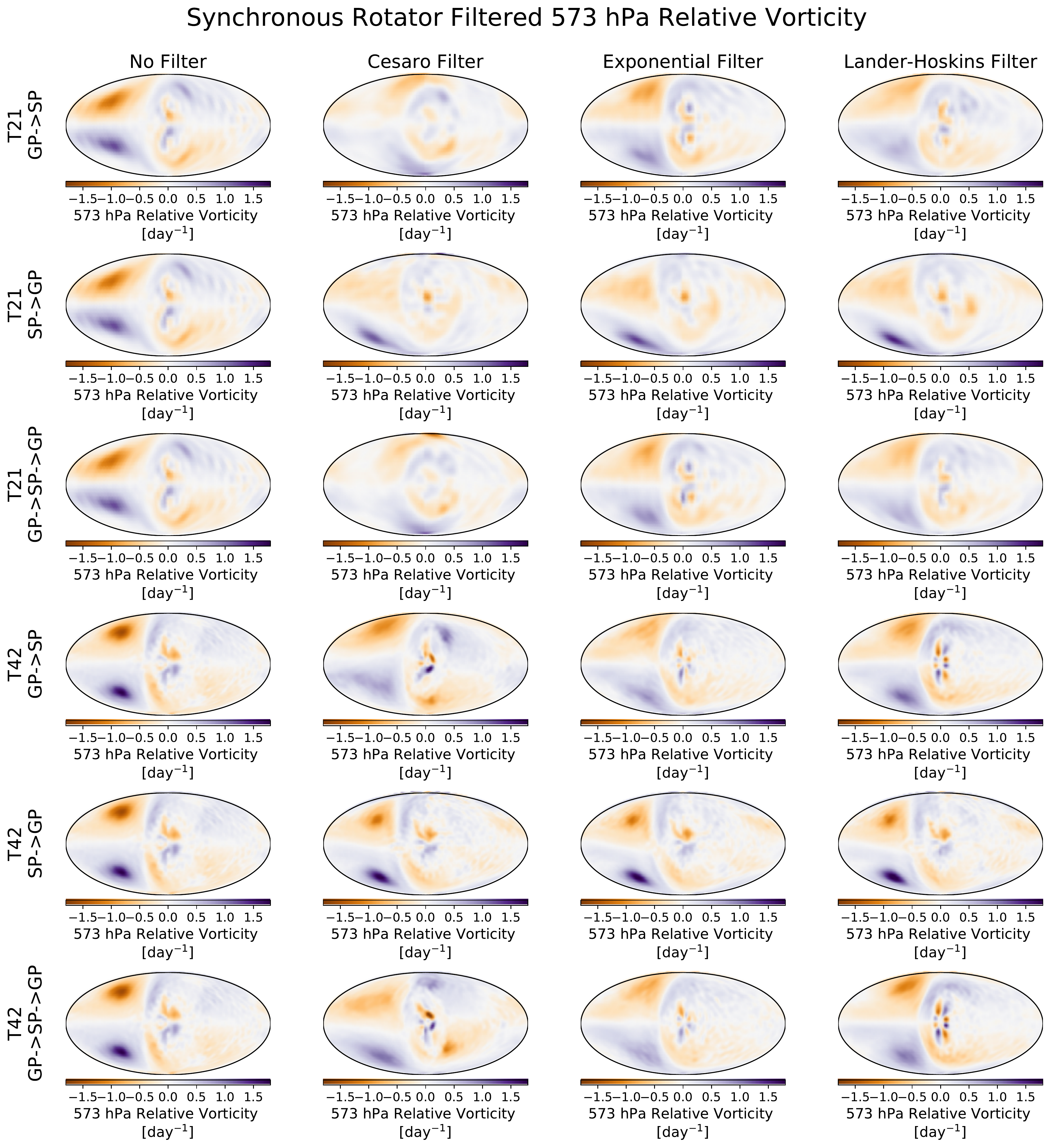}
\end{center}
\caption{Mean annual mid-atmosphere (573 hPa) relative vorticity of synchronously-rotating aquaplanets with Earth bulk parameters, 60-day rotation, 3400 K incident spectrum, 1360 W/m$^2$, and no radiative effect from sea ice beyond latent heat flux. The substellar point is centered. Both T21 and T42 resolutions are shown. Three different filters are shown, as well as the case with no filter, with 3 different filter configurations for each filter: `GP$\rightarrow$SP', corresponding to a filter during the transform from the gridpoint domain to the spectral domain, `SP$\rightarrow$GP', corresponding to a filter at the transform from the spectral domain to the gridpoint domain, and `GP$\rightarrow$SP$\rightarrow$GP', indicating a filter at both transforms. Vorticity structures are sensitive to filter choice, but applying an exponential filter at both transforms appears to give the greatest consistency between resolutions while preserving small-scale structures.}\label{pyfig:benchtl_vort5}
\end{figure*}

\begin{figure*}
\begin{center}
\includegraphics[width=6.5in]{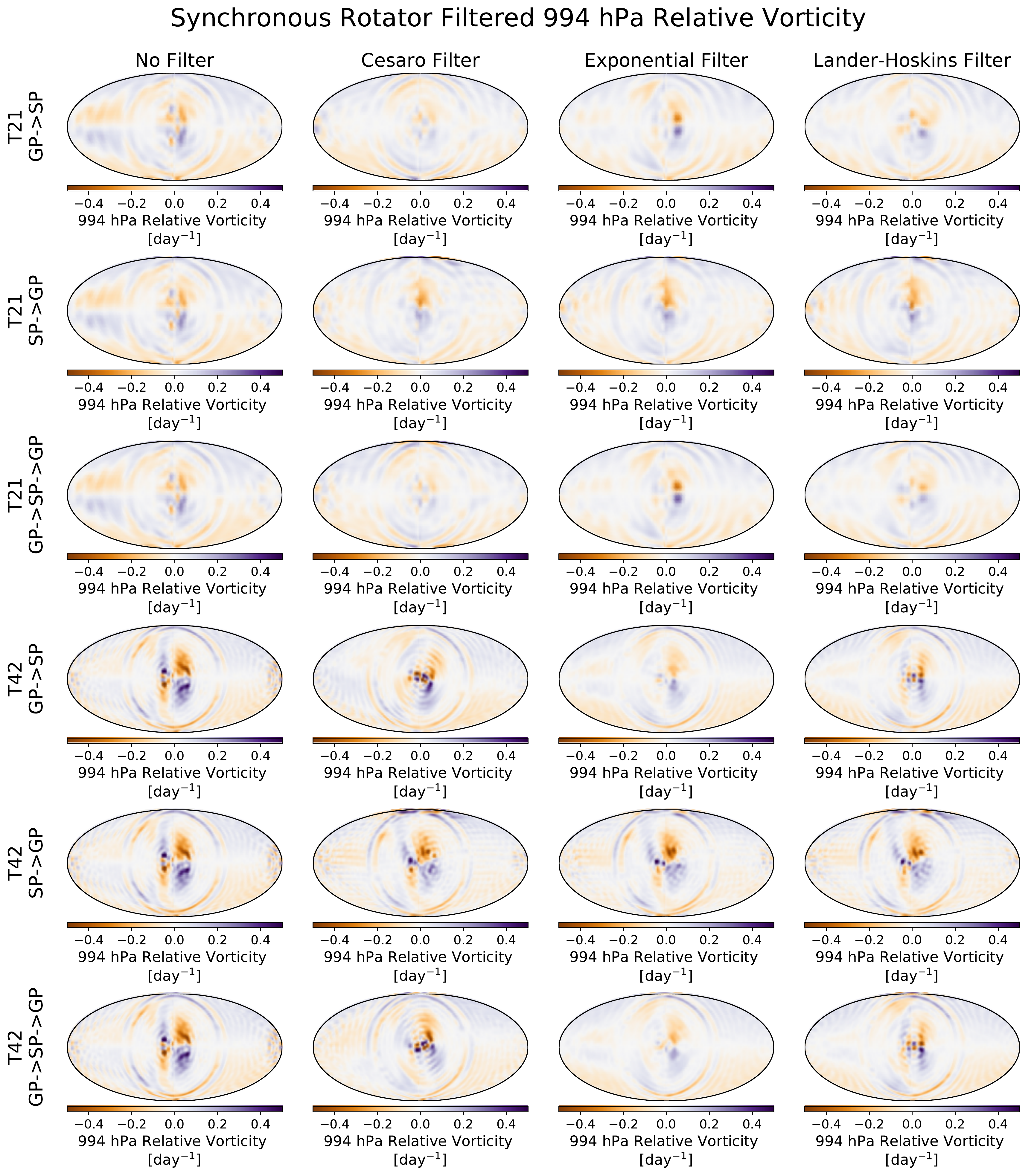}
\end{center}
\caption{Mean annual near-surface (994 hPa) relative vorticity of synchronously-rotating aquaplanets with Earth bulk parameters, 60-day rotation, 3400 K incident spectrum, 1360 W/m$^2$, and no radiative effect from sea ice beyond latent heat flux. The substellar point is centered. Both T21 and T42 resolutions are shown. Three different filters are shown, as well as the case with no filter, with 3 different filter configurations for each filter: `GP$\rightarrow$SP', corresponding to a filter during the transform from the gridpoint domain to the spectral domain, `SP$\rightarrow$GP', corresponding to a filter at the transform from the spectral domain to the gridpoint domain, and `GP$\rightarrow$SP$\rightarrow$GP', indicating a filter at both transforms. Gibbs effects in near-surface vorticity are most-reduced by an exponential filter applied at both transforms.}\label{pyfig:benchtl_vort9}
\end{figure*}

In contrast, the overall performance of the exponential and Lander-Hoskins filters in both YB tests is generally positive when the filter is used at both spectral transforms, as shown in \autoref{pyfig:benchtl_ts} through \autoref{pyfig:benchtl_vort9}, which show surface temperature, cloud maps, integrated vertical liquidity, wind speed, and relative vorticity. While Gibbs ripples are more apparent in the cloud fields even with filtration than in the THAI-B models, their impact on fields such as temperature and water vapor are almost completely removed when the exponential or Lander-Hoskins filters are applied at both transforms. Similarly, atmospheric circulation, as measured by wind speeds and relative vorticity in the mid-atmosphere and near the surface, is less-affected by Gibbs ripples with exponential or Lander-Hoskins filtration at both transforms. The exponential filter appears to be the most effective and most consistent across resolutions, particularly with respect to relative vorticity.


\subsection{Exponential filter tuning}

\begin{figure*}
\begin{center}
\includegraphics[width=6.5in]{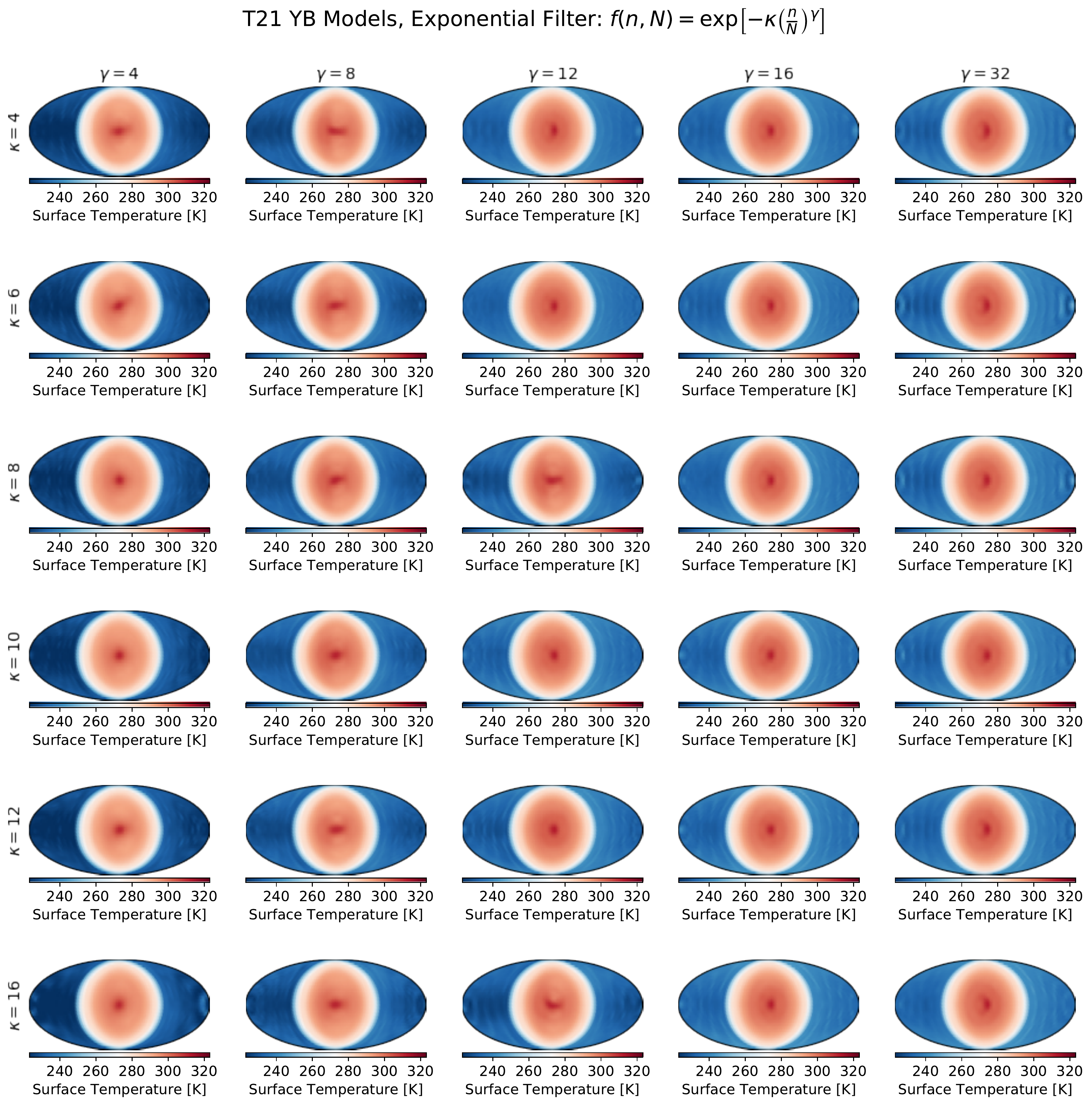}
\end{center}
\caption{Mean annual surface temperature of a grid of synchronously-rotating aquaplanets with exponential filters at both transforms and varying tuning parameters $\kappa$ and $\gamma$. These models have Earth bulk parameters, 60-day rotation, 3400 K incident spectrum, 1360 W/m$^2$, and no radiative effect from sea ice beyond latent heat flux. The substellar point is centered. The models are at T21 resolution, or 32 latitudes and 64 longitudes, corresponding to approximately 5.6$^\circ$ horizontal resolution. The strongest filtration is in the bottom-left corner of the grid. In the upper-right part of the grid, at T21, this filter does not sufficiently remove Gibbs ripples from the temperature field.}\label{pyfig:tune_ts21}
\end{figure*}

\begin{figure*}
\begin{center}
\includegraphics[width=6.5in]{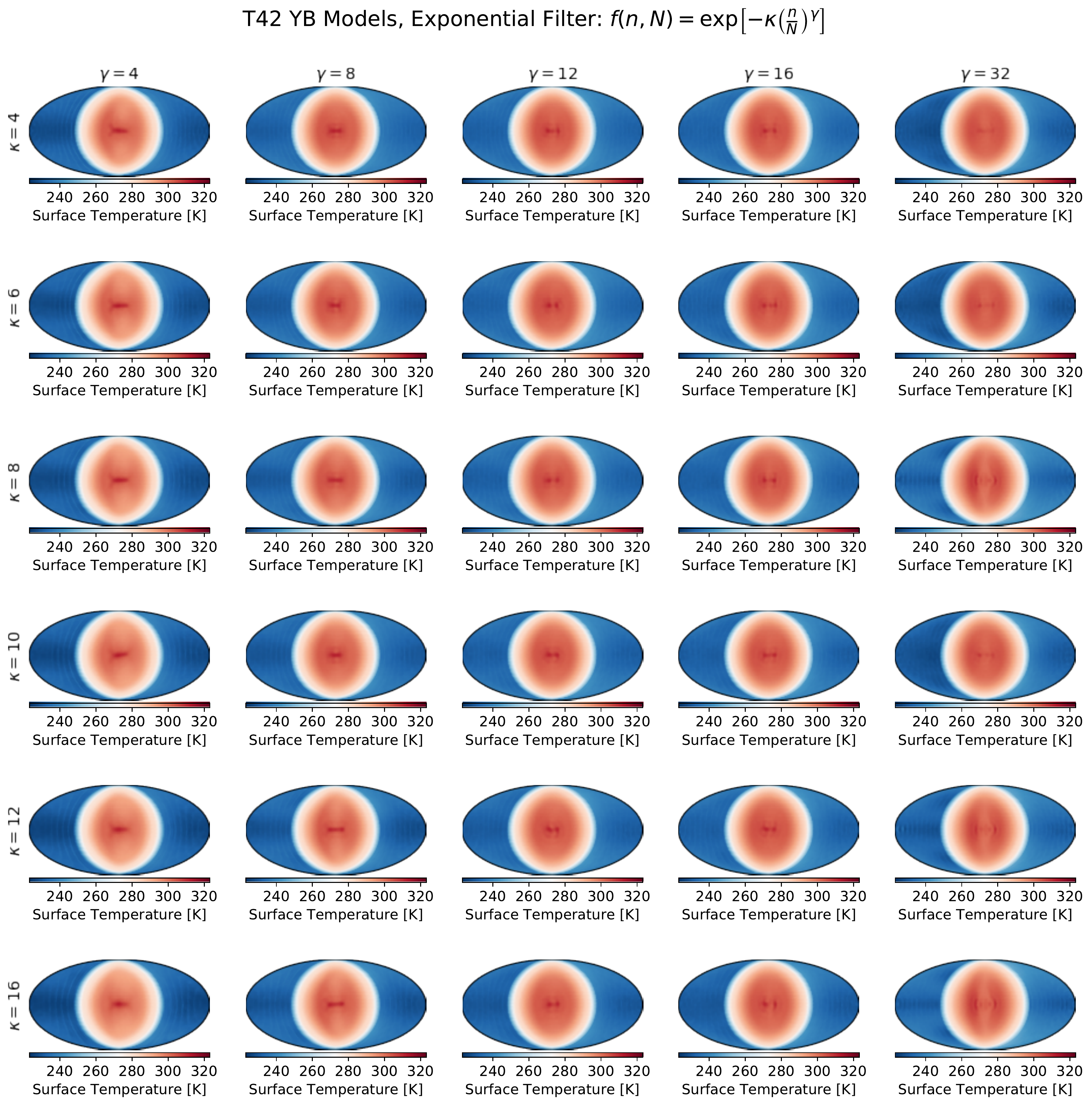}
\end{center}
\caption{Mean annual surface temperature of a grid of synchronously-rotating aquaplanets with exponential filters at both transforms and varying tuning parameters $\kappa$ and $\gamma$. These models have Earth bulk parameters, 60-day rotation, 3400 K incident spectrum, 1360 W/m$^2$, and no radiative effect from sea ice beyond latent heat flux. The substellar point is centered. The models are at T42 resolution, or 64 latitudes and 128 longitudes, corresponding to approximately 2.8$^\circ$ horizontal resolution. The strongest filtration is in the bottom-left corner of the grid. Modestly shallow ($\gamma=8$) and moderately-amplified ($\kappa=8$) filtration provides good agreement between T21 and T42.}\label{pyfig:tune_ts42}
\end{figure*}

\begin{figure*}
\begin{center}
\includegraphics[width=6.5in]{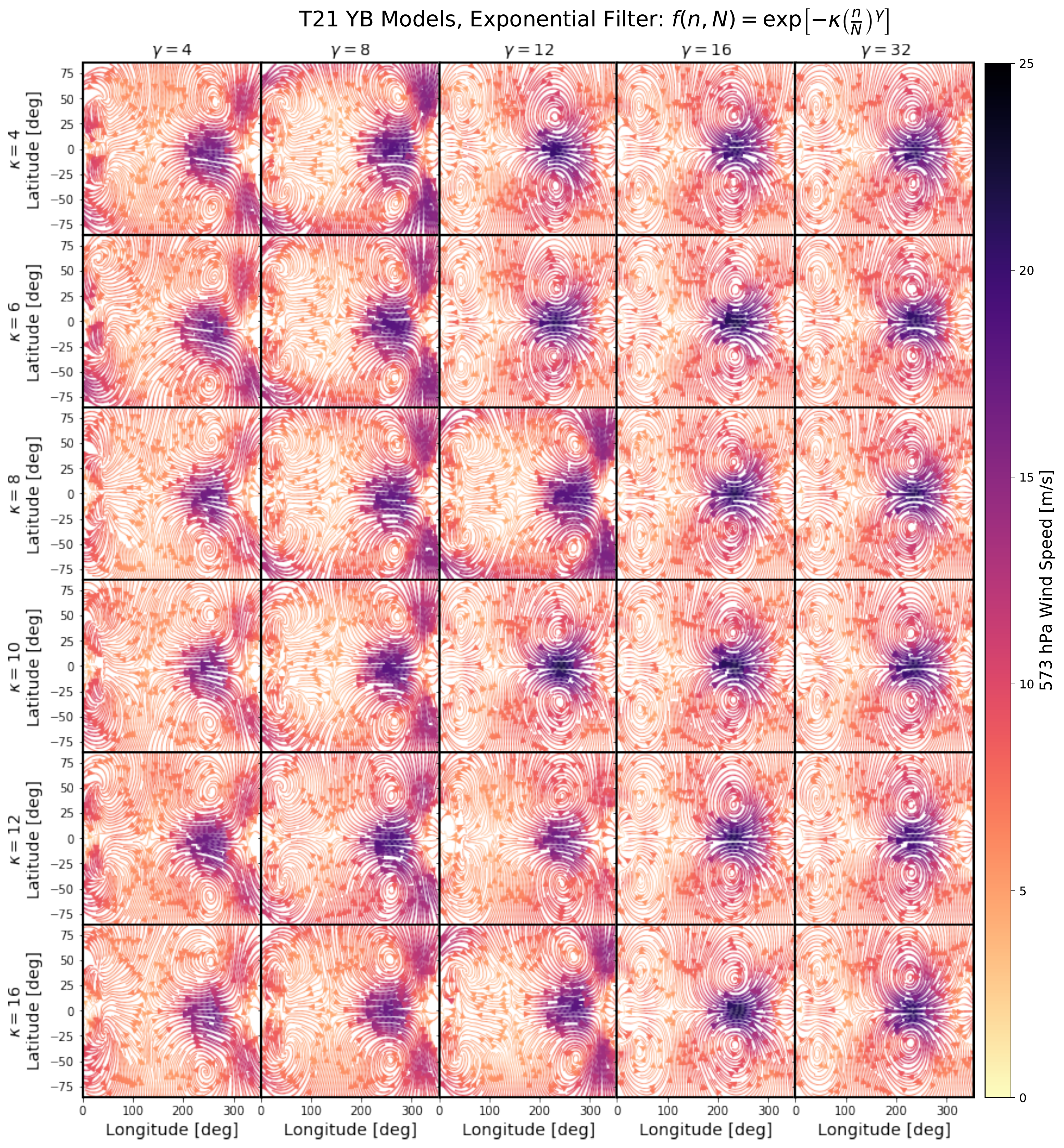}
\end{center}
\caption{Mean annual mid-atmosphere (573 hPa) wind speed and circulation patterns for a grid of synchronously-rotating aquaplanets with exponential filters at both transforms and varying tuning parameters $\kappa$ and $\gamma$. These models have Earth bulk parameters, 60-day rotation, 3400 K incident spectrum, 1360 W/m$^2$, and no radiative effect from sea ice beyond latent heat flux. The antistellar point is centered. The models are at T21 resolution, or 32 latitudes and 64 longitudes, corresponding to approximately 5.6$^\circ$ horizontal resolution. Windspeeds on the eastern edges of the nightside gyres appear faster with stronger filtration (lower-left).}\label{pyfig:tune_wind21}
\end{figure*}

\begin{figure*}
\begin{center}
\includegraphics[width=6.5in]{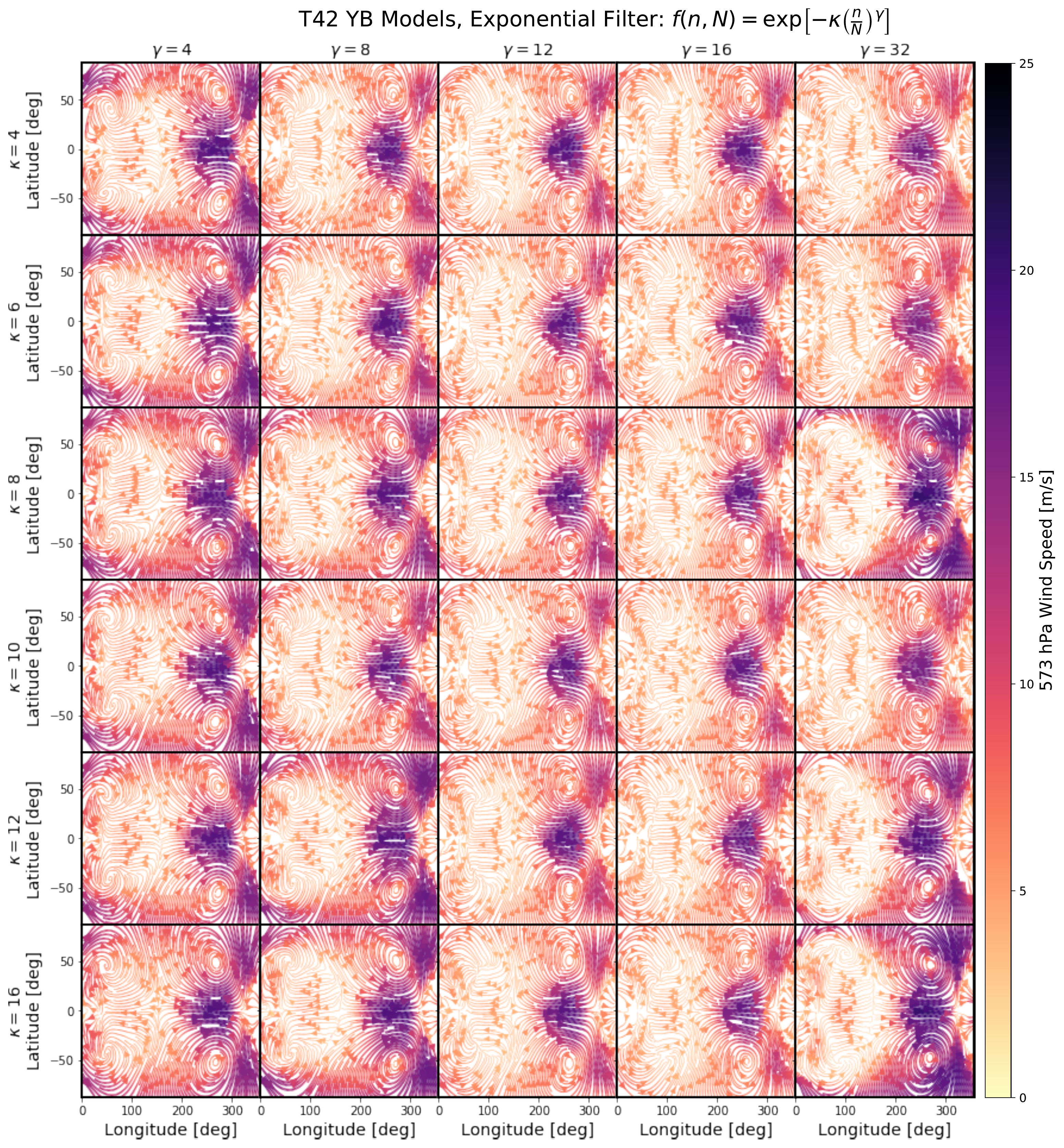}
\end{center}
\caption{Mean annual mid-atmosphere (573 hPa) wind speed and circulation patterns for a grid of synchronously-rotating aquaplanets with exponential filters at both transforms and varying tuning parameters $\kappa$ and $\gamma$. These models have Earth bulk parameters, 60-day rotation, 3400 K incident spectrum, 1360 W/m$^2$, and no radiative effect from sea ice beyond latent heat flux. The antistellar point is centered. The models are at T42 resolution, or 64 latitudes and 128 longitudes, corresponding to approximately 2.8$^\circ$ horizontal resolution. Night-side gyre wind speeds are consistent between resolutions around $\kappa=8$ and $\gamma=8$.}\label{pyfig:tune_wind42}
\end{figure*}

\begin{figure*}
\begin{center}
\includegraphics[width=6.5in]{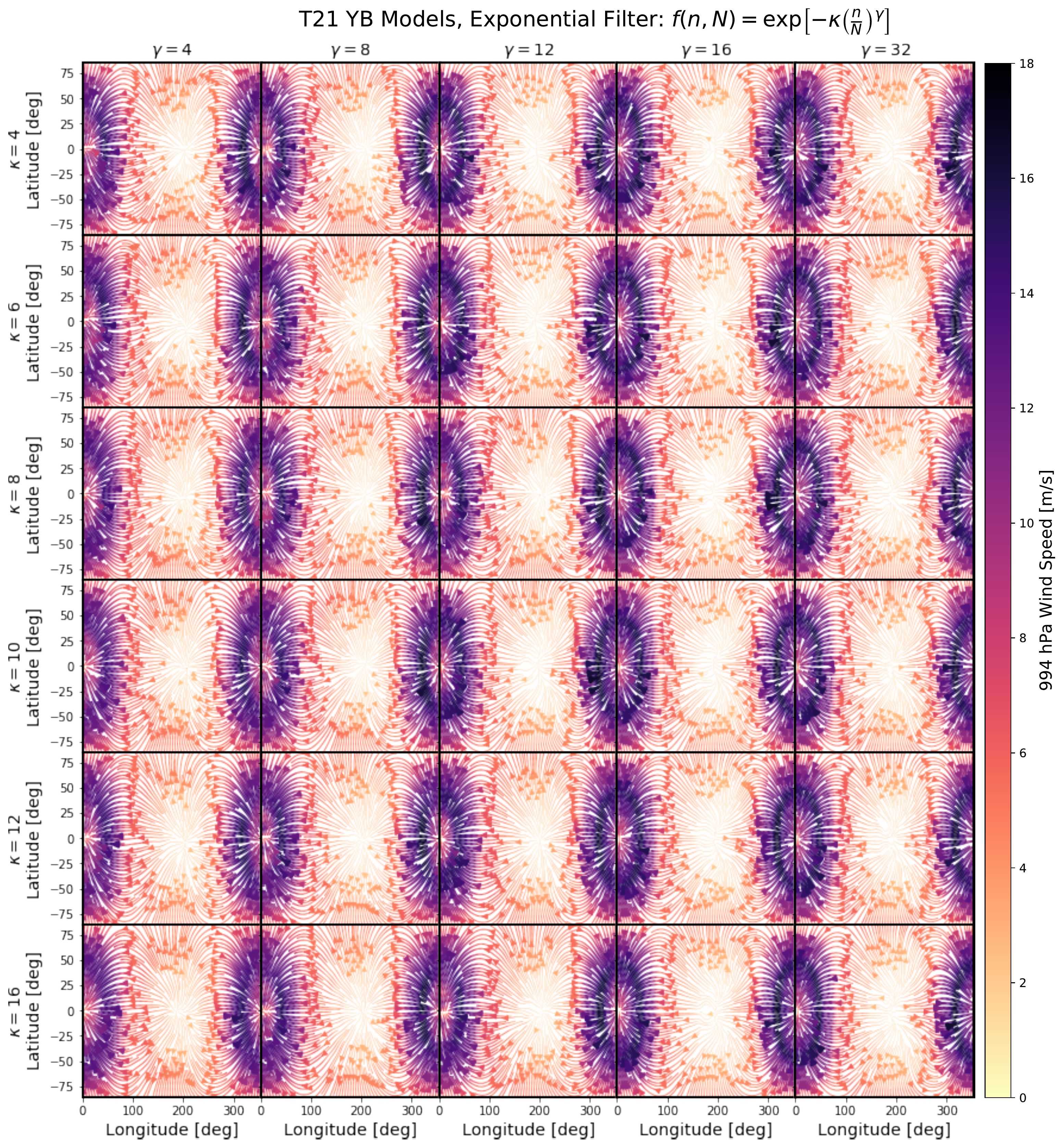}
\end{center}
\caption{Mean annual near-surface (994 hPa) wind speed and circulation patterns for a grid of synchronously-rotating aquaplanets with exponential filters at both transforms and varying tuning parameters $\kappa$ and $\gamma$. These models have Earth bulk parameters, 60-day rotation, 3400 K incident spectrum, 1360 W/m$^2$, and no radiative effect from sea ice beyond latent heat flux. The antistellar point is centered. The models are at T21 resolution, or 32 latitudes and 64 longitudes, corresponding to approximately 5.6$^\circ$ horizontal resolution. Similar to mid-atmosphere winds, strong filtration tends to produce higher wind speeds, here most noticeably in the substellar upwelling region.}\label{pyfig:tune_swind21}
\end{figure*}

\begin{figure*}
\begin{center}
\includegraphics[width=6.5in]{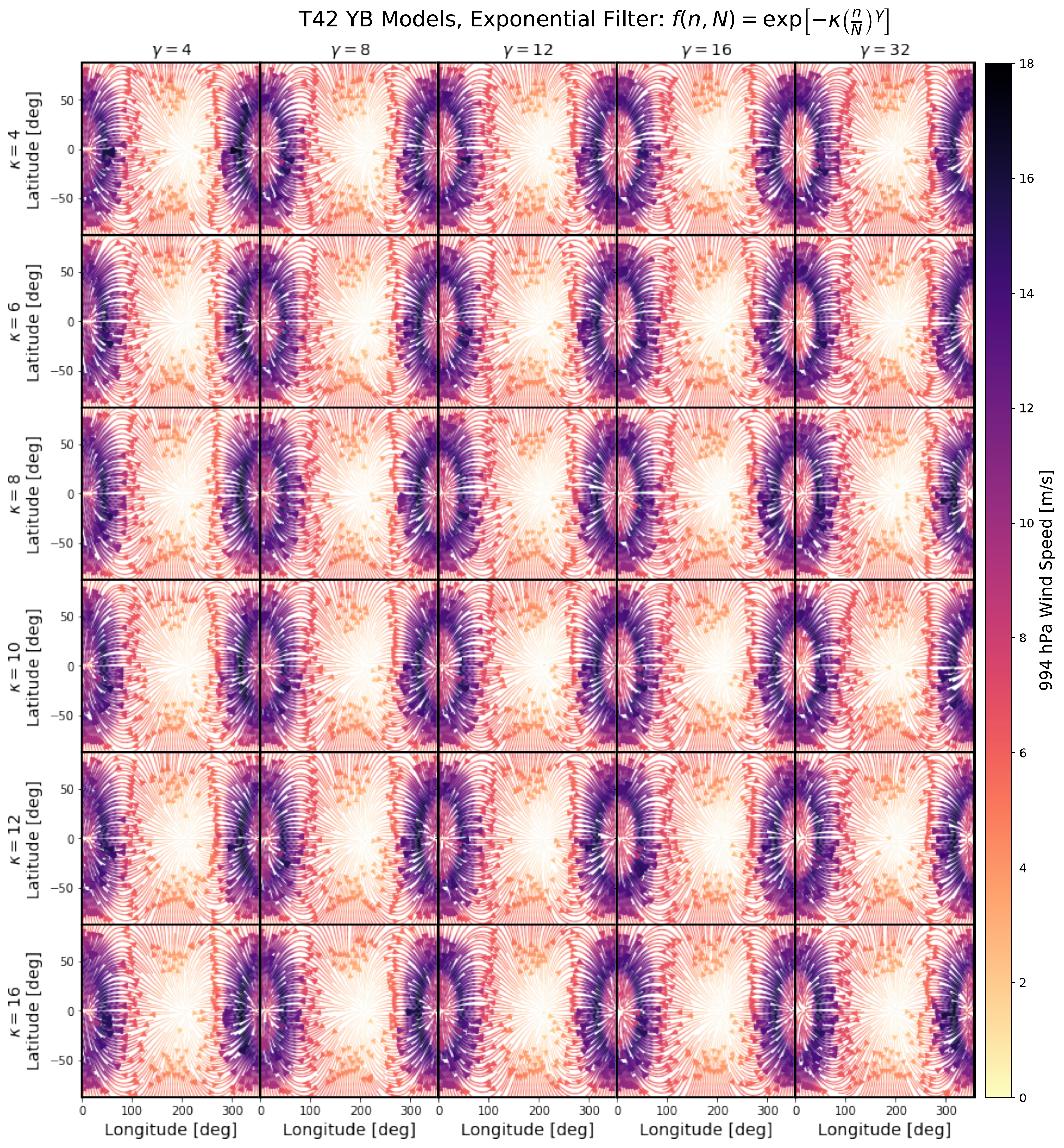}
\end{center}
\caption{Mean annual near-surface (994 hPa) wind speed and circulation patterns for a grid of synchronously-rotating aquaplanets with exponential filters at both transforms and varying tuning parameters $\kappa$ and $\gamma$. These models have Earth bulk parameters, 60-day rotation, 3400 K incident spectrum, 1360 W/m$^2$, and no radiative effect from sea ice beyond latent heat flux. The antistellar point is centered. The models are at T42 resolution, or 64 latitudes and 128 longitudes, corresponding to approximately 2.8$^\circ$ horizontal resolution. As with mid-atmosphere winds, surface winds are reasonably consistent between resolutions around $\kappa=8$ and $\gamma=8$.}\label{pyfig:tune_swind42}
\end{figure*}

\begin{figure*}
\begin{center}
\includegraphics[width=6.5in]{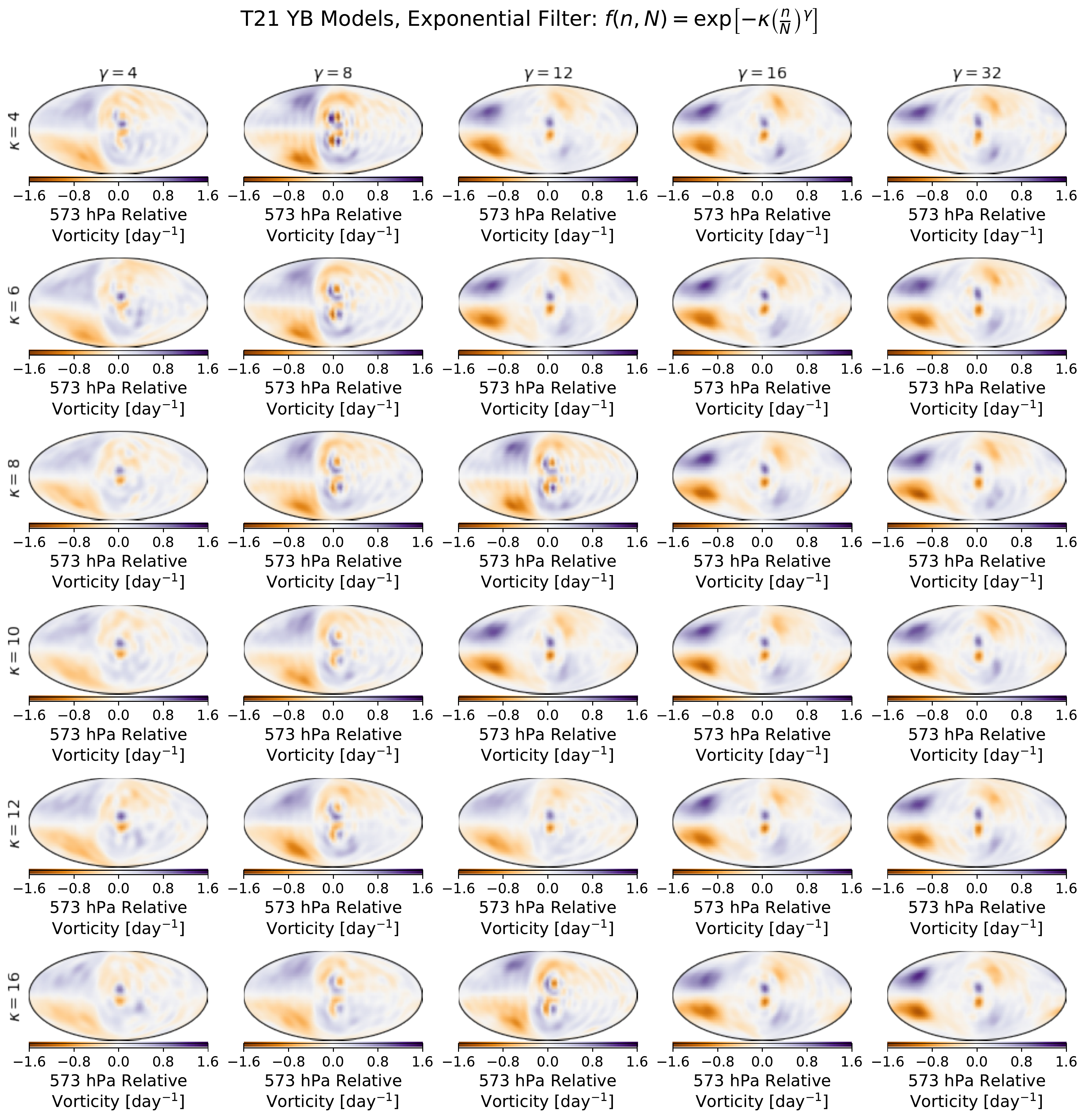}
\end{center}
\caption{Mean annual mid-atmosphere (573 hPa) relative vorticity for a grid of synchronously-rotating aquaplanets with exponential filters at both transforms and varying tuning parameters $\kappa$ and $\gamma$. These models have Earth bulk parameters, 60-day rotation, 3400 K incident spectrum, 1360 W/m$^2$, and no radiative effect from sea ice beyond latent heat flux. The substellar point is centered. The models are at T21 resolution, or 32 latitudes and 64 longitudes, corresponding to approximately 5.6$^\circ$ horizontal resolution. Ripples are not fully-removed with weak filtration (upper-right), while small-scale features are damped out with strong filtration (lower-left). $\gamma=8$ filtration appears to preserve structural complexity at the substellar point.}\label{pyfig:tune_5vort21}
\end{figure*}

\begin{figure*}
\begin{center}
\includegraphics[width=6.5in]{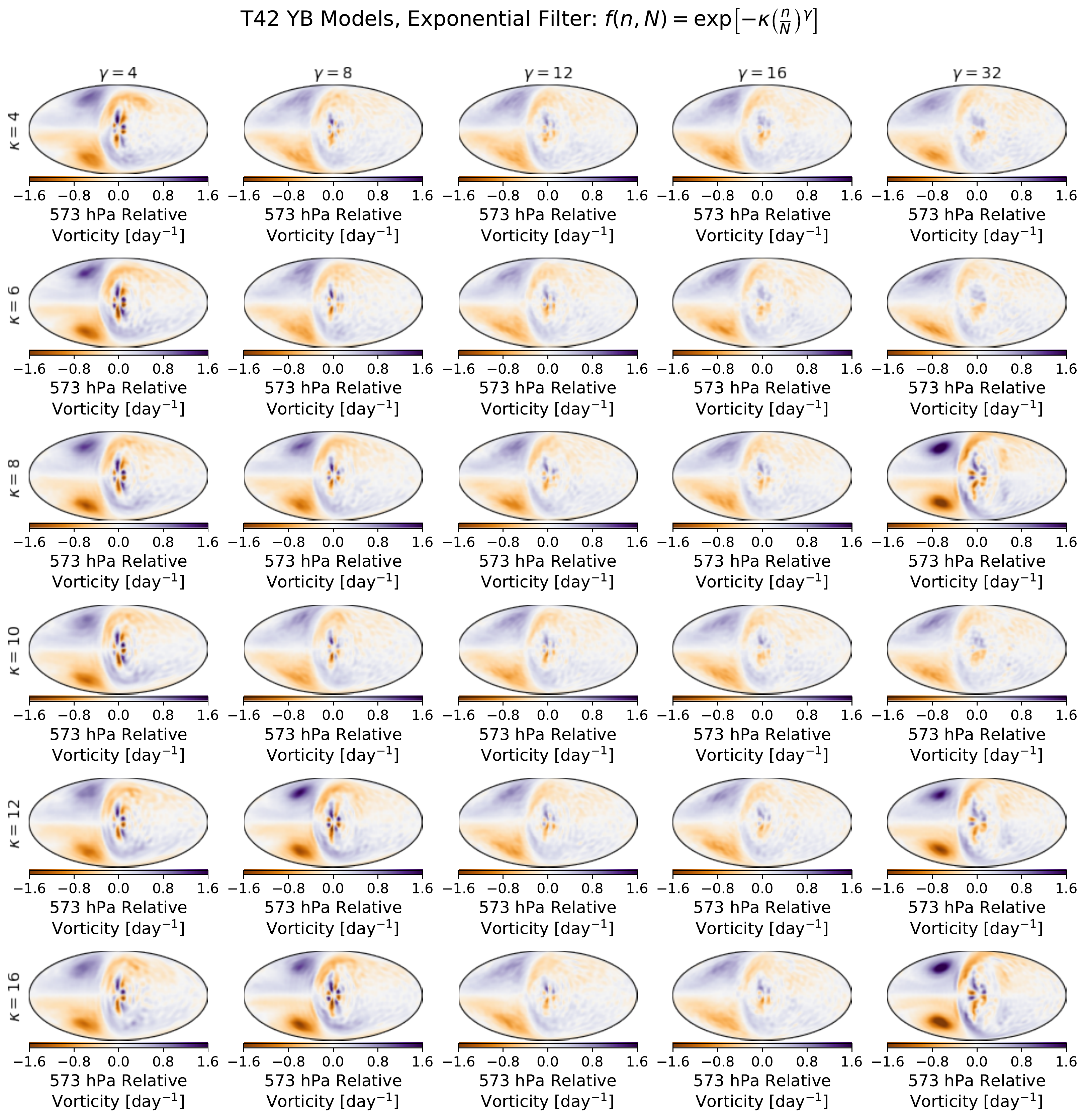}
\end{center}
\caption{Mean annual mid-atmosphere (573 hPa) relative vorticity for a grid of synchronously-rotating aquaplanets with exponential filters at both transforms and varying tuning parameters $\kappa$ and $\gamma$. These models have Earth bulk parameters, 60-day rotation, 3400 K incident spectrum, 1360 W/m$^2$, and no radiative effect from sea ice beyond latent heat flux. The substellar point is centered. The models are at T42 resolution, or 64 latitudes and 128 longitudes, corresponding to approximately 2.8$^\circ$ horizontal resolution. Small-scale structures and gyre size and strength are sensitive to filter tuning, but are most consistent between resolutions at $\kappa=8$ and $\gamma=8$.}\label{pyfig:tune_5vort42}
\end{figure*}

\begin{figure*}
\begin{center}
\includegraphics[width=6.5in]{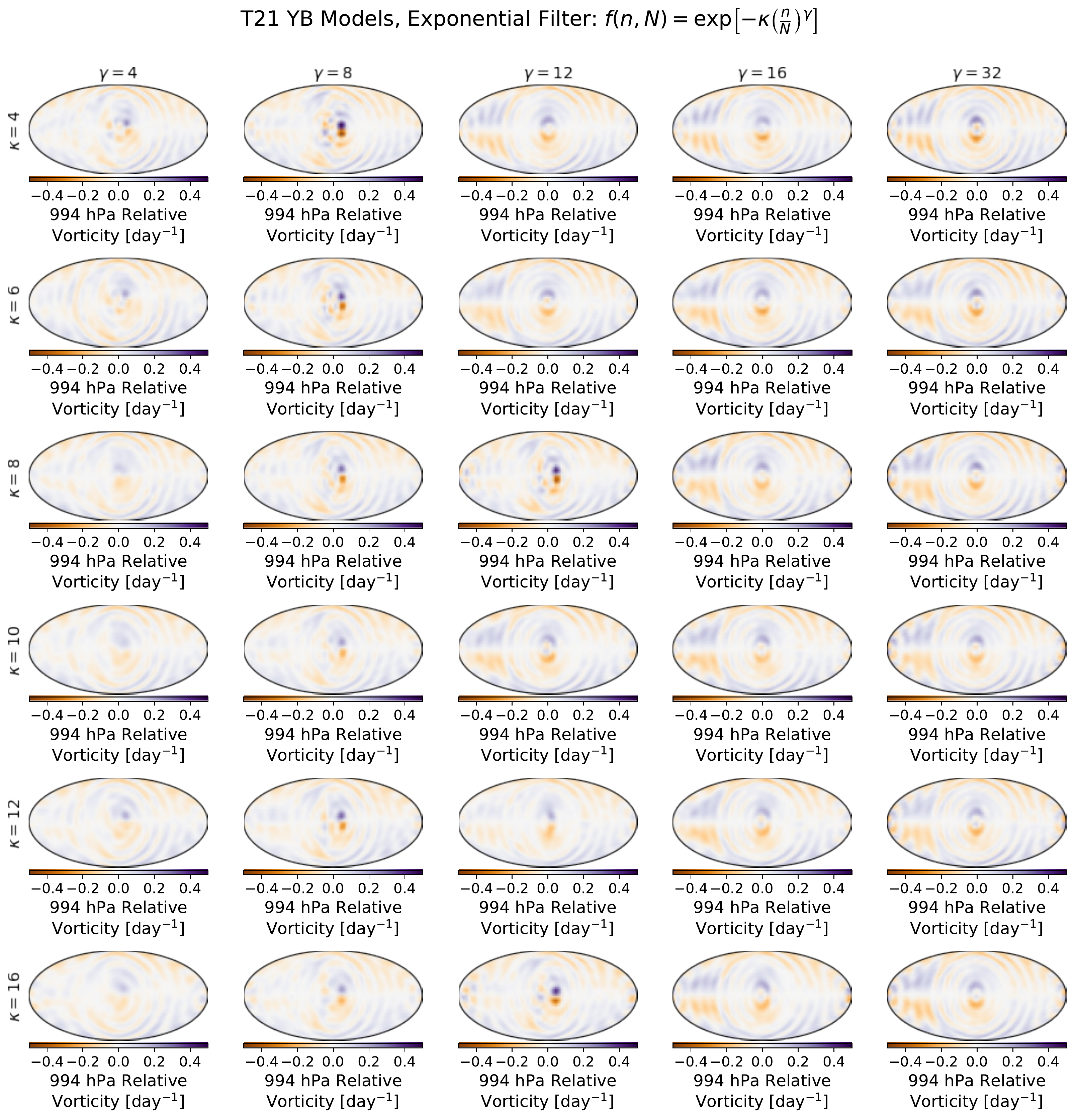}
\end{center}
\caption{Mean annual near-surface (994 hPa) relative vorticity for a grid of synchronously-rotating aquaplanets with exponential filters at both transforms and varying tuning parameters $\kappa$ and $\gamma$. These models have Earth bulk parameters, 60-day rotation, 3400 K incident spectrum, 1360 W/m$^2$, and no radiative effect from sea ice beyond latent heat flux. The substellar point is centered. The models are at T21 resolution, or 32 latitudes and 64 longitudes, corresponding to approximately 5.6$^\circ$ horizontal resolution. Ripples are not fully-removed at any tuning, but the substellar upwelling vortices are masked by Gibbs ripples with weak filtration (upper right), and damped out by strong filtration (lower-left). They do appear however with $\gamma=8$.}\label{pyfig:tune_9vort21}
\end{figure*}

\begin{figure*}
\begin{center}
\includegraphics[width=6.5in]{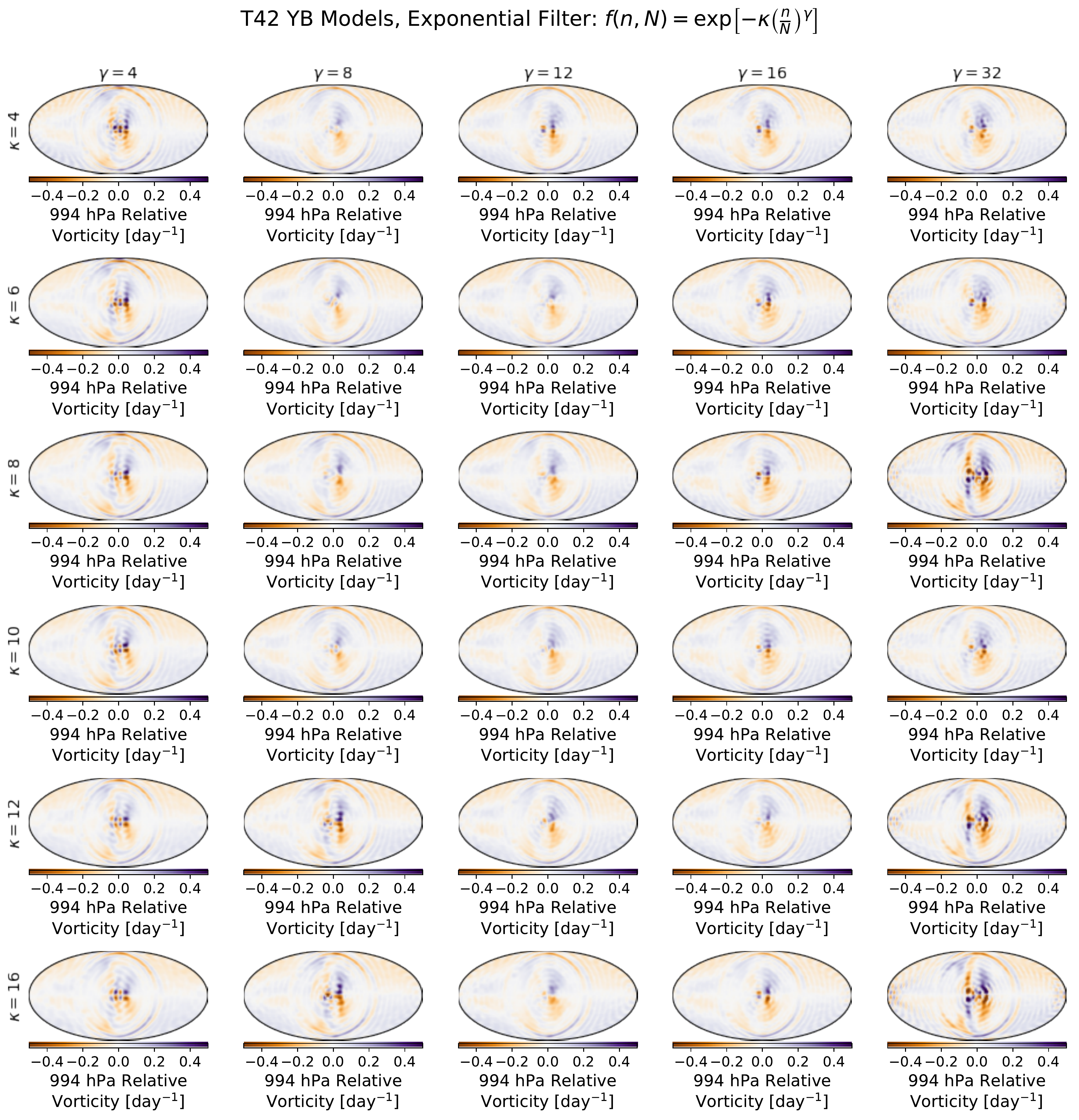}
\end{center}
\caption{Mean annual near-surface (994 hPa) relative vorticity for a grid of synchronously-rotating aquaplanets with exponential filters at both transforms and varying tuning parameters $\kappa$ and $\gamma$. These models have Earth bulk parameters, 60-day rotation, 3400 K incident spectrum, 1360 W/m$^2$, and no radiative effect from sea ice beyond latent heat flux. The substellar point is centered. The models are at T42 resolution, or 64 latitudes and 128 longitudes, corresponding to approximately 2.8$^\circ$ horizontal resolution. Upwelling vortices are well-resolved without ripples primarily at moderate filtration strengths, and are fairly consistent between resolutions at $\kappa=8$ and $\gamma=8$.}\label{pyfig:tune_9vort42}
\label{lastpage}
\end{figure*}

While the exponential filter appears to perform well in preliminary tests alongside the other filter types, it has two free parameters which can be used to adjust both the severity of the damping (via the function's steepness) and its overall strength (via amplification of the damping at every wavenumber). To identify optimal tunings for the exponential filter, we ran a grid of models at both T21 and T42 resolutions, varying $\kappa$ and $\gamma$. The weakest filters have high $\gamma$, implying a very sharp dropoff at high wavenumbers, and low $\kappa$, meaning low amplification of the damping function at each wavenumber. These filters concentrate most of the damping power only over the largest wavenumbers. Conversely, the strongest filters have shallow filter functions due to low $\gamma$, and high amplification from large $\kappa$, meaning the damping power is spread out over a wide range of wavenumbers.  The results are shown in \autoref{pyfig:tune_ts21} through \autoref{pyfig:tune_9vort42}. 

The strength of the exponential filter is reflected in the surface temperature most-obviously through nightside temperatures (\autoref{pyfig:tune_ts21} and \autoref{pyfig:tune_ts42}). The stronger the filter, the colder the nightside appears to be. This behavior is stronger at T21 than at T42, indicating that it is a numerical artifact of the filter, not a physical response that is emerging. With only weak filtration, however, Gibbs ripples are not fully-removed from the temperature field. Moderate filter strengths are therefore optimal.

Similar dependence on filter strength is found in atmospheric circulation, as measured by wind speed, circulation pattern, and relative vorticity at both the mid-atmosphere (573 hPa in our output) and near the surface (994 hPa). Wind speed and circulation patterns are shown in \autoref{pyfig:tune_wind21} through \autoref{pyfig:tune_swind42}. We generally find that wind speeds are higher with stronger filtration, seen most-clearly in the mid-atmosphere through the wind speeds of the eastern sides of the nightside gyres, and near the surface through the wind speeds of the substellar upwelling region. This trend however is also resolution-dependent, with a larger effect at T42. The combination of $\kappa=8$ and $\gamma=8$ appears to minimize the discrepancy between resolutions. 

The trend in gyre wind speeds is reflected in the mid-atmosphere relative vorticity, in which the strength of the nightside high-latitude vortices is sensitive to the filter strength, as shown in \autoref{pyfig:tune_5vort21} and \autoref{pyfig:tune_5vort42}. Similarly to wind speed and distribution, $\kappa=8$ and $\gamma=8$ gives good agreement between resolutions. In addition to the nightside gyres, however, the mid-atmosphere vorticity also probes the structure of the upwelling column at the substellar point. Modest filtration at T21 reveals multiple counterrotating columns; with weak filtrations, these are washed out by Gibbs ripples, and with strong filtration, they are damped out by the filter. Moderate to strong filtration preserves the existence of multiple counterrotating columns at T42, and the distribution and strength is most consistent with T21 at $\kappa=8$ and $\gamma=8$.

Near the surface, filter strength primarily affects whether the two dominant upwelling vortices (on either side of the equator) are resolved, as shown in \autoref{pyfig:tune_9vort21} and \autoref{pyfig:tune_9vort42}. The Gibbs ripples are not compelely damped at any filter strength at T21, and are slightly discernible even in the best cases at T42. At moderate filter strengths and T21 resolution, two strong substellar vortices emerge, but with weak filtration, they are replaced by Gibbs ripples, and with strong filtration, they are damped out and unresolved. At T42, Gibbs effects can become noticeable with weak filtration, and seem to reappear with strong filtration, such that Gibbs-like effects are only nearly-completely removed with moderate-strength filtration. This may be because the structures being resolved are near the filtration length scale, and if allowed to grow, are able to further-smooth Gibbs ripples in the atmosphere. If the filtration is too strong, however, these features are not allowed to grow, and the remaining Gibbs ripples come to dominate the surface-level dynamics. 

Overall, moderate filtration such as that provided by $\kappa=8$ and $\gamma=8$ appears to provide the best and most consistent performance across T21 and T42 resolutions. The optimal tuning may however be scenario-dependent, and users are encouraged to run preliminary tests to identify the best tuning parameters before beginning large experiments. For quick preliminary surveys of synchronously-rotating planetary climates, however, we recommend $\kappa=8$ and $\gamma=8$ as reasonable defaults.


\bibliographystyle{mnras}








\bsp	
\label{lastpage}
\end{document}